\documentclass{w-edbk}

\usepackage{w-edbkps}

\usepackage{graphicx} % WW: standard LaTeX graphics tool
\newcommand{\VT}[1]{\ensuremath{{V_{T#1}}}}
\newbox\sectsavebox
\setbox\sectsavebox=\hbox{\VT{xyz}}
\begin{document}
\subtitle{Classical and quantum magnetization 
          reversal studied in nanometer-sized\\ 
          particles and clusters}
\editionstatement{}
\editor{to be published in: Advances in Chemical Physics \\
invitation from Stuart A. Rice \\
Second version of 31 Dec. 2000 \\
   \\
   \\
   {\bf Author}
   \\
   \\
   Wolfgang Wernsdorfer \\
   Lab. L. N\'eel - CNRS, BP166, \\
   38042 Grenoble Cedex 9, France, \\
   e-mail : wernsdor@labs.polycnrs-gre.fr}
\titlepage
\tableofcontents
\introduction
\label{intro}

Since the late 1940s, nanometer-sized magnetic 
particles have generated continuous 
interest as the study of their properties has 
proved to be scientifically and technologically 
very challenging. In particular it was recognized 
that the ferromagnetic state, with a given 
orientation of the particle moment, has a remanent 
magnetization if the particle is small 
enough. This was the starting point of huge 
permanent magnets and magnetic recording 
industries. However, despite intense activity 
during the last few decades, the difficulties 
in making nanoparticles of good enough quality 
has slowed the advancement of this field. 
As a consequence, for 50 years, these applications 
concentrated above and then near the 
micrometer scale. In the last decade, this 
has no longer been the case because of the 
emergence of new fabrication techniques which 
have led to the possibility of making 
small objects with the required structural 
and chemical qualities. In order to study these 
objects new techniques were developed 
such as magnetic force microscopy, 
magnetometry based on micro-Hall probes or
micro-SQUIDs. This led to a new 
understanding of the magnetic behavior of nanoparticles, 
which is now very important for 
the development of new fundamental theories 
of magnetism and in modeling new 
magnetic materials for permanent magnets 
or high density recording.

\begin{figure}
\begin{center}\leavevmode
\includegraphics[width=1.0\textwidth]{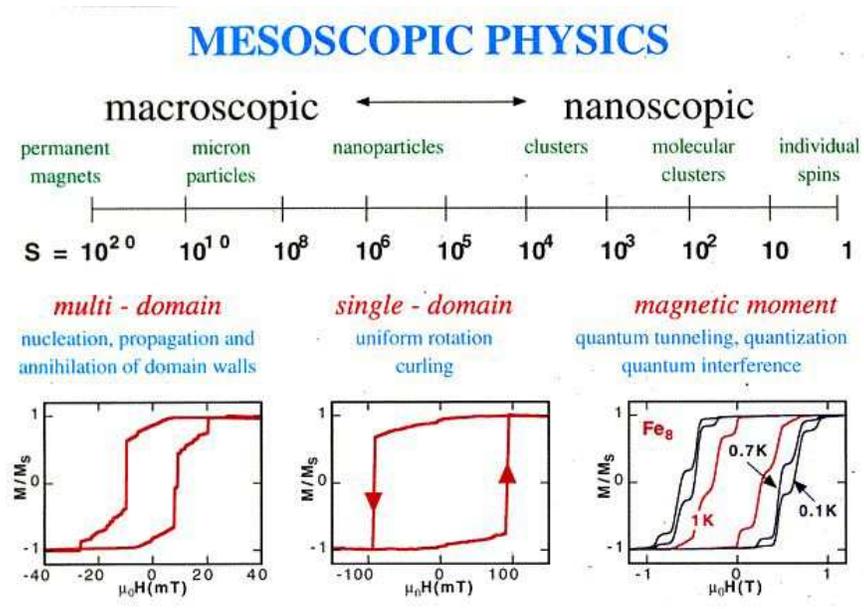}
\caption{Scale of size which goes from macroscopic 
down to nanoscopic sizes. 
The unit of this scale is the number of magnetic 
moments in a magnetic system (roughly corresponding 
to the number of atoms). The hysteresis loops are 
typical examples of magnetization reversal via 
nucleation, propagation and annihilation of domain 
walls ({\it left}), via uniform rotation ({\it middle}), 
and quantum tunneling ({\it right}).}
\label{fig_scale}
\end{center}
\end{figure}

In order to put this review into perspective, let us consider 
Fig.~\ref{fig_scale} which presents a scale 
of size ranging from macroscopic down to 
nanoscopic sizes. The unit of this scale is the 
number of magnetic moments in a magnetic 
system. At macroscopic sizes, a magnetic system
is described by magnetic domains (Weiss 1907)~\cite{Hubert98} 
which are separated by domain walls. 
magnetization reversal occurs via nucleation, 
propagation and annihilation of domain walls 
(see the hysteresis loop on the left in Fig.~\ref{fig_scale} 
which was measured on an individual elliptic CoZr particle 
of 1~$\mu$m~$\times$ 0.8~$\mu$m, and a thickness of 50~nm). 
Shape and width of domain walls depend on the material 
of the magnetic system, on its size, shape and surface, 
and on its temperature~\cite{Aharoni96}. 
The material  dependence of the domain walls has 
motivated the definition of two length 
scales: (i)~the domain wall width $\delta$ defined by 
$\delta = \sqrt{A / K}$ and (ii)~the exchange 
length $\lambda$ defined by 
$\lambda = \sqrt{A} / M_{\rm S}$ 
where $A$ is the exchange energy, $K$ is the 
crystalline anisotropy constant and $M_{\rm S}$ 
is the spontaneous magnetization. Qualitatively, 
the first definition shows that anisotropy energy favors 
a thin wall, while the exchange energy favors a 
thick wall. For very small crystalline 
anisotropy, the first definition suggests an 
infinite domain wall width which has a large 
total energy. This is due to the magnetostatic 
energy term which can be reduced by 
subdividing the ferromagnetic crystal into domains. 
Therefore, for very small crystalline 
anisotropy, the domain wall width is of the 
order of magnitude of the exchange length 
$\lambda$. Both length scales can range from 
sub-micrometer scales in alloys to atomic 
scales in rare earth systems.

When the system size is of the order of magnitude 
of $\delta$ or $\lambda$, the 
formation of domain walls requires too much energy.
Therefore, the magnetization remains in 
the so-called single domain state. 
\footnote{In the theory of micromagnetism,
the single domain state describes the state where
the magnetization is perfectly aligned~\cite{Aharoni96}
whereas experimentalists mean often a state without
domain wall.}
Hence, the magnetization might reverse by 
uniform rotation, curling or other
non uniform modes (see hysteresis loop 
in the middle of Fig.~\ref{fig_scale}). 
In this review, we discuss mainly this size range where the 
physics is rather simple (Sects.~\ref{chap_reversal_T_0} 
and~\ref{chap_reversal_clas}).

For system sizes well below $\delta$ and $\lambda$, 
one must take into account explicitly the magnetic moments
(spins) and their couplings. The theoretical 
description is complicated by 
the particle's boundaries~\cite{Freemann91,Pastor95,Bertsch99}.

At the smallest size (below which one must 
consider individual atoms and spins) there are 
either free clusters made of several 
atoms~\cite{Billas97,Bloomfield96} 
or molecular clusters which are macro-
molecules with a central complex 
containing magnetic atoms. In the last 
case, measurements on the Mn$_{12}$ acetate 
and Fe$_8$ molecular 
clusters showed that the 
physics can be described by a collective moment 
of spin $S$~= 10 (Sect.~\ref{mol_cluster}). 
By means of simple 
hysteresis loop measurements, the quantum 
character of these molecules showed up in 
well defined steps which are due to resonance 
quantum tunneling between energy 
levels (see hysteresis loop on the right 
in Fig.~\ref{fig_scale}). 

In the following sections, we review the most 
important theories and experimental results 
concerning the magnetization reversal of 
single-domain particles and clusters. 
Special emphasis is laid 
on single particle measurements avoiding 
complications due to distributions of particle 
size, shape etc. Measurements on particle 
assemblies has been reviewed in~\cite{Dormann97}. 
We mainly discuss the low temperature 
regime in order to avoid spin excitations.

In Sect.~\ref{chap_mes}, we briefly review the commonly 
used measuring techniques. Among them, 
electrical transport measurements, Hall probes and 
micro-SQUID techniques seem to be the 
most convenient techniques for low temperature measurements. 
Sect.~\ref{chap_reversal_T_0} discusses the 
mechanisms of magnetization reversal in single domain 
particles at zero kelvin. 
The influence of temperature on the magnetization reversal 
is reported in Sect.~\ref{chap_reversal_clas}. Finally, 
Sect.~\ref{chap_MQT} shows that for very small systems or 
very low temperature, magnetization can 
reverse via tunneling.

%%%%%%%%%%%%%%%%%%%%%%%%%%%%%%%%
\title[Single particle measurement techniques]
{Single particle measurement techniques}
\label{chap_mes}
%%%%%%%%%%%%%%%%%%%%%%%%%%%%%%%%

The following sections review commonly used 
single particle measuring techniques
avoiding complications due to distributions 
of particle size, shape etc.
which are always present in particle 
assemblies~\cite{Dormann97}.
Special emphasis is laid on the 
micro-SQUID technique and the developed methods
which allowed the most detailed 
studies at low temperatures.

%%%%%%%%%%%%%%%%%%%%%%%%%%%%%%%%
\section{Overview of single particle measurement techniques}
\label{mes_techniques}
%%%%%%%%%%%%%%%%%%%%%%%%%%%%%%%%

The dream of measuring the magnetization reversal of 
an individual magnetic particle goes 
back to the pioneering work of N\'eel~\cite{Neel49a,Neel49b}. 
The first realization was published by 
Morrish and Yu in 1956~\cite{Morrish56}. 
These authors employed a quartz-fiber torsion 
balance to perform magnetic measurements 
on individual micrometer 
sized $\gamma-$Fe$_2$O$_3$ particles. 
With their technique, they wanted to avoid 
the complication of 
particle assemblies which are due to different 
orientations of the particle's easy axis of 
magnetization and particle--particle dipolar 
interaction. They aimed to show the existence 
of a single-domain state in a magnetic particle. 
Later on, other groups tried to study single 
particles but the experimental precision did 
not allow a detailed study. A first 
breakthrough came via the work of Knowles~\cite{Knowles78} 
who developed a simple 
optical method for measuring the switching field, 
defined as the minimum applied field 
required to reverse the magnetization of a particle. 
However, the work of Knowles failed 
to provide quantitative information on well 
defined particles. More recently, insights into 
the magnetic properties of individual and 
isolated particles were obtained with the help of 
electron holography~\cite{Tonomura86}, 
vibrating reed magnetometry~\cite{Richter89}, 
Lorentz microscopy~\cite{Heffermann91,Salling91}, 
magneto-optical Kerr effect~\cite{Bardou96},
and magnetic force microscopy 
~\cite{Chang94,Ledermann94}.
Recently, magnetic nano-structures have 
been studied by the technique of magnetic 
linear dichroism in the angular distribution 
of photoelectrons or 
by photoemission electron microscopy
~\cite{Bansmann99,Meiweis99}. 
In addition to magnetic domain observations, 
element-specific information 
is available via the characteristic absorption 
levels or threshold photoemission
\footnote{We refer to the literature 
concerning other domain observation techniques~\cite{Hubert98}.}.
Among all mentioned techniques,
most of the studies have been carried out using 
magnetic force microscopy at room temperature. 
This technique has an excellent spatial resolution but 
time dependent measurements are difficult 
due to the sample--tip interaction.

Only a few groups were able to study the magnetization 
reversal of individual nanoparticles or 
nanowires at low temperatures. The first 
magnetization measurements of individual 
single-domain nanoparticles and nanowires 
at very low temperatures were presented by 
Wernsdorfer et al.~\cite{WW_JMMM95a}. 
The detector (a Nb micro-bridge-DC-SQUID) and 
the studied particles were fabricated using 
electron-beam lithography. 
Coppinger et al.~\cite{Coppinger95} investigated 
the magnetic properties of nanoparticles by 
resistance measurements. They observed the 
two-level fluctuations in the conductance of 
a sample containing self-organizing ErAs 
quantum wires and dots in a semi-insulating 
GaAs matrix. By measuring the electrical 
resistance of isolated Ni wires with diameters 
between 20 and 40~nm, Giordano and Hong 
studied the motion of magnetic domain 
walls~\cite{Hong95,Wegrowe98}. Other low 
temperature techniques which may be 
adapted to single particle measurements are 
Hall probe magnetometry~\cite{Kent94,Geim98,Schweinbock00}, 
magnetometry based on magnetoresistance
~\cite{Cros97,Gallagher97,Wegrowe99} 
or spin-dependent tunneling with Coulomb 

blockade~\cite{Schelp97,Gueron99}. 
At the time of writing, the micro-SQUID 
technique allows the most detailed study of 
the magnetization reversal of 
nanometer-sized particles
~\cite{WW_JAP95,WW_PRL96_Ni,WW_PRL97_Co,
WW_PRL97_BaFeO,Bonet99,Jamet01a}. 
The following section reviews the basic 
ideas of the micro-SQUID technique.

%%%%%%%%%%%%%%%%%%%%%%%%%%%%%%%%
\section{Micro-SQUID magnetometry}
\label{micro_SQUID}
%%%%%%%%%%%%%%%%%%%%%%%%%%%%%%%%

The Superconducting Quantum Interference Device (SQUID) has been used very 
successfully for magnetometry and voltage or current measurements in the fields of 
medicine, metrology and science~\cite{Clarke88,Ketchen93}. SQUIDs are mostly 
fabricated from a Nb$-$AlO$_x-$Nb trilayer, several hundreds of nanometers thick. The 
two Josephson junctions are planar tunnel junctions with an area of at least 
$0.5~\mu$m$^2$. In order to avoid flux pinning in the superconducting 
film the SQUID is 
placed in a magnetically shielded environment. The sample's flux is transferred via a 
superconducting pick up coil to the input coil of the SQUID. Such a device is widely used 
as the signal can be measured by simple lock-in techniques. However, this kind of 
SQUID is not well suited for measuring the magnetization of single submicron-sized 
samples as the separation of SQUID and pickup coil leads to a relatively small coupling 
factor. A much better coupling factor can be achieved by coupling the sample directly with 
the SQUID loop. In this arrangement, the main difficulty arises from the fact that the 
magnetic field applied to the sample is also applied to the SQUID. 
The lack of sensitivity to a high field applied in the SQUID plane 
led us to the development of the micro-bridge-DC-SQUID
technique~\cite{WW_JMMM95a} which allows us to apply 
several Tesla in the plane of 
the SQUID without dramatically reducing the SQUID's sensitivity.

%%%%%%%%%%%%%%%%%%%%%%%%%%%%%%%%
\subsection{Choice of SQUID configuration}
%%%%%%%%%%%%%%%%%%%%%%%%%%%%%%%%

The main criteria for the choice of the micro-SQUID configuration 
were an easy coupling to a mesoscopic sample,
a simple fabrication, a simple mode of operation, robustness and 
stability, the desired temperature range, 
and operation in high magnetic fields (in particular
for fields applied in the SQUID plane).
These criteria led to the use of micro bridge junctions instead 
of the commonly used tunnel junctions. 

The Josephson effect in micro bridge junctions has first been 
suggested in 1964 by Anderson and 
Dayem \cite{Anderson64}. These superconducting 
weak links seemed to be very 
promising in order to design planar DC-SQUIDs 
with a one step thin film technology. 
However, Dayem bridges exhibit a Josephson 
current-phase relation only when their 
dimensions are small compared to the coherence length $\xi$.
Such dimension were difficult to reach at those days.
Nowadays, electron beam lithography allows one 
to directly fabricate reliable micro-bridge 
Josephson Junctions made of materials like Al, Nb and Pb.

%%%%%%%%%%%%%%%%%%%%%%%%%%%%%%%%
\subsection{Fabrication of micro-SQUIDs}
%%%%%%%%%%%%%%%%%%%%%%%%%%%%%%%%

\begin{figure}
\begin{center}\leavevmode
\includegraphics[width=.7\textwidth]{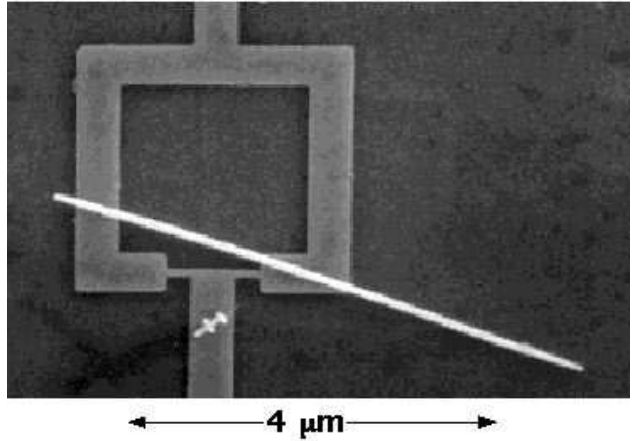}
\caption{Scanning electron micrograph of a Nb 
micro-bridge-DC-SQUID fabricated by electron beam lithography.
A Ni wire of diameter of about 90 nm was deposit on
the SQUID (Sect.~\ref{curling_exp}).}
\label{SQUID_SEM}
\end{center}
\end{figure}

By using electron beam lithography, planar Al or Nb 
micro-bridge-DC-SQUIDs (of 0.5 to 4 $\mu$m in diameter) 
can be constructed (Fig.~\ref{SQUID_SEM}) 
\cite{Chapelier91,Mailly93,WW_PhD96}. Al SQUIDs can be
obtained by evaporating a 20 to 50 nm thin film onto
a PMMA mask, followed by standard lift-off techniques.
The fabrication of Nb SQUIDs revealed to be 
more difficult. The direct evaporation of Nb onto
a PMMA mask led to Nb SQUIDs of poor quality which
manifested itself in a very low critical temperature
of the superconductivity. Better Nb qualities were
achieved by using UHV facilities and substrate temperatures
of about 800 K, or higher, in order to grow
20 nm thin Nb films on Si substrates. 
\footnote{In order to have a SQUID 
which can be exposed 
to a high field applied in the SQUID 
plane, the SQUID was made out of a very 
thin layer preventing flux trapping.
In most cases, we used mostly 20 nm 
thick Nb layers allowing measurements of 
hysteresis loops in magnetic fields of up to 2 T. 
Such SQUIDs might work at fields higher than 10 T 
when using very thin ($<$ 10 nm) Nb layers of high quality.}
The Nb films were covered
by 5 nm of Si. The Nb SQUIDs were then patterned 
with Reactive Ion Etching (RIE) using an AlO mask made
by a electron beam lithography via a standard PMMA technique.

An alternative method to fabricate micro-bridge-DC-SQUIDs
has recently been proposed by Bouchiat 
{\it et al.}~\cite{Bouchiat01}.
The new method is based on local anodization of 
3 to 6.5 nm thick Nb strip lines under the voltage 
biased tip of an Atomic Force Microscope.
Micro-bridges junctions and SQUID loops were obtained 
either by partial or total oxidation of
the Nb layer. The first fabricated devices had 
about the same performance than micro-SQUIDs 
fabricated by electron beam lithography.
The AFM-made SQUIDs should offer new
features such as the fabrication at a chosen 
position allowing an optimized coupling to magnetic
signals. In addition, we expect 
an increased intrinsic sensitivity: 
In the case of small magnetic clusters which are
placed very close to the micro-bridge junctions, 
an improvement of the sensitivity of one to
two orders of magnitude might be achieved due to the reduction 
of the micro-bridges size. It might allow us to detect
the spin flips of about 100 magnetic moments.

\begin{figure}
\begin{center}\leavevmode
\includegraphics[width=.7\textwidth]{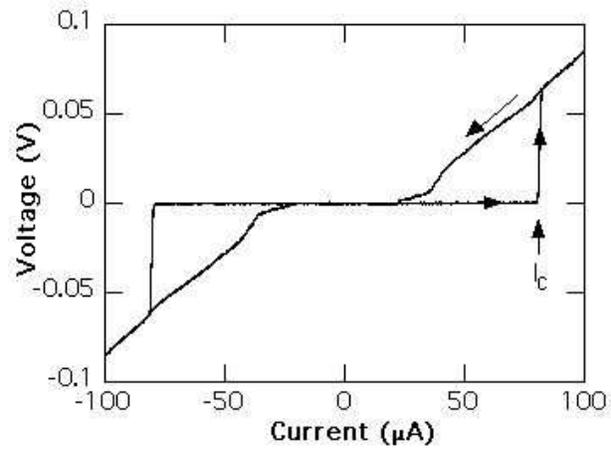}
\caption{Oscilloscope reading of a voltage versus current curve of a Nb 
micro-bridge-DC-SQUID.
The SQUID transits from the superconducting to the
normal state at the critical current $I_{\rm c}$.}
\label{SQUID_V_I}
\end{center}
\end{figure}

\begin{figure}
\begin{center}\leavevmode
\includegraphics[width=.7\textwidth]{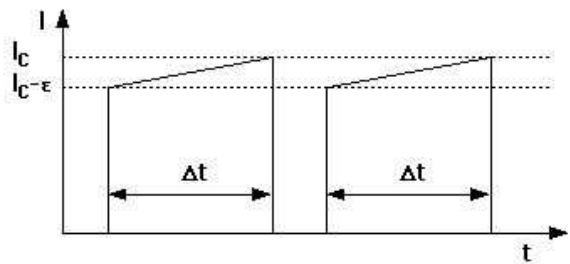}
\caption{Current injected into SQUID loop.
First, the current is increased quickly up
to a current $I_{\rm c} - \epsilon$ which is 
close to the critical current $I_{\rm c}$.
Then, the current is ramped up to $I_{\rm c}$.
As soon as a $dV/dt$ pulse is detected at the 
SQUID due to the transition
from the superconducting to the normal state 
the current is set to zero. 
$I_{\rm c}$ is proportional to
the duration of the current ramp $\Delta t$. 
The repetition rate of the cycle is up to 10 kHz.}
\label{I_ramp}
\end{center}
\end{figure}

\begin{figure}
\begin{center}\leavevmode
\includegraphics[width=.7\textwidth]{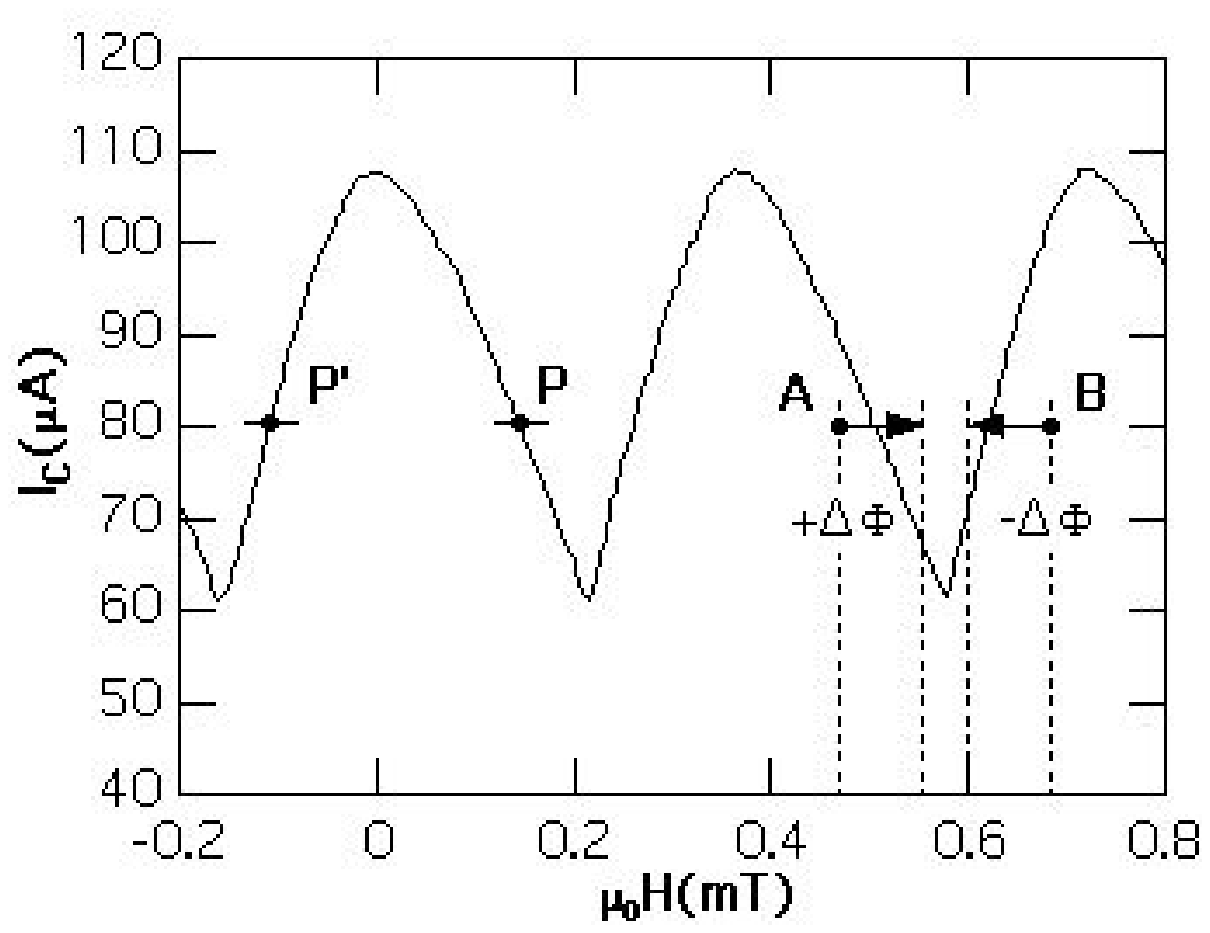}
\caption{Critical current $I_{\rm c}$ of a Al micro-SQUID 
as a function of a magnetic field applied 
perpendicular to the SQUID plane. 
For the feedback mode an external
field perpendicular to the SQUID loop is applied 
which keeps the critical current constant
at a working point, for example point {\bf P} or {\bf P'}.
For the cold mode 
method the SQUID is biased close to the 
critical current so that it is in state {\bf A} or {\bf B}, 
respectively of positive or negative flux 
jumps induced by the magnetization reversal.}
\label{fig_caract}
\end{center}
\end{figure}

%%%%%%%%%%%%%%%%%%%%%%%%%%%%%%%%
\subsection{Magnetization measurements via critical current 
measurements}
\label{Ic_direct}
%%%%%%%%%%%%%%%%%%%%%%%%%%%%%%%%

The micro-bridge-DC-SQUID have a hysteretic 
$V-I$ curve (Fig.~\ref{SQUID_V_I}):
Ramping the current up from zero,
the SQUID transits from the superconducting to the
normal state at the critical current $I_{\rm c}$. 
Due to Joule heating in the normal state,
the SQUID stays resistive down to currents
much smaller than $I_{\rm c}$.
This hysteretic $V-I$ curve made it impossible to use 
standard SQUID electronics or lock-in amplifier 
to read out the SQUID. Therefore, Benoit {\it et al.}
developed a method consisting in measuring the critical 
current of the micro-SQUID \cite{Chapelier91,Mailly93}.
A computer-controlled circuit triggers 
simultaneously a current ramp (Fig.~\ref{I_ramp})
and a 40-MHz quartz clock. As soon as a $dV/dt$
pulse is detected at the SQUID due to the transition
from the superconducting to the normal state 
\footnote{It is important to 
mention that the $dV/dt$ pulse can be 
detected directly on the current biasing lead 
of the SQUID, {\it i.e.} for most cases
it is sufficient to connect the SQUID with
a single wire and the mass
of the cryostat.},
the clock stops and the current is set to zero. The clock
reading is transferred to the computer, and the cycle
begins again. The critical current is proportional to
the duration of the current ramp. The repetition
rate is about 10 kHz, limited by the time needed to
settle the current. 
As the critical 
current $I_{\rm c}$ is a periodic function of the 
flux going through the SQUID loop 
(Fig.~\ref{fig_caract}), one can easily 
deduce the flux change in the SQUID loop by 
measuring the critical current.
The sensitivity achieved by the critical current measurement 
technique was about $10^{-5} \Phi_0/\sqrt{Hz}$ 
for Al SQUIDs and $10^{-4} \Phi_0/\sqrt{Hz}$ for
Nb SQUIDs ($\Phi_0$ = h/2e = $2 \times 10^{-15}$ Wb).

\begin{figure}
\begin{center}\leavevmode
\includegraphics[width=.7\textwidth]{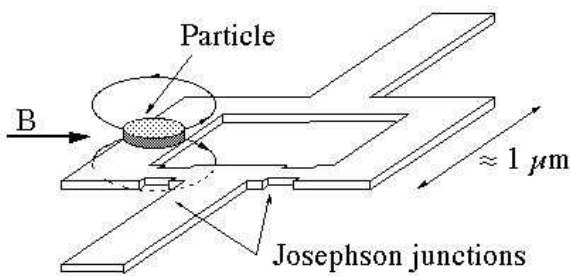}
\caption{Schematic drawing of a planar micro-bridge-DC-SQUID on which 
a ferromagnetic particle is placed. 
The SQUID detects the flux through its loop 
produced by the sample magnetization. Due to the close 
proximity between sample and 
SQUID a very efficient and direct flux coupling is achieved.}
\label{fig_SQUID}
\end{center}
\end{figure}

In order to have good magnetic flux coupling,
the mesoscopic systems, for instant the magnetic particles,
is directly placed on the SQUID loop
(Fig.~\ref{fig_SQUID}) \cite{WW_PhD96}. 
The SQUID detects the flux through its loop produced by the 
sample's magnetization.  For hysteresis 
loop measurements, the external field 
is applied in the plane of 
the SQUID (Fig.~\ref{fig_SQUID}), 
thus the SQUID is only sensitive 
to the flux induced by the stray field of the 
sample's magnetization. 
The flux sensitivity of the critical current
mode allowed us to detect the
magnetization reversals corresponding to 
$10^{4} \mu_{\rm B}/\sqrt{Hz} (10^{-16} 
emu/\sqrt{Hz})$, {\it i.e.} the magnetic 
moment of a Co nanoparticle with a 
diameter of 5 nm.
The time resolution was given by the time between 
two measurements of the critical current. In this 
case, the achieved time resolution was 100 $\mu$s.

%%%%%%%%%%%%%%%%%%%%%%%%%%%%%%%%
\subsection{Feedback mode for hysteresis loop measurements}
\label{feedback}
%%%%%%%%%%%%%%%%%%%%%%%%%%%%%%%%

For the detection of flux variations larger than
$\Phi_0/2$, the direct critical current method 
(Sect.~\ref{Ic_direct}) is
complicated by the fact that $I_{\rm c}$ is a non monotonous 
function of the flux (Fig.~\ref{fig_caract}) 
which goes through the SQUID loop.
In this cases, we used
a feedback mode consisting in applying an external
field perpendicular to the SQUID loop 
which compensates the stray field variations
keeping the critical current constant
at a working point as shown in Fig.~\ref{fig_caract}. 
After each $I_{\rm c}$ measurement,
the computer calculates the difference between
a working point and $I_{\rm c}$, multiplies it
with a feedback factor, and adds it to the 
feedback (or compensation) field.
The feedback field gives then the stray field 
variation of the sample's magnetization.
It can easily be given
in unites of $\Phi_0$. Knowing the  coupling
factor between particle and SQUID loop, absolute
magnetic moment measurements are possible.

The time resolution of the feedback mode depends
on the rate of the $I_{\rm c}$ measurements.
For rates of about 10 kHz, we are able to follow
flux variations in the millisecond range.

%%%%%%%%%%%%%%%%%%%%%%%%%%%%%%%%
\subsection{Cold mode method for magnetization switching measurements}
\label{cold}
%%%%%%%%%%%%%%%%%%%%%%%%%%%%%%%%

For studying the magnetization switching
of nanoparticles, we developed a special mode,
called the {\it cold mode method}
\cite{WW_PRL97_Co,WW_PRL97_BaFeO,Bonet99,Jamet01a},
which are much faster and more sensitive than
the  previous presented modes. 
The achieved flux sensitivity allowed us to detect the
magnetization switching corresponding to 
$10^{3} \mu_{\rm B} (10^{-17} emu)$, {\it i.e.} the switching 
of a Co nanoparticle with a diameter of 2 to 3 nm.
The time resolution of the switching detection
reached the nanosecond range. 

The cold mode method is
also specially adapted for studying the temperature 
dependence of the magnetization reversal and 
macroscopic quantum tunneling of magnetization.
Indeed, the main difficulty associated with the 
SQUID detection technique lies in the Joule heating 
when the critical current is reached. 
After the normal state transition at the critical 
current, the SQUID dissipates for about 100 ns 
which slightly heats the magnetic particle 
coupled to the SQUID. This problem can be 
solved by the cold mode method which
uses the SQUID only as a trigger 
\cite{WW_PRL97_Co,WW_PRL97_BaFeO,Bonet99,Jamet01a}. 

The cold mode consists in biasing the SQUID 
close to the critical current while a field 
is applied perpendicular to the SQUID plane 
so that the SQUID is in state {\bf A} ({\bf B}) 
for a positive (negative) flux jump (Fig.~\ref{fig_caract}) 
which can be induced by the magnetization 
reversal of a particle coupled to the SQUID loop. 
The magnetization reversal then triggers a 
transition of the SQUID from the superconducting 
to the normal state, {\it i.e.} a $dV/dt$ pulse can
be detected on the current lead biasing 
the SQUID. Our SQUID electronics 
allows us to detect the $dV/dt$ pulse some 
nanoseconds after the magnetization reversal,
allowing very precise switching field measurements.
Because this method only heats the sample 
after the magnetization reversal, 
we called it the {\it cold mode method}.

Another advantage of this mode is that
the sample does not interfere with the rf-noise 
which is induced in oxide layer Josephson junctions
of conventional SQUIDs because the hysteretic micro-SQUID 
is in the superconducting state 
before the magnetization reversal. 

\begin{figure}
\begin{center}\leavevmode
\includegraphics[width=.7\textwidth]{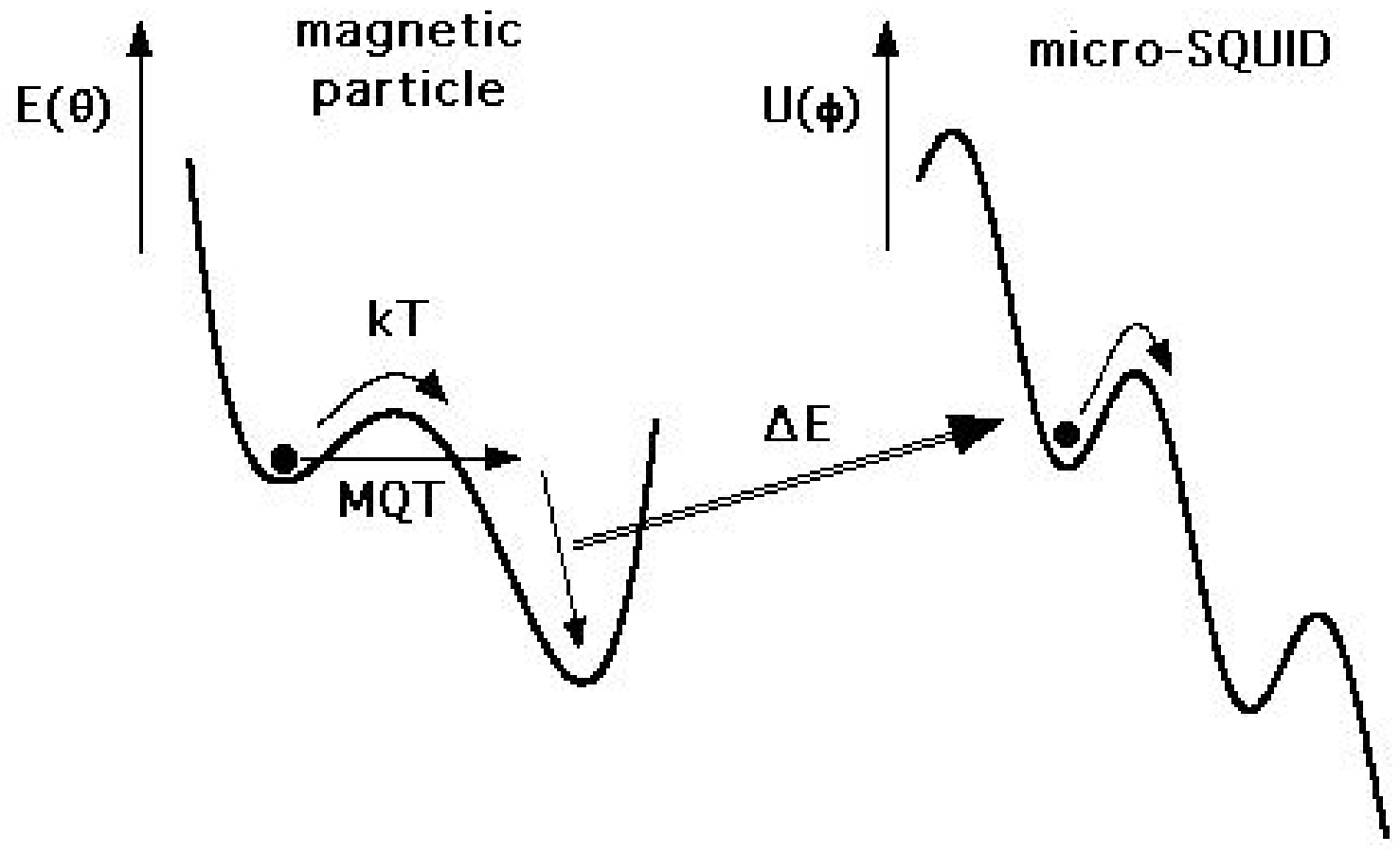}
\caption{Energy scheme of the cold mode method. 
After the particle overcomes the 
saddle point or tunnels through the energy barrier, 
it falls into the lower potential well 
releasing energy. A very small fraction $\Delta E$ of 
this energy is transferred to the SQUID and 
drives the SQUID out of its metastable 
superconducting state.}
\label{cold_mode}
\end{center}
\end{figure}

Finally, the cold mode method is very important 
for studying macroscopic quantum 
tunneling of magnetization. Quantum theory predicts 
that the escape rate from a metastable 
potential well by quantum tunneling is strongly 
reduced by the coupling of the magnetic 
system with its environment. Therefore, the measuring 
device must be weakly coupled to 
the magnetic particle. However, in order to measure 
the magnetization reversal, the SQUID 
must be strongly coupled to the magnetic particle 
which hinders the possibility of 
quantum tunneling. This problem can be solved by 
using the cold mode method. In order 
to show this schematically, Fig.~\ref{cold_mode} 
represents two energy potentials, one is the 
double well potential of the particle, the other 
is the periodic potential of the SQUID. 
Before the magnetization reversal, both systems 
are in a metastable state: the particle 
because of an applied field which is close to the 
switching field and the SQUID because 
of a current through the SQUID loop which is 
close to the critical current. 
When the particle overcomes the saddle point or tunneling 
through the energy barrier, its 
magnetization rotates by only few degrees. 
For this starting process of the magnetization reversal, 
the coupling between particle 
and SQUID can be arranged to be very small. 
\footnote{In order to illustrate the dipolar couplings, 
let us consider the energy scales involved. For most 
of the particles measured so far below 1 K, the energy 
barrier height from the metastable state up to 
the saddle point is of the order of a few Kelvins 
whereas that from the lower 
state up to the saddle point is between 
$10^3$ and $10^6$ K. These energy scales 
should be compared with the energy necessary 
to drive the SQUID out of its metastable 
superconducting state which is of the order of a few Kelvin. 
Therefore, only a small energy 
transfer is necessary to measure the magnetization 
reversal. In addition, a proper 
orientation of the easy axis of magnetization 
with respect to the SQUID loop can further 
reduce the coupling during the first stage of 
the magnetization reversal. In the case of an 
easy axis of magnetization perpendicular to 
the current direction in the 
SQUID wire (Fig.~\ref{fig_SQUID}), the 
coupling factor between SQUID loop and particle 
is proportional to (1 - $\cos{\varphi}$), 
where $\varphi$ is the angle between the direction 
of magnetization and its easy axis.
Therefore, the coupling is very weak at the first 
stage of magnetization reversal where $\varphi$ is small.}
Afterwards, the particle falls 
into the lower well which implies a rotation 
of magnetization of up to 180$^{\circ}$. 
During this process, the coupling between 
particle and SQUID should be strong 
enough to drive the SQUID out of its metastable state. 
The corresponding transition from 
the superconducting into the normal state is 
easily measurable for a hysteretic SQUID.

The main 
disadvantage of the cold mode is that only the 
switching field of magnetization reversal can 
be measured and not the magnetization before 
and during the magnetization reversal.

%%%%%%%%%%%%%%%%%%%%%%%%%%%%%%%%
\subsection{Blind mode method for three dimensional 
switching field measurements}
\label{blind}
%%%%%%%%%%%%%%%%%%%%%%%%%%%%%%%%

A disadvantage of the micro-SQUID technique is
that it does not function properly when a significant field 
is applied perpendicular to the SQUID loop. 
It works also only below the superconducting
critical temperature of Nb ($T_{\rm c} \approx 7$ K for our SQUIDs). 
These facts limited us to 2D measurements and to $T < T_{\rm c}$ 
in the first experiments~\cite{WW_PRL96_Ni,WW_PRL97_Co,WW_PRL97_BaFeO}. 
However, we showed recently that 
full-three-dimensional measurements can be done by
using an indirect method~\cite{Bonet99}.
In addition, this technique allows us to study the magnetization
reversal for $T > T_{\rm c}$.

\begin{figure}
\begin{center}\leavevmode
\includegraphics[width=.7\textwidth]{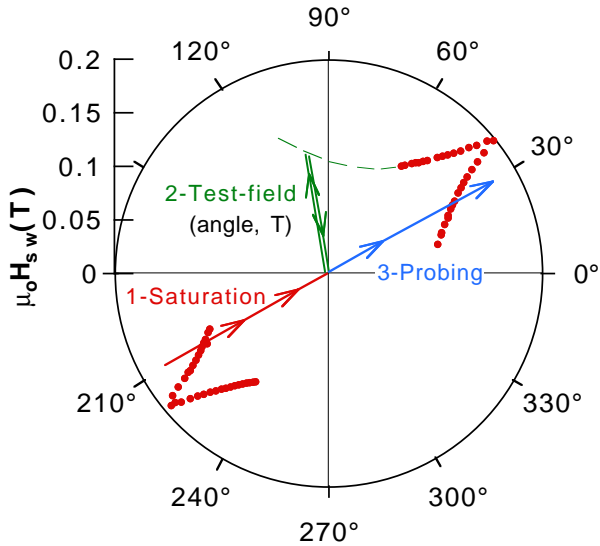}
\caption{Schematics for the {\it blind mode} method
presenting the angular dependence of the switching field near 
the easy axis of a 3 nm Co cluster (full circles) 
(Sect.~\ref{St_W_exp}). 
First, the magnetization of the particle is saturated
in a given direction (at T = 35 mK).
Then, a test-field is applied, at a temperature between 35 mK
and 30 K, which may or may not cause a magnetization
switching. Finally, the SQUID is switched on (at T = 35 mK)
and a field is applied in the plane of the SQUID to probe
the resulting magnetization state.}
\label{blind_mode}
\end{center}
\end{figure}

Let us consider Fig.~\ref{blind_mode} showing switching 
field measurements (Sect.~\ref{chap_reversal_T_0})
for in-the-SQUID-plane applied fields, 
detected using the {\it cold mode} (Sect.~\ref{cold}).
The three dimensional switching field measurements
and the studies as a function of temperature can be done
using a three step method which we call 
the {\it blind mode} (Fig.~\ref{blind_mode}):

\begin{enumerate}
\item
{\bf Saturation:} The magnetization of the particles is saturated
in a given direction (at T = 35 mK).

\item
{\bf Testing:} A test-field is applied at a temperature between 35 mK
and 30 K 
\footnote{Our highest temperatures of 30 K 
was only limited by the cooling time.
Below 30 K, we achieved cooling rates of few
Kelvins per second.}
which may or may not cause a magnetization switching. 

\item
{\bf Probing:} After cooling to $T$ = 35 mK, the SQUID is switched on
and a field is swept in the plane of the SQUID to probe
the resulting magnetization state using direct critical
current measurements, the feedback mode, or the cold mode.
\end{enumerate}

If the SQUID detects a magnetization jump in step (3), 
this means that the previously applied test-field was  
weaker than the switching field for the direction being 
probed in step (2). 
The next interation will then be done with a stronger test-field. 
On the other hand, if the SQUID does not detect any 
magnetization jump in step (3), this means that the 
reversal occurred during step (2). The next interation will 
then be done with a weaker field.
When choosing the new test-field with the help of a 
bisection algorithm, we needed about 8 repetitions of the 
three steps in order to get the switching field with good precision.
This method allows us to scan the entire field space.
\footnote{It is worth mentioning that special precautions are necessary for
anisotropies which are more complex than 
uniaxial (Sect.~\ref{chap_reversal_T_0}).}

%%%%%%%%%%%%%%%%%%%%%%%%%%%%%%%%
\subsection{Micro-SQUID arrays}
\label{SQUID_array}
%%%%%%%%%%%%%%%%%%%%%%%%%%%%%%%%

\begin{figure}
\begin{center}\leavevmode
\includegraphics[width=.8\textwidth]{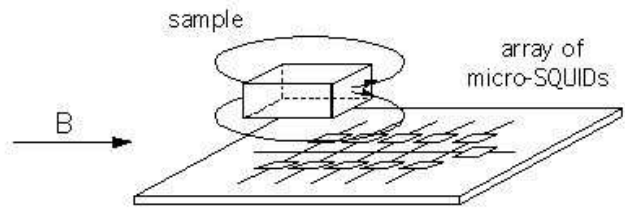}
\caption{Schematic representation of our magnetometer which is an array of micro-
SQUIDs. Its high sensitivity allows us to study single crystals of the order of 10 to 
500~$\mu$m which are placed directly on the array.}
\label{fig_SQUID_array}
\end{center}
\end{figure}

We also use arrays of micro-SQUIDs as a magnetometer for macroscopic 
samples. There are three applications which were particular
interesting: crystals of magnetic 
molecular clusters~\cite{WW_Science99} (Sect.~\ref{mol_cluster}),
nucleation and depinning of magnetic domain walls in 
thin films~\cite{Mangin97},
and arrays of magnetic dots~\cite{Fruchart99_Fe}. 
The procedure consists in placing a sample 
on top of an array of micro-SQUIDs 
so that some SQUIDs are directly under the sample, some SQUIDs are 
at the border of the sample, and some SQUIDs are beside the sample  
(Fig.~\ref{fig_SQUID_array}).
When a SQUID is very close to the sample, it is sensing 
locally the magnetization reversal whereas when the SQUID is far away, 
it integrates over a bigger sample volume. 
Therefore, depending on the sample, one can obtain more insight in the 
magnetization reversal than with conventional techniques
which measure only the total magnetization.

Our magnetometer works in the 
temperature range between 0.035 and 6~K and 
in fields up to 5~T with sweeping rates 
as high as 30~T/s, and a field stability better 
than a microtesla. The time resolution is about 
1~ms (Sect.~\ref{feedback}) allowing short-time measurements. The field 
can be applied in any direction of the 
micro-SQUID plane with a precision much better 
than 0.1$^{\circ}$ by separately 
driving three orthogonal coils \cite{WW_PhD96}. 

%%%%%%%%%%%%%%%%%%%%%%%%%%%%%%%%
\subsection{Scanning SQUID microscope}
%%%%%%%%%%%%%%%%%%%%%%%%%%%%%%%%

The micro-SQUID technique has recently been 
used to build a scanning SQUID 
microscope~\cite{Hasselbach00}.
The SQUID is designed by electron beam lithography 
at the apex of a silicon
cantilever. The lever is attached to a force sensor, 
allowing to image magnetically, 
as well as topographically with a spatial resolution of 
50 nm and a flux resolution 
of about $10^{-4} \phi_0$.
The first application of this technique concerned the imaging of 
vortices in artificial networks.

%%%%%%%%%%%%%%%%%%%%%%%%%%%%%%%%
\subsection{Outlook}
%%%%%%%%%%%%%%%%%%%%%%%%%%%%%%%%

After the development of micro-SQUID technique in the early 90's
~\cite{Chapelier91,Mailly93}, the study of magnetization
reversal in magnetic nano-structures
began in 1993~\cite{WW_PhD96}. 
The first studied systems were micrometer sized
particles containing about 10$^{10}$ magnetic moments.
During the following years, the micro-SQUID
technique has been improved to study smaller and 
smaller systems. In 2000, clusters containing 
about 10$^{3}$ magnetic moments could be studied.
This achievement rises the question whether further
improvements might be possible. The
fundamental limit of a SQUID is the quantum limit
which correspond to a sensitivity of one magnetic moment
for a SQUID with 1 $\mu$m$^2$.
One might come close to this limit by using shunted SQUIDs 
~\cite{Cernicchiaro96}. Another possibility could be a
reduction of the section of the micro-bridges~\cite{Bouchiat01}.
Finally, the micro-SQUID technique could be
improved by using superconducting materials 
with higher critical temperatures allowing measurements at 
higher temperatures.

%%%%%%%%%%%%%%%%%%%%%%%%%%%%%%%%
\title[Mechanisms of Magnetization Reversal at Zero Kelvin]
{Mechanisms of Magnetization Reversal at Zero Kelvin}
\label{chap_reversal_T_0}
\markboth{Mechanisms of Magnetization Reversal at $T = 0$ K}
{Mechanisms of Magnetization Reversal at $T = 0$ K}
%%%%%%%%%%%%%%%%%%%%%%%%%%%%%%%%

As already briefly discussed in the introduction, 
for a sufficiently small magnetic sample it 
is energetically unfavorable to form a stable magnetic domain wall. 
The specimen then behaves as a single magnetic domain. 
For the smallest single domain particles
\footnote{In the theory of micromagnetism,
the single domain state describes the state where
the magnetization is perfectly aligned~\cite{Aharoni96}
whereas experimentalists mean often a state without
domain wall.}, the 
magnetization is expected to 
reverse by uniform rotation of magnetization (Sec.~\ref{St_W}). 
For somewhat larger ones, non uniform reversal modes 
are more likely, for example the curling 
reversal mode (Sec.~\ref{curling}). 
For larger particles, 
magnetization reversal occurs via a domain wall 
nucleation process starting in a rather 
small volume of the particle. For even larger
particles, the nucleated domain wall can be
stable for certain fields. The magnetization reversal
happens then via nucleation and annihilation 
processes (Sec.~\ref{nucleation}).
In these sections, we neglect 
temperature and quantum effects.

The following section discusses in detail the uniform rotation 
mode which is used in many theories, 
in particular in N\'eel, Brown and 
Coffey's theory of magnetization reversal by 
thermal activation (Sect.~\ref{chap_reversal_clas}) 
and in the theory of 
macroscopic quantum tunneling of magnetization 
(Sect.~\ref{chap_MQT}). 

%%%%%%%%%%%%%%%%%%%%%%%%%%%%%%%%
\section[Magnetization Reversal by Uniform Rotation] 
%\string\hfill\string\break\space (Stoner--Wohlfarth model)]
{Magnetization Reversal by Uniform Rotation \\ (Stoner--Wohlfarth 
Model)}
\label{St_W}
%%%%%%%%%%%%%%%%%%%%%%%%%%%%%%%%

The model of uniform rotation of magnetization, developed by 
Stoner and Wohlfarth~\cite{St_W48}, and N\'eel~\cite{Neel47}, 
is the simplest classical model describing 
magnetization reversal. One considers a particle of an 
ideal magnetic material where exchange 
energy holds all spins tightly parallel to each other, and the 
magnetization magnitude does not depend 
on space. In this case, the exchange energy 
is constant, and plays no role in the energy 
minimization. Consequently, there is competition only 
between the anisotropy energy of the 
particle and the effect of the applied field. The original study by 
Stoner and Wohlfarth 
assumed only uniaxial shape anisotropy which is the anisotropy of the 
magnetostatic 
energy of the sample induced by its non-spherical shape.
Thiaville has generalized this model for an arbitrary effective 
anisotropy which includes 
any magnetocrystalline anisotropy and even surface 
anisotropy~\cite{Thiaville98}.

\begin{figure}
\begin{center}\leavevmode
\includegraphics[width=.7\textwidth]{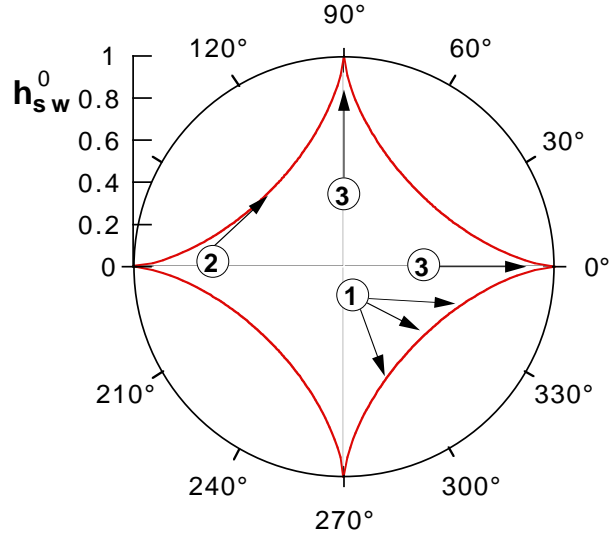}
\caption{Angular dependence of the Stoner--Wohlfarth 
switching field $h_{\rm sw}^0 = H_{\rm sw}^0/H_{\rm a}$ 
(Eq.~\ref{eq_H_St_W}). This curve is often called the \lq 
Stoner--Wohlfarth astroid'. The cases (1) to (3) correspond to 
Eqs.~\ref{eq_E_1a} to~\ref{eq_E_4} concerning the field dependence of 
the anisotropy barrier height.}
\label{ast_ST_W}
\end{center}
\end{figure}

In the simplest case of uniaxial anisotropy, 
the energy of a Stoner--Wohlfarth particle is given by
\begin{equation}
E = K V \sin^2\phi - \mu_0 M_{\rm S} V H \cos(\phi - \theta) \;,
\label{eq_E_pot}
\end{equation}
where $KV$ is the uniaxial anisotropy energy which depends 
on the shape of 
the particle, $V$ is the volume of the particle, $M_{\rm S}$ is the 
spontaneous magnetization, 
$H$ the magnitude of the applied field, and 
$\phi$ and $\theta$ are the angles of 
magnetization and applied field respectively, 
with respect to the easy axis of magnetization. The 
potential energy of (\ref{eq_E_pot}) has two minima separated 
by an energy barrier. For given 
values of $\theta$ and $H$, the magnetization lies at an angle 
$\phi$ which locally minimizes 
the energy. This position can by found by equating to zero the first 
derivative with respect 
to $\phi$ of (\ref{eq_E_pot}): $\partial E/\partial \phi = 0$. 
The second derivative provides the criteria 
for maxima and minima. The magnetization reversal is defined by the 
minimal field value 
at which the energy barrier between the metastable minimum 
and the stable one vanishes, {\it i.e.}, 
at $\partial E/\partial \phi = \partial^2E/\partial \phi^2 = 0$. 
A short analysis yields the angular dependence of this 
field, called the switching field $H_{\rm sw}^0$,
or in dimensionless units:

\begin{equation}
h_{\rm sw}^0 = \frac {H_{\rm sw}^0}{H_{\rm a}} = \frac {1} 
{\left(\sin^{2/3}\theta + \cos^{2/3}\theta \right)^{3/2}} \;,
\label{eq_H_St_W}
\end{equation}
where $H_{\rm a} = 2 K/(\mu_0 M_{\rm S})$ is the anisotropy field. 
The angular dependence of $h_{\rm sw}^0$ is plotted in 
Fig.~\ref{ast_ST_W}.

\begin{figure}
\begin{center}\leavevmode
\includegraphics[width=.7\textwidth]{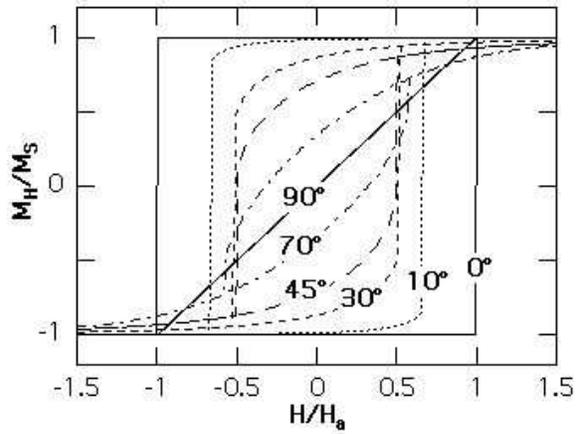}
\caption{Hysteresis loops of a Stoner--Wohlfarth particle for 
different field angles 
$\theta$. The component of magnetization along the
direction of the applied field is 
plotted, {\it i.e.}, $M_{\rm H} = M_{\rm S}\cos(\phi - \theta)$.}
\label{hyst_St_W}
\end{center}
\end{figure}

\begin{figure}
\begin{center}\leavevmode
\includegraphics[width=.8\textwidth]{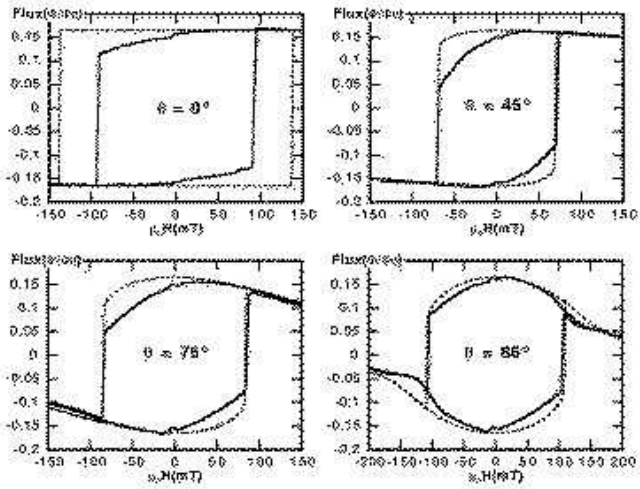}
\caption{Hysteresis loops of a nanocrystalline elliptic Co particle 
of $70 \times 50 \times 25$~nm$^3$. 
The {\it dashed line} is the prediction 
of the Stoner--Wohlfarth 
model of uniform 
rotation of magnetization.
The deviations are due to non uniform magnetization states.}
\label{fig_hyst_Co}
\end{center}
\end{figure}

Contrary to $h_{\rm sw}^0$, the hysteresis loops cannot be expressed 
analytically and have to be 
calculated numerically. The result is seen in Fig.~\ref{hyst_St_W} 
showing the component of 
magnetization projected along the direction of the applied field, {\it i.e.}, 
$M_{\rm H} = M_{\rm S}\cos(\phi - \theta)$. Such loops are 
often called Stoner--Wohlfarth hysteresis loops. 
\footnote{It is important to 
note that single particle 
measurement techniques do not measure this component $M_{\rm H}$. 
For example for the micro-SQUID technique, 
with the easy axis of magnetization in the plane of the SQUID and 
perpendicular to the current direction in the SQUID wire 
(Fig.~\ref{fig_SQUID}), one 
measures a 
magnetic flux which is proportional to $M_{\rm S} \sin\phi$ 
(Fig.~\ref{fig_hyst_Co}).}

The main advantage of this classical theory is that it is 
sufficiently simple to add some 
extra features to it, as presented in the following.

%%%%%%%%%%%%%%%%%%%%%%%%%%%%%%%%
\subsection{Generalization of the Stoner--Wohlfarth model}
\label{St_W_general}
%%%%%%%%%%%%%%%%%%%%%%%%%%%%%%%%

The original model of Stoner and Wohlfarth assumed only uniaxial 
shape anisotropy with one anisotropy constant (one second order term,
see Eq.~\ref{eq_E_pot}). 
This is sufficient to describe highly symmetric cases like a prolate 
spheroid of revolution or an infinite cylinder. However, real systems 
are often quite complex and the anisotropy is a sum of mainly shape 
(magnetostatic), magnetocrystalline, magnetoelastic and surface anisotropy. 
One additional 
complication arises because the different contributions of 
anisotropies are often aligned in an arbitrary way one
with respect to each other. 
All these facts motivated a generalization of the Stoner--Wohlfarth 
model for an arbitrary effective anisotropy which was done by 
Thiaville~\cite{Thiaville98,Thiaville00}. 

Similar to the Stoner--Wohlfarth model, one
supposes that the exchange interaction in the cluster
couples all the spins strongly together to form a
giant spin whose direction is described by the unit vector $\vec{m}$.
The only degrees of freedom of the particle's magnetization are the
two angles of orientation of $\vec{m}$.
The reversal of the magnetization is described by the
potential energy:
\begin{equation}
E(\vec{m},\vec{H}) = E_0(\vec{m}) - \mu_0 V M_{\rm s} \vec{m}.\vec{H}
\label{eq_E}
\end{equation}
where $V$ and $M_{s}$ are the magnetic volume
and the saturation magnetization of the particle respectively, 
$\vec{H}$ is
 the external magnetic field,
and $E_0(\vec{m})$ the magnetic anisotropy energy which is given by:
\begin{equation}
E_0(\vec{m}) = E_{\rm shape}(\vec{m})
	+ E_{\rm MC}(\vec{m})
	+ E_{\rm surface}(\vec{m})
	+ E_{\rm ME}(\vec{m})
\label{eq_E_0}
\end{equation}

$E_{\rm shape}$ is the magnetostatic energy related to
the cluster shape. 
$E_{\rm MC}$ is the magnetocrystalline anisotropy (MC) 
arising from the coupling
of the magnetization with the crystalline 
lattice, similar as in bulk.
$E_{\rm surface}$ is due to the symmetry breaking and
surface strains.
In addition, if the particle experiences an external stress, the 
volumic
relaxation inside the particle induces a magneto-elastic (ME) 
anisotropy energy $E_{\rm ME}$.
All these anisotropy energies can be developed in a power
series of $m_{x}^{a}m_{y}^{b}m_{z}^{c}$ 
with $p = a + b + c = 2, 4, 6,\ldots$ 
giving the order of the anisotropy term. 
Shape anisotropy can be written as a biaxial anisotropy with two 
second order terms. 
Magnetocrystalline anisotropy is in most 
cases either uniaxial (hexagonal systems) or cubic, yielding 
mainly second and forth order terms. Finally, in the simplest 
case, surface and magneto-elastic anisotropies are of second order. 

Thiaville proposed a geometrical method 
to calculate the particle's energy and
to determine the switching field for all angles of the applied 
magnetic field yielding the critical surface of switching fields
which is analogous to the Stoner--Wohlfarth 
astroid (Fig.~\ref{ast_ST_W}).

The main interest of Thiaville's calculation is that measuring 
the critical surface of the switching field allows one to find 
the effective anisotropy of the nanoparticle. The knowledge of 
the latter is important for temperature dependent studies 
(Sect.~\ref{chap_reversal_clas}) 
and quantum tunneling investigations (Sect.~\ref{chap_MQT}). 
Knowing precisely the particle's shape and the crystallographic axis 
allows one to determine the different contributions to the 
effective anisotropy.

Thiaville's calculation predicts also the field dependence of the 
energy barrier height $\Delta E$ close to switching 
($\varepsilon = \left(1 - H/H_{\rm sw}^0\right) << 1$) 
which is important to know for temperature dependent 
studies (Sects.~\ref{chap_reversal_clas} and \ref{chap_MQT}). 
Three cases have to be distinguished:

\begin{enumerate}
\item
In the majority of cases except the two following cases 
(see case 1 in Fig.~\ref{ast_ST_W}):
\begin{equation}
	\Delta E = 
	4 KV \frac {\sqrt{2}}{3} \frac {\cos\gamma}{\sqrt{\rho}} 
\varepsilon^{3/2} = 
	E_0 \varepsilon^{3/2}
\label{eq_E_1a}
\end{equation}
where $KV$ is the anisotropy energy constant, 
$\gamma$ is the angle of incidence between the local normal 
to the critical surface and the field sweeping direction, 
$\rho$ is the radius of curvature of the focal 
curve~\cite{Thiaville00} 
at $H_{\rm sw}^0$. 
It is important to emphasize that all these variables 
can be found experimentally by measuring the critical 
surface of the switching field. For uniaxial anisotropy, 
{\it i.e.} the 2D Stoner--Wohlfarth case, 
equation~\ref{eq_E_1a} becomes:
\begin{equation}
	\Delta E = 
	4 KV \left(\frac {2}{3}\right)^{3/2} 
	\frac {\mid \cos\theta \mid^{1/3}}{1+\mid \cos\theta \mid^{2/3}}
	\varepsilon^{3/2}
\label{eq_E_1b}
\end{equation}
where $\theta$ is the angle of the applied field 
with respect to the easy axis of magnetization (Eq.~\ref{eq_E_pot}).

\item
At glancing incidence (see case 2 in Fig.~\ref{ast_ST_W}) with respect 
to the critical surface ($\gamma = \pi / 2$), the power of 
$\varepsilon$ is different, yielding: 
\begin{equation}
	\Delta E = E'_0 \varepsilon^3
\label{eq_E_2}
\end{equation}
where $E'_0$ has been calculated only in the 
two dimensional case~\cite{Thiaville98}.

\item
At a cusp point where $\gamma = \pi / 2$ 
(see case 3 in Fig.~\ref{ast_ST_W}): 
\begin{equation}
	\Delta E = E"_0 \varepsilon^2
\label{eq_E_3}
\end{equation}
where $E"_0$  has been calculated only in the 
two dimensional case~\cite{Thiaville98}. 
In the case of uniaxial anisotropy, 
{\it i.e.} the 2D Stoner--Wohlfarth case), 
equation~\ref{eq_E_3} becomes:
\begin{equation}
	\Delta E = 
	KV \varepsilon^2 = KV \left(1 - \frac{H}{H_{\rm a}} \right)^2
\label{eq_E_4}
\end{equation}
where $H_{\rm a} = \frac{2K}{\mu_0 M_{\rm s}}$ (Eq.~\ref{eq_E_pot}). 
This equation is only valid for $\theta = 0$ and $\pi /2$, and it
is valid for $0 \leq H \leq H_{\rm a}$.
This famous result of N\'eel~\cite{Neel49a,Neel49b} has 
often been wrongly used for assemblies of nanoparticles
where it is very difficult to achieve the conditions 
$\theta = 0$ and $\pi /2$~\cite{Victora89}. 
Up to now, only the power 3/2 and 2 (cases (1) and (3)) 
has been found by single particle measurements
~\cite{WW_PRL97_Co,WW_PRL97_BaFeO} 
(see also Figs.~\ref{Hsw_T} and \ref{Hsw_T_a}).
\end{enumerate} 

Another simple analytical 
approximation for the field dependence of the 
energy barrier $\Delta E(H)$ was derived numerically 
by Pfeiffer~\cite{Pfeiffer90a,Pfeiffer90b}:
\begin{equation}
	\Delta E = KV \left(1 - H/H_{\rm sw}^0\right)^a = E_0 \varepsilon^a 
\;,
\label{eq_Pfeiffer}
\end{equation}
where $a = 0.86 + 1.14  h_{\rm sw}^0$, 
and $h_{\rm sw}^0$ is given by 
Eq.~\ref{eq_H_St_W}. This approximation 
is good for the intermediate
field regime, {\it i.e.} for fields $H$ not too close 
to $H_{\rm a}$ and not too small.

%%%%%%%%%%%%%%%%%%%%%%%%%%%%%%%%
\subsection{Experimental evidence for magnetization 
reversal by uniform rotation}
\label{St_W_exp}
%%%%%%%%%%%%%%%%%%%%%%%%%%%%%%%%

In order to demonstrate experimentally the uniform rotation mode, the 
angular 
dependence of the magnetization reversal has often been studied 
(see references in~\cite{Aharoni96}). However, a comparison of 
theory with experiment is difficult because magnetic 
particles often have a non uniform magnetization state which is due to 
rather complicated 
shapes and surfaces, crystalline defects, and surface anisotropy. In 
general, for many 
particle shapes the demagnetization fields inside the particles are 
non uniform leading to 
non uniform magnetization states~\cite{Aharoni96}. An example is 
presented in 
Fig.~\ref{fig_hyst_Co} which compares 
typical hysteresis loop measurements of an elliptical Co particle, 
fabricated by electron beam lithography, with the prediction of the 
Stoner--Wohlfarth 
model (Fig.~\ref{fig_hyst_Co}). 
Before magnetization reversal, the magnetization decreases more 
strongly than 
predicted because the magnetic configuration is not collinear as in 
the Stoner--Wohlfarth 
model, but instead presents deviations mainly near the particle surface. 
The angular dependence of the switching field 
agrees with the Stoner--Wohlfarth model only for angles 
$\theta \leq 30^{\circ}$ where non-linearities and defects play a less 
important role~\cite{WW_JMMM95a,WW_JMMM95b}. 

Studies of magnetization reversal processes in ultrathin 
magnetic dots 
with in-plane uniaxial anisotropy 
showed also switching fields which are very 
close to the Stoner--Wohlfarth model although magnetic 
relaxation experiments clearly showed
that nucleation volumes are by far smaller than an 
individual dot volume~\cite{Fruchart99_Fe}. 
These studies show clearly that switching field measurements 
as a function of the angles of the applied field cannot be
taken unambiguously as a proof of a Stoner--Wohlfarth reversal.

The first clear demonstration of the uniform reversal mode has been
found with Co nanoparticles~\cite{WW_PRL97_Co}, and 
BaFeO nanoparticles~\cite{WW_PRL97_BaFeO}, the latter having a dominant 
uniaxial 
magnetocrystalline anisotropy. The three dimensional angular 
dependence
of the switching field measured on BaFeO 
particles of about 20~nm could be explained with the
Stoner--Wohlfarth model taking into account the shape 
anisotropy and hexagonal crystalline anisotropy of 
BaFeO~\cite{Bonet99}. 
This explication is supported by temperature and time 
dependent measurements yielding activation volumes
which are very close to the particle volume (Sect.~\ref{chap_reversal_clas}).

\begin{figure}
\sidebyside{
\includegraphics[width=.45\textwidth]{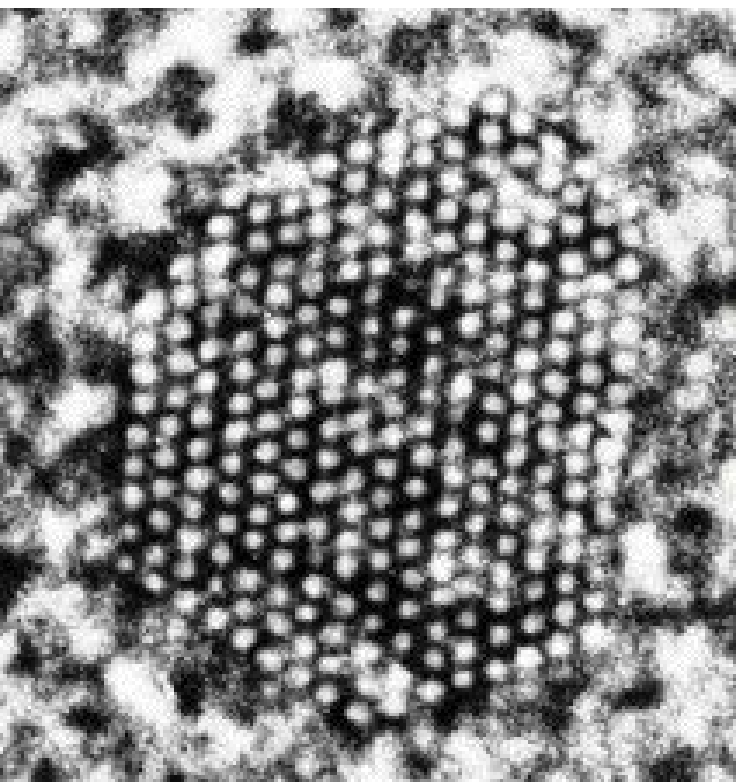}
\caption{High Resolution Transmission Electron
Microscopy observation along 
a [110] direction of a 3 nm cobalt cluster
exhibiting a {\it f.c.c}--structure.}
\label{fig_TEM}
}
{
\includegraphics[width=.45\textwidth]{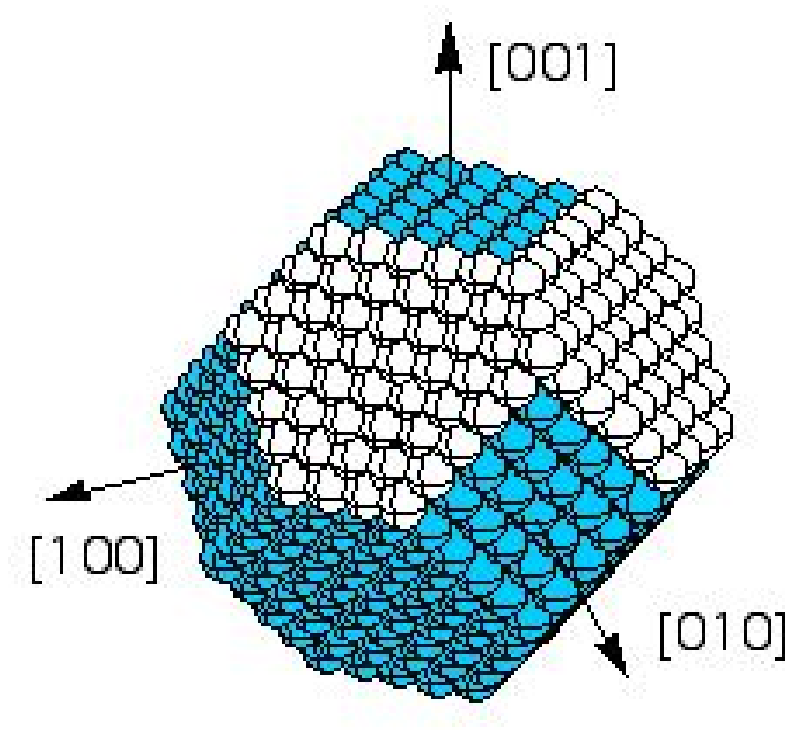}
\caption{Scheme of a typical cluster shape 
with light-gray atoms belonging to
the 1289-atoms truncated octahedron basis and 
dark-gray atoms belonging to the (111)
and (001) added facets.}
\label{cluster_scheme}
}
\end{figure}

We present here the first measurements on 
individual cobalt clusters
of 3 nm in diameter containing about a thousand 
atoms (Fig.~\ref{fig_TEM})~\cite{Jamet01a}.
In order to achieve the needed
sensitivity, Co clusters preformed in the gas phase are 
directly embedded in a co-deposited thin
Nb film which is subsequently used to pattern
micro-SQUIDs. 
A laser vaporization and inert gas 
condensation source is used to
produce an intense supersonic beam 
of nanosized Co clusters which can be
deposited in various matrices under 
Ultra High Vacuum (UHV) conditions. 
Due to the low energy deposition regime, clusters
do not fragment upon impact on the substrate~\cite{Perez97}.
The niobium matrix is simultaneously deposited
from a UHV electron gun evaporator
leading to continuous films with a low 
concentration of embedded Co clusters~\cite{JametPRB00}.
These films are used to pattern
planar micro-bridge-DC-SQUIDs
by electron beam lithography.
The later ones allow us to detect the 
magnetization reversal of a single
Co cluster for an applied magnetic field in any direction
and in the temperature range between 0.03 and 30 K
(Sect.~\ref{micro_SQUID}).
However, the desired sensitivity is only achieved for Co
clusters embedded into the micro-bridges where the magnetic 
flux coupling is high enough. Due to the low 
concentration of embedded Co clusters, we have a maximum
of 5 non-interacting particles in a micro-bridge 
which is 300 nm long and 50 nm wide. 
We can separately detect the magnetization switching for
each cluster. Indeed they are clearly different in intensity 
and orientation because of the random distribution
of the easy magnetization directions.
The {\it cold mode} method (Sect.~\ref{cold}) 
in combination with 
the {\it blind} method (Sect.~\ref{blind}) allows us 
to detect separately the magnetic signal for each cluster.

High Resolution Transmission Electron Microscopy observations
showed that the Co clusters are well crystallized 
in a {\it f.c.c}--structure (Fig.~\ref{fig_TEM}) 
with a sharp size distribution~\cite{JametPRB00}. 
They mainly form truncated
octahedrons (Fig.~\ref{cluster_scheme})~\cite{Jamet01a}.

\begin{figure}
\begin{center}\leavevmode
\includegraphics[width=.55\textwidth]{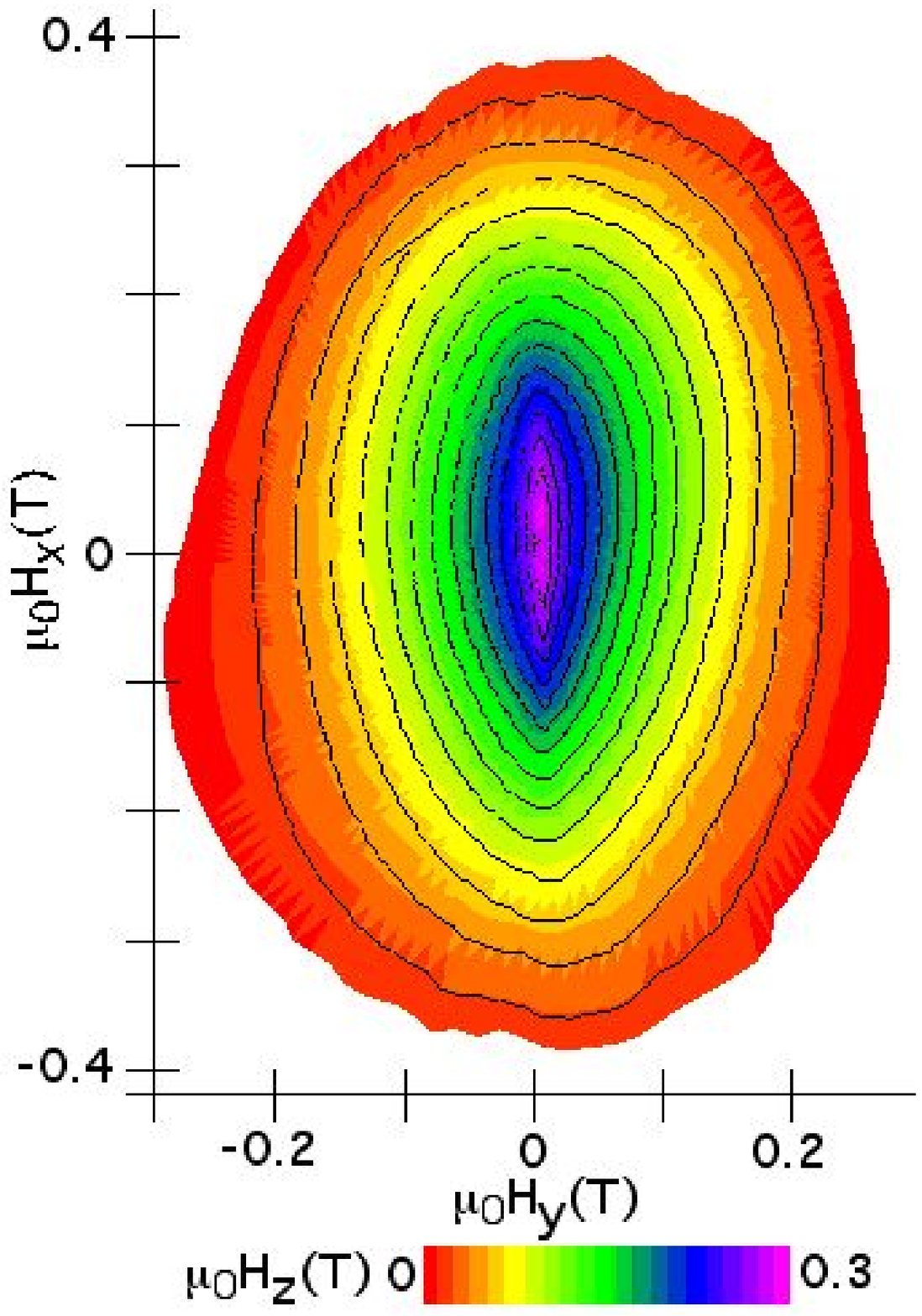}
\includegraphics[width=.7\textwidth]{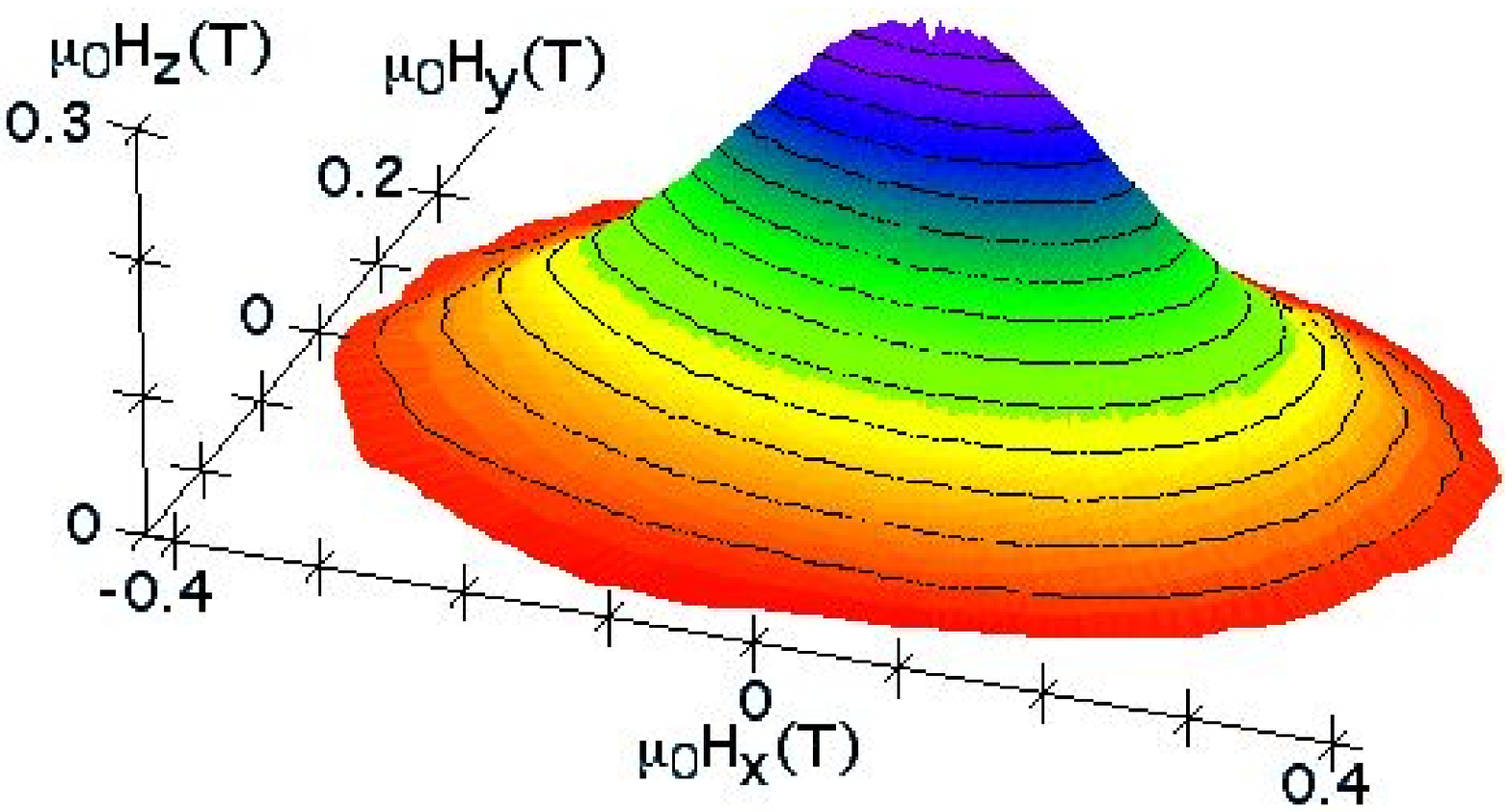}
\caption{Top view and side view of the experimental 
three dimensional angular dependence of the switching field 
of a 3 nm Co cluster at 35 mK. 
This surface is symmetrical with respect to 
the $H_x$--$H_y$--plane and only the upper part ($\mu_0 H_z >$ 0 T) is shown.
Continuous lines on the surface are contour lines 
on which $\mu_0 H_z$ is constant.}
\label{ast_Co_cluster}
\end{center}
\end{figure}

\begin{figure}
\begin{center}\leavevmode
\includegraphics[width=.55\textwidth]{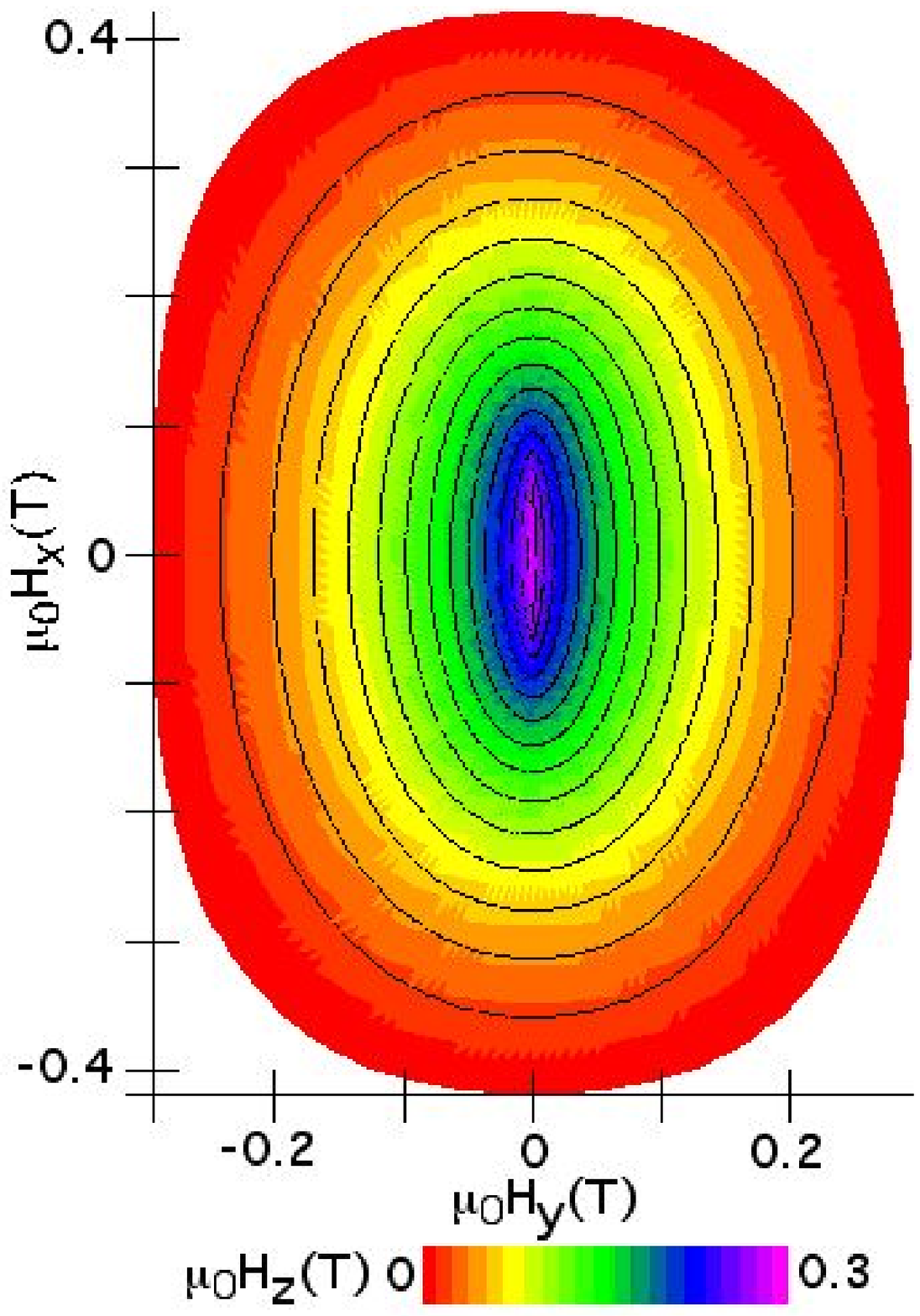}
\includegraphics[width=.7\textwidth]{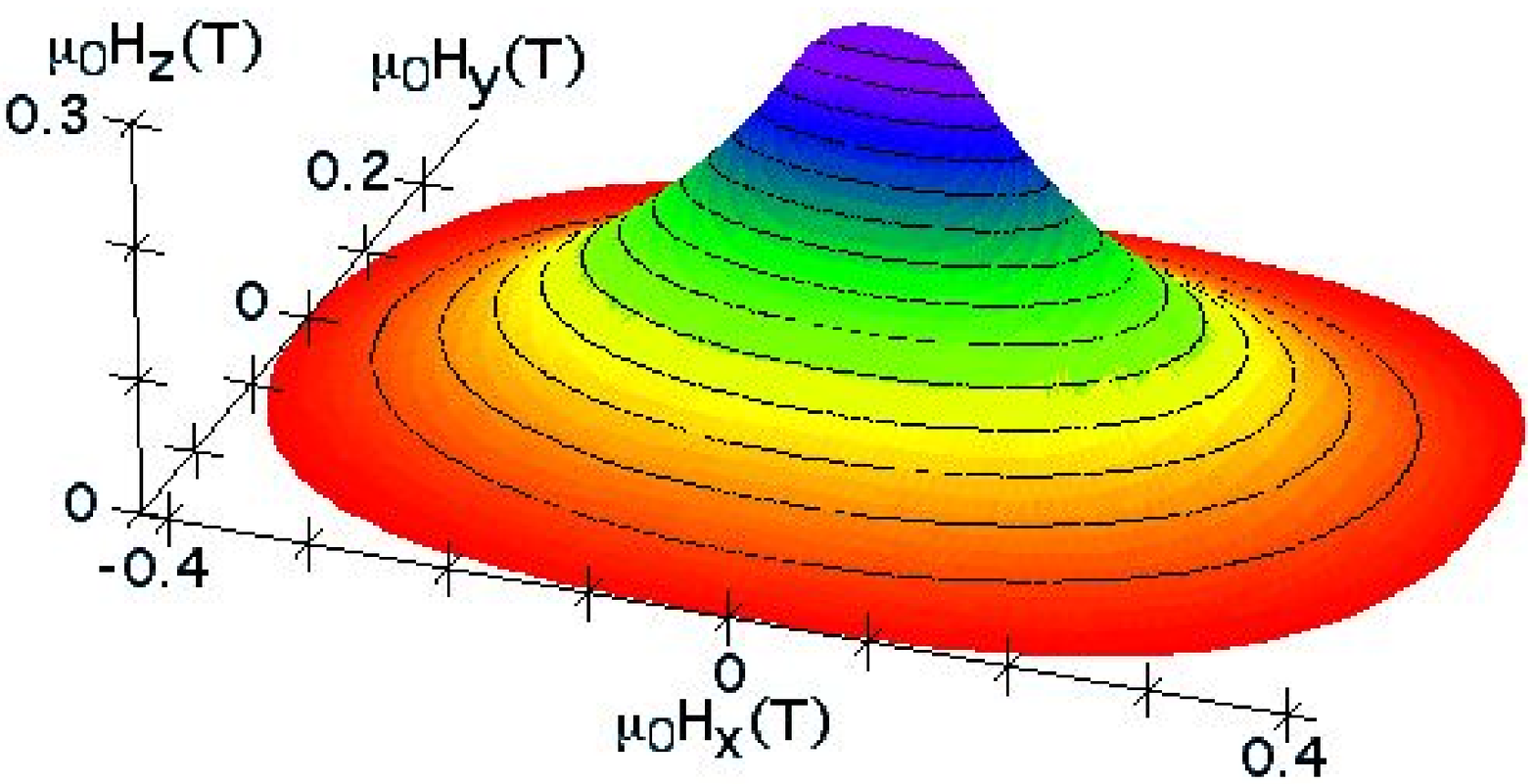}
\caption{Top view and side view of the theoretical switching field 
surface considering second and forth order terms 
in the anisotropy energy. }
\label{ast_Co_cluster_fit}
\end{center}
\end{figure}

Fig.~\ref{ast_Co_cluster} displays a typical 
measurement of switching fields in
three dimensions of a 3 nm Co cluster at $T$ = 35 mK.
This surface is a three-dimensional picture directly 
related to the anisotropy involved in the
magnetization reversal of the particle (Sect.~\ref{St_W}).
It can be reasonably fitted with the generalized 
Stoner and Wohlfarth model~\cite{Thiaville98}
(Sect.~\ref{St_W_general}).
We obtain the following anisotropy energy: 
\begin{equation}
E_{0}(\vec{m})/v = -K_{1}m_{z}^{2} + K_{2}m_{x}^{2}
    -K_{4}(m_{x'}^{2}m_{y'}^{2} + m_{x'}^{2}m_{z'}^{2} + m_{y'}^{2}m_{z'}^{2}) 
\label{eq_biaxial}
\end{equation}
where $K_{1}$ and $K_{2}$ are the anisotropy constants along $z$ and
$x$, the easy and hard magnetization axis, respectively.
$K_{4}$ is the forth order anisotropy constant and 
the $(x'y'z')$ coordinate system is deduced from $(xyz)$ 
by a 45$^{\circ}$ rotation around the $z$ axis. 
We obtained $K_{1}=2.2 \times 10^{5}$ J/m$^{3}$,
$K_{2}=0.9 \times 10^{5}$ J/m$^{3}$, and
$K_{4}=0.1 \times 10^{5}$ J/m$^{3}$.
The corresponding theoretical surface is showed in 
Fig.~\ref{ast_Co_cluster_fit}. 
Furthermore, we measured the temperature dependence of the 
switching field distribution (Sect.~\ref{cluster_T}). 
We deduced the blocking temperature of 
the particle $T_{\rm B} \approx$ 14 K, 
and the number of magnetic atoms in 
this particle: $N \approx$ 1500 atoms (Sect.~\ref{cluster_T}). 
Detailed measurements on about 20 different particles 
showed similar three dimensional switching field distributions 
with comparable anisotropy 
($K_{1}=(2.0 \pm 0.3) \times 10^{5}$ J/m$^{3}$,
$K_{2}=(0.8 \pm 0.3) \times 10^{5}$ J/m$^{3}$, and
$K_{4}=(0.1 \pm 0.05) \times 10^{5}$ J/m$^{3}$)
and size ($N = 1500 \pm 200$ atoms).

In the following, we analyze various contributions to the 
anisotropy energy of the Co clusters. Fine structural studies 
using EXAFS measurements~\cite{JametPRB00} were performed 
on 500 nm thick niobium films containing a very low 
concentration of cobalt clusters. They showed that niobium 
atoms penetrate the cluster surface to almost two atomic 
monolayers because cobalt and niobium are miscible elements. 
Further magnetic measurements~\cite{JametPRB00} on the 
same samples showed that these two atomic monolayers 
are magnetically dead. For this reason, we estimated 
the shape anisotropy of the typical nearly spherical 
deposited cluster in Fig.~\ref{cluster_scheme} after removing 
two atomic monolayers from the surface. By calculating all 
the dipolar interactions inside the particle assuming 
a bulk magnetic moment of $\mu_{at} = 1.7 \mu_{\rm B}$, 
we estimated the shape anisotropy constants: 
$K_{1} \approx 0.3 \times 10^{5}$ J/m$^{3}$
along the easy magnetization axis and 
$K_{2} \approx 0.1 \times 10^{5}$ J/m$^{3}$
along the hard magnetization axis. These values are much smaller
than the measured ones which means that $E_{\rm shape}$ is 
not the main cause of the second order 
anisotropy in the cluster. 

The forth order term $K_{4}=0.1 \times 10^{5}$ J/m$^{3}$ 
should arise from the cubic magnetocrystalline 
anisotropy in the {\it f.c.c} 
cobalt clusters. However, this value is smaller 
than the values reported 
in previous works~\cite{Lee90,Chuang94}.
This might by due to the different atomic 
environment of the surface atoms
in respect to that of bulk {\it f.c.c.} Co.
Taking the value of the bulk \cite{Lee90,Chuang94}: 
$K_{\rm bulk} = 1.2 \times 10^{5}$ J/m$^{3}$
only for the core atoms in the cluster,
we find $K_{\rm MC} \approx 0.2 \times 10^{5}$ J/m$^{3}$
which is in reasonable agreement with our measurements.

We expect that the contribution of the magnetoelastic 
anisotropy energy $K_{\rm MC}$ coming from the matrix-induced 
stress on the particle is also small. Indeed, using the 
co-deposition technique, niobium atoms cover uniformly 
the cobalt cluster creating an isotropic distribution of stresses. 
In addition, they can relax preferably inside the matrix 
and not in the particle volume since niobium is less rigid than cobalt. 
We believe therefore that only interface anisotropy $K_{\rm surface}$ 
can account for the experimentally observed second order anisotropy terms. 
Niobium atoms at the cluster surface might enhance this 
interface anisotropy through surface strains and 
magnetoelastic coupling. This emphasizes the
dominant role of the surface in nanosized systems. 

In conclusion, the three dimensional switching
field measurements of individual clusters 
giving access to its magnetic
anisotropy energy. A quantitative understanding of the latter
is still difficult but it seems that the cluster-matrix 
interface provides the main contribution to the magnetic
anisotropy.
Such interfacial effects could be promising
to control the magnetic anisotropy in small particles in
order to increase their blocking temperature up to the
required range for applications.

%%%%%%%%%%%%%%%%%%%%%%%%%%%%%%%%
\subsection{Uniform Rotation with cubic anisotropy}
%%%%%%%%%%%%%%%%%%%%%%%%%%%%%%%%

We have seen in the previous section that the magnetic anisotropy is 
often dominated by second order anisotropy terms. However for 
nearly symmetric shapes, forth order terms 
\footnote{For example, the forth order terms of {\it f.c.c.} 
magnetocrystalline anisotropy.} can be 
comparable or even dominant with respect to the second order terms. 
Therefore, it is interesting to discuss further the features of forth 
order terms. We restrict the discussion to the 2D 
problem~\cite{Ching91,Thiaville98} (see~\cite{Thiaville00} for 3D).

The reversal of the magnetization is described by equation~\ref{eq_E} 
which can be rewritten in 2D:
\begin{equation}
E(\theta) = E_0(\theta) - 
               \mu_0 v M_{\rm s} (H_x \cos(\theta) + H_y \sin(\theta))
\label{eq_E_2D}
\end{equation}
where $v$ and $M_{s}$ are the magnetic volume
and the saturation magnetization of the particle respectively, $\theta$ is
the angle between the magnetization direction and $x$, $H_x$ and $H_y$ 
are the components of the external magnetic field along $x$ and $y$,
and $E_0(\theta)$ is the magnetic anisotropy energy.
The conditions of critical fields ($\partial E/\partial \theta = 0$ and 
$\partial E^2/\partial \theta^2 = 0$) yield a parametric form of the locus of 
switching fields:
\begin{eqnarray}
	H_x = - \frac{1}{2 \mu_0 v M_{\rm s}} 
	        \left( \sin(\theta) \frac{dE}{d\theta} +
			       \cos(\theta) \frac{d^2E}{d\theta^2} \right)
	\label{eq_hx}
\\
	H_y = + \frac{1}{2 \mu_0 v M_{\rm s}} 
	        \left( \cos(\theta) \frac{dE}{d\theta} -
			       \sin(\theta) \frac{d^2E}{d\theta^2} \right) 
	\label{eq_hy}
\end{eqnarray}

\begin{figure}
\begin{center}\leavevmode
\includegraphics[width=.7\textwidth]{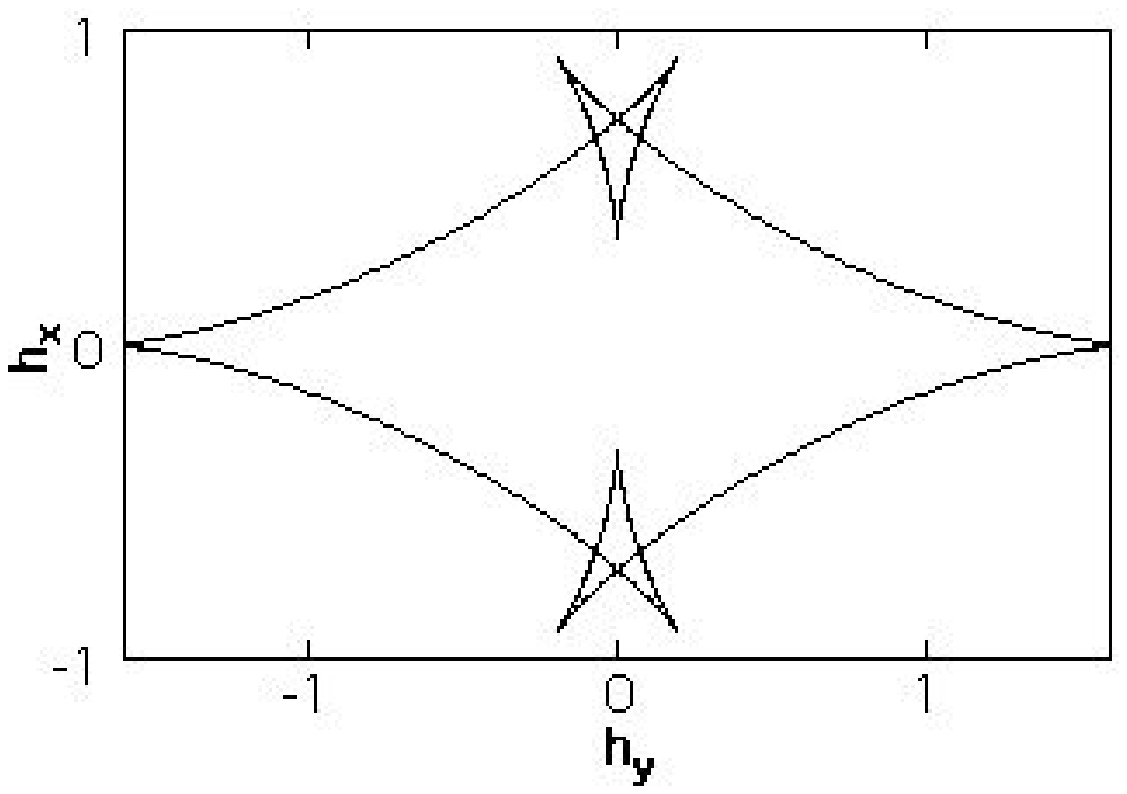}
\caption{Angular dependence of the 
switching field obtained from 
Eqs.~\ref{eq_hx} to~\ref{eq_E0_fcc} with $K_1 > 0$ 
and $K_2 = - 2/3 K_1$. The field is 
normalized by the factor $2 K_1 / (\mu_0 M_{\rm s})$.}
\label{ast_fcc}
\end{center}
\end{figure}

\begin{figure}
\sidebyside{
\includegraphics[width=.45\textwidth]{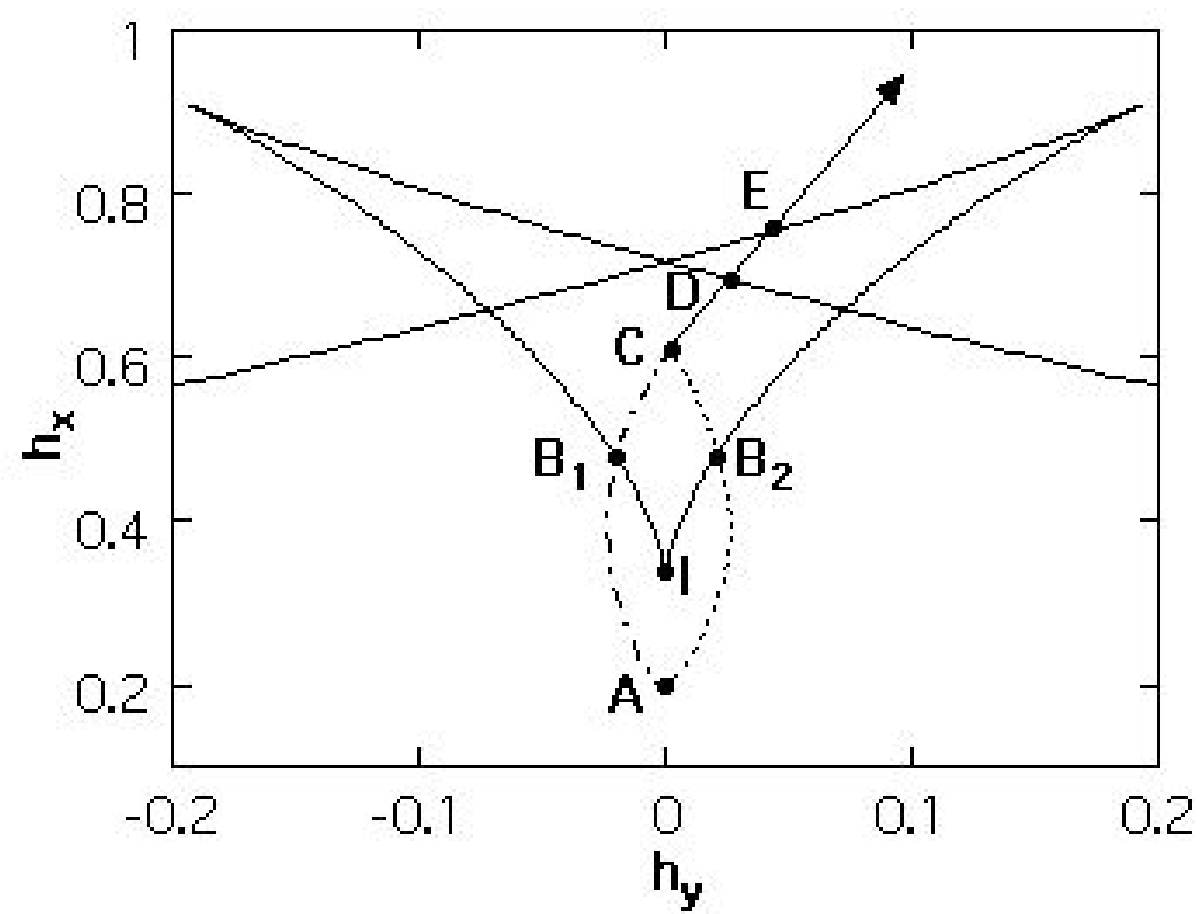}
\caption{Enlargement of angular dependence 
of the switching field of 
Fig.~\ref{ast_fcc}. Two possible paths 
of the applied field are 
indicated: starting from point {\bf A} and going 
over the point {\bf B$_1$} leads to magnetization reversal 
in {\bf E} whereas going over the point {\bf B$_2$} leads to reversal 
in {\bf D}.}
\label{ast_fcc_zoom}
}
{
\includegraphics[width=.45\textwidth]{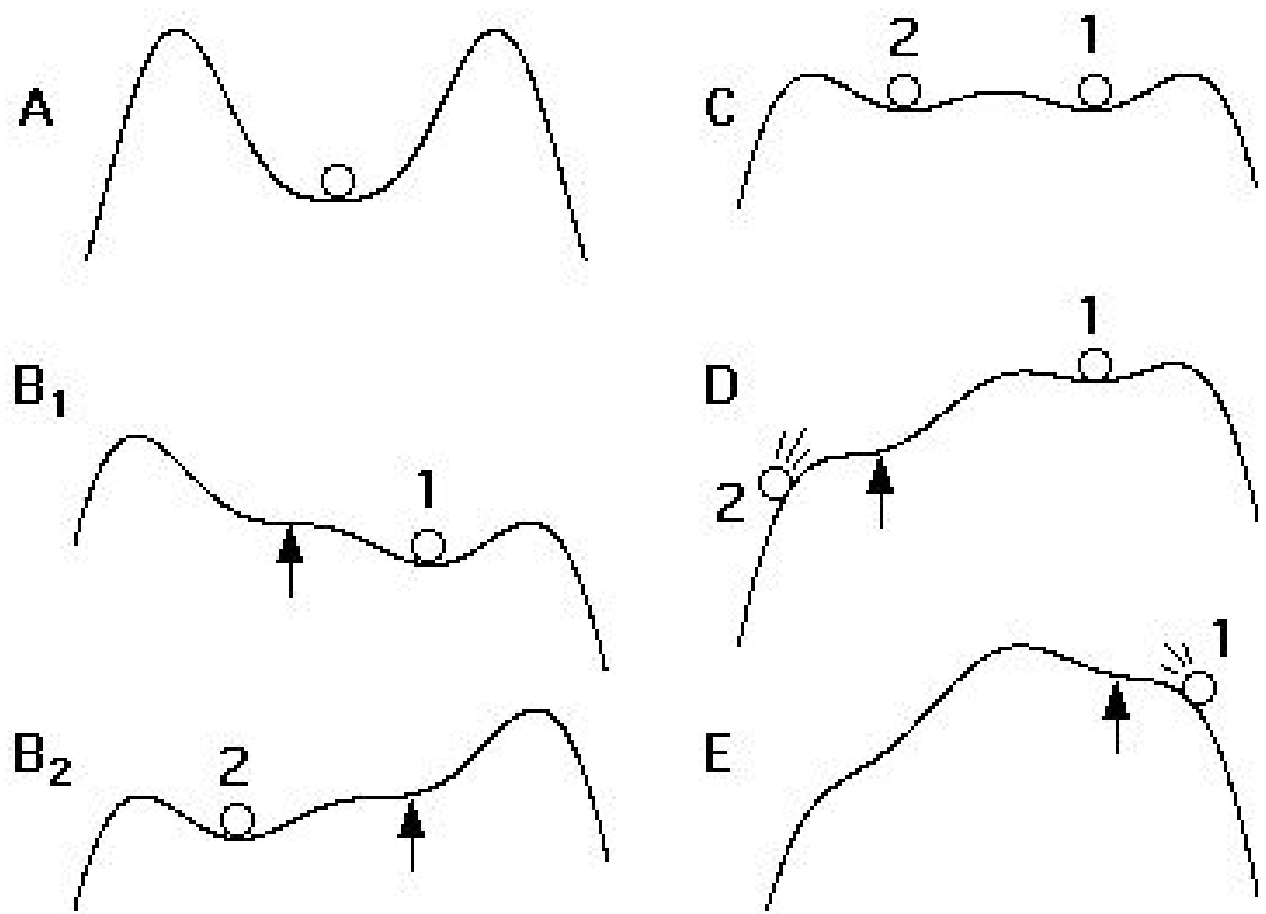}
\caption{Scheme of the potential energy near the metastable state 
for different applied fields as indicated in Fig.~\ref{ast_fcc_zoom}.
The ball represent the state of the magnetization and the arrow 
locate the appearing or disappearing well.}
\label{potential_fcc}
}
\end{figure}

As an example we study a system with uniaxial shape 
anisotropy and cubic anisotropy.
The total magnetic anisotropy energy can be described by:
\begin{equation}
E_0(\theta) = v K_1 \sin^2(\theta + \theta_0) + 
              v K_2 \sin^2(\theta)\cos^2(\theta) 
\label{eq_E0_fcc}
\end{equation}
where $K_1$ and $K_2$ are anisotropy constants ($K_1$ could be a 
shape anisotropy and $K_2$ the cubic crystalline anisotropy of a {\it 
f.c.c.}--crystal). $\theta_0$ is a constant which allows to turn one 
anisotropy contribution with respect to the other one.
Fig.~\ref{ast_fcc} displays an example of a critical curve 
which can easily be calculated from 
Eqs.~\ref{eq_hx} to \ref{eq_E0_fcc}.
When comparing the standard Stoner--Wohlfarth astroid 
(Fig.~\ref{ast_ST_W}) with Fig.~\ref{ast_fcc}, we can realize that the 
critical curve can cross itself several times. In this case, the 
switching field of magnetization depends on 
the path followed by the applied 
field. In order to understand this point, let us follow the energy 
potential (\ref{eq_E_2D} and \ref{eq_E0_fcc}) when sweeping the 
applied field as indicated in Fig.~\ref{ast_fcc_zoom}. When the field 
is in {\bf A}, the energy $E$ has two minima and the magnetization is in 
the metastable potential well. As the field increases, the metastable 
well becomes less and less stable. Let us compare two paths, one 
going along {\bf A} $\rightarrow$ {\bf B$_1$} $\rightarrow$ 
{\bf C} $\rightarrow$ {\bf D} $\rightarrow$ {\bf E}, 
the other over {\bf B$_2$} instead of {\bf B$_1$}.
Fig.~\ref{potential_fcc} shows $E$ in the vicinity of the 
metastable well for different field values along 
the considered paths (the stable potential well
is not presented).
One can realize that the state of the 
magnetization in {\bf C} depends on 
the path followed by the field: going over 
{\bf B$_1$} leads to the magnetization state in the 
left metastable well (1), whereas going 
over {\bf B$_2$} leads to the right 
metastable well (2). The latter path leads to magnetization switching 
in {\bf D}, and the former one to a switching in {\bf E}. Note that a small
magnetization switch happens when reaching {\bf B$_1$} or {\bf B$_2$}.
Point {\bf I} is a supercritical bifurcation.

\begin{figure}
\begin{center}\leavevmode
\includegraphics[width=.7\textwidth]{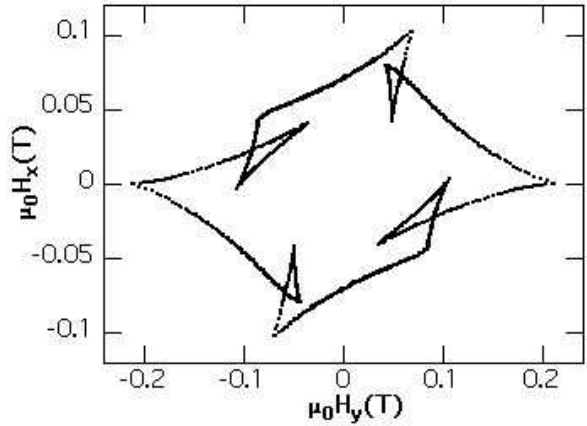}
\caption{Angular dependence of the switching field of a 3 nm Co 
cluster showing a strong influence of crystalline anisotropy.}
\label{part_fcc}
\end{center}
\end{figure}

The first measurement of such a field path dependence of the switching 
field were performed on single--domain FeCu nanoparticles of about 
15 nm with a cubic crystalline anisotropy and a small 
arbitrarily oriented shape anisotropy~\cite{Bonet98}. 
Fig.~\ref{part_fcc} presents switching field measurements of a 15 nm 
cobalt nanoparticle showing clearly a contribution 
of cubic crystalline anisotropy 
and the field path dependence of the switching 
field.

%%%%%%%%%%%%%%%%%%%%%%%%%%%%%%%%
\section{Non uniform magnetization reversal}
\label{non_uniform}
%%%%%%%%%%%%%%%%%%%%%%%%%%%%%%%%

We have seen in the previous sections that for 
extremely small particles, magnetization should 
reverse by uniform rotation. 
For somewhat larger single domain particles, 
non uniform reversal modes 
are more likely.
The simplest one is the curling which is discussed 
in the following section. 

%%%%%%%%%%%%%%%%%%%%%%%%%%%%%%%%
\subsection{Magnetization reversal by curling}
\label{curling}
%%%%%%%%%%%%%%%%%%%%%%%%%%%%%%%%

The simplest non uniform reversal mode is the curling 
reversal mode~\cite{Frei57,Aharoni96}.
The critical parameter is the exchange length 
$\lambda = \sqrt{A}/M_{\rm s}$, 
delimiting the region of uniform rotation and curling 
($A$ is the exchange constant). 
Therefore in the case of the size $R > \lambda$ 
($R$ is for example the radius of a cylinder), magnetization reversal 
via curling is more favorable. 
In the following, we review the analytical result of an ellipsoid of 
rotation,
which can by applied approximately to most of the shapes of 
nanoparticles
or nanowires~\cite{Aharoni99}. 

The variation of the switching field with the angle $\theta$ 
(defined between the applied field and the long axis of the ellipsoid) 
is given by~\cite{Aharoni97}:
\begin{equation}
H_{\rm sw}^0 = 
	\frac{M_{\rm s}}{2} \frac {a_x a_z} 
	{\sqrt{a_z^2\sin^{2}\theta + a_x^2\cos^{2}\theta}} \;,
\label{eq_curl_ell}
\end{equation}
where $a_{x,z} = 2 N_{x,z} - k/S^2$, 
$N_{x,z}$ are the demagnetization factors,
$S = R/\lambda$, and $R$ is the minor semi-axis of the ellipsoid. 
The parameter $k$ is a monotonically decreasing 
function of the aspect ratio of the ellipsoid. 
This function is plotted in Fig.~1 of 
Ref.~\cite{Aharoni86}. 
The smallest and highest value of k is that for an infinite 
cylinder (k = 1.079) and a sphere (k = 1.379), respectively. 

For a long ellipsoid of rotation, 
the demagnetization factors are given by:
\begin{equation}
N_z = \frac{1}{n^2 - 1} 
	\frac {n} {\sqrt{n^2 - 1}}
	\ln\left(n-1+\sqrt{n^2 - 1}\right)
	 \;,  \;
N_x = \frac{1 - N_z}{2} \;,
\label{eq_N_z}
\end{equation}
where $n$ is the ratio of the length to the diameter. 

\begin{figure}
\begin{center}\leavevmode
\includegraphics[width=.7\textwidth]{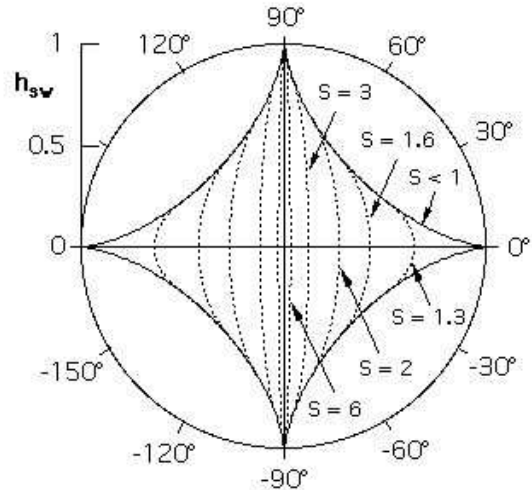}
\caption{Angular dependence of the switching field of an infinite 
cylinder 
for several reduced cylinder radii $S$. For $S < 1$, the switching 
field is
given by the uniform rotation mode (Sect.~\ref{St_W}).}
\label{fig_curling}
\end{center}
\end{figure}

For an infinite cylinder Eq.~\ref{eq_curl_ell} 
becomes~\cite{Frei57,Aharoni66}:
\begin{equation}
H_{\rm sw}^0 = 
	\frac{M_{\rm s}}{2} 
	\frac {h_t (1+h_t)} {\sqrt{h_t^2 + (1+2 h_t)\cos^{2}\theta}} \;,
\label{eq_curl_cy}
\end{equation}
where $h_t = -1.079/S^2$. 
Eq.~\ref{eq_curl_cy} is a good approximation for a very long ellipsoid 
of rotation. 
It is plotted in Fig.~\ref{fig_curling} for several radii of an 
infinite cylinder.

The case of uniform rotation of magnetization was generalized 
by Thiaville to an arbitrary anisotropy energy 
function and to three dimensions (Sect.~\ref{St_W_general}). 
For the curling mode, this generalization is not possible. 
However, approximated calculations were 
proposed~\cite{Ishii91,Aharoni99,Aharoni00}
and micro-magnetic simulations were performed~\cite{Ferre97}.

%%%%%%%%%%%%%%%%%%%%%%%%%%%%%%%%
\subsection{Experimental evidence for magnetization reversal by 
curling}
\label{curling_exp}
%%%%%%%%%%%%%%%%%%%%%%%%%%%%%%%%

We report here the first studies of isolated nanoscale wires 
with diameters smaller than 100 nm, for which single-domain 
states could be expected~\cite{WW_PRL96_Ni,WW_PRB97_Ni}. 
The cylindrical geometry, 
with its large shape anisotropy, is well suited for comparison with 
theory.

Ni wires were produced by filling electrochemically the pores 
of commercially available nanoporous track-etched polycarbonate 
membranes of thicknesses of 10 $\mu$m. 
The pore size was chosen in the range 
of 30 to 100 nm~\cite{Doudin95,Meier96}. 
In order to place one wire on the SQUID detector, 
we dissolved the membrane in chloroform and put a drop on a chip 
of some hundreds of SQUIDs. 
Magnetization measurements were performed on SQUIDs 
with a single isolated wire (Fig.~\ref{SQUID_SEM}). 

\begin{figure}
\begin{center}\leavevmode
\includegraphics[width=.7\textwidth]{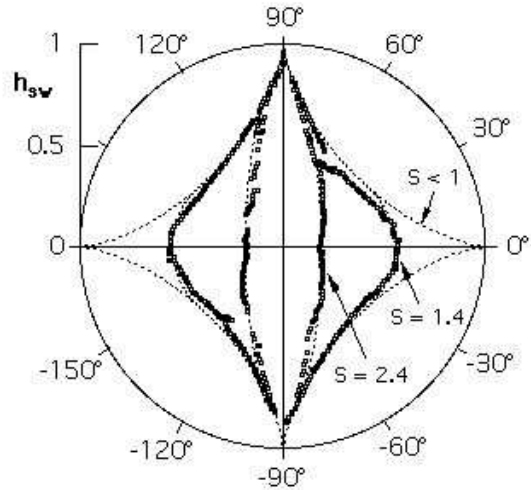}
\caption{Angular dependence of the switching field of two Ni wires
with a diameter of 45 nm and 92 nm, {\it i.e.} $S$ = 1.4 and 2.4. 
The switching fields are
normalized by 125 mT and 280 mT respectively.}
\label{fig_Ni_wires}
\end{center}
\end{figure}

The angular dependence of the switching field of 
wires with 45 and 92 nm in diameter are shown in 
Fig.~\ref{fig_Ni_wires}. 
These measurements are in quantitative agreement with the
curling mode (Eq.~\ref{eq_curl_ell}). 
Nevertheless, dynamical measurements
showed a nucleation volume which is much
smaller than the wire volume~\cite{WW_PRL96_Ni,WW_PRB97_Ni} 
(Sect.~\ref{Ni_wire_T}). Therefore, we believe that
the magnetization reversal starts close to
curling instability but the nucleation happens 
in a small fraction of the wire only,
then rapidly propagating along the
whole sample. This picture is also in good agreement with 
micromagnetic simulations~\cite{Ferre97} and 
the micromagnetic model of Braun~\cite{Braun99}.
\footnote{Note that the curling model 
only predicts an instability. There has never been
a claim to describe what happens 
afterwards~\cite{Aharoni96}.}

The angular dependence of the switching field 
of Ni wires with larger diameters (270 - 450 nm) 
were measured at room temperature by 
Lederman {\it et al.}~\cite{Ledermann95}. 
Their results could roughly be explained by the curling mode.

%%%%%%%%%%%%%%%%%%%%%%%%%%%%%%%%
\subsection{Magnetization reversal by nucleation and 
annihilation of domain walls}
\label{nucleation}
%%%%%%%%%%%%%%%%%%%%%%%%%%%%%%%%

For magnetic particles which have at 
least two dimensions much larger
than the domain wall width, the magnetization reversal 
may occur via nucleation/propagation and annihilation
of one or several domain walls happening at two
or more applied fields.
We focus here on a 30-nm-thick elliptic Co particle
defined by electron beam lithography and 
lift-off techniques out of sputtered thin films~\cite{WW_PRB96}.
The Co film has a nanocrystalline structure leading to
a magnetically soft material with 
a coercive field value of 3 mT at 4 K. 
Therefore we neglected the magneto-crystalline anisotropy. 
The nanofabricated particles have an elliptic contour with
in-plane dimensions of 300 nm $\times$ 200 nm and a thickness of
30 nm. 

\begin{figure}
\begin{center}\leavevmode
\includegraphics[width=.7\textwidth]{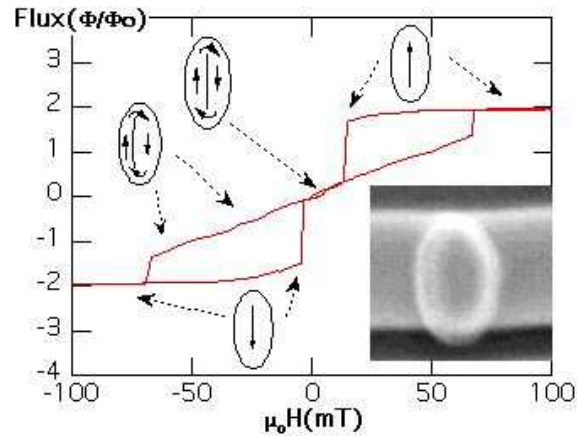}
\caption{Hysteresis loops  at 0.1 K of the elliptic 
Co particle seen in 
the inset which shows the electron 
micrograph of the wire of a 
micro-bridge SQUID with the Co particle 
(300 nm $\times$ 200 nm $\times$ 30 nm). 
The in-plane field is applied along 
the long axis of the particles.
The domain wall structure in an
elliptical particle is also presented schematically 
as proposed by van den Berg. Arrows indicate the
spin direction. The two small magnetization jumps 
near $\approx \pm5$mT might be due to the reversal
of the center vortex of the domain structure.}
\label{fig_Co_domain}
\end{center}
\end{figure}

In order to study the domain structure of our particles, we
measure the angular dependence of hysteresis loops. 
Fig.~\ref{fig_Co_domain}
shows a typical hysteresis loop of an individual Co
particle. The magnetic
field is applied in the plane of the particle. 
The hysteresis loop is mainly characterized
by two magnetization jumps. Starting from a
saturated state, the first jump can be associated with
domain wall nucleation and the second jump with domain wall
annihilation. During these jumps, the magnetization
switches in less than 100 $\mu$s (our time resolution, 
see Sect.~\ref{feedback}). The reversible
central region of the hysteresis loops is evidence for
the motion of the domain wall through the particle.

The simplest domain structure, showing such a hysteresis
loop, has been proposed by van den Berg~\cite{Berg87}
in zero field and
calculated for fields smaller than the saturation field by Bryant
and Suhl~\cite{Bryant89}. 
\footnote{Note that these models are 2D 
and neglect the domain wall width.
Therefore, they can give only a qualitative description.}
This domain structure has been observed 
experimentally
on low anisotropy circular thin film disks (100 $\mu$m in diameter) 
using high-resolution Kerr techniques~\cite{Ruhrig90}. 
In zero field, the particle has a vortex-like domain wall
as shown in Fig.~\ref{fig_Co_domain}. 
When a magnetic
field is applied, this domain wall is pushed to the border of the
particle. For higher fields the domain wall is annihilated
and the particle becomes single domain.
The main conditions of the model of 
van den Berg model are (i) that the magnetic
material is very soft and (ii) that the system is two dimensional.
The first condition is satisfied by our particles as they
are made of a randomly oriented nanocrystallized Co, being
very soft. The second condition is not quite well satisfied, 
however MFM measurements confirmed 
this domain structure~\cite{Fernandez98}. Furthermore,
we obtained similar results for thinner particles (10 and 20 nm). 
More complicated domain structures as proposed by 
van den Berg~\cite{WW_PRB96} may be excluded 
by the fact that similar Co particles
of length smaller than 200 nm are single 
domain~\cite{WW_JMMM95a,WW_JMMM95b}.

\begin{figure}
\begin{center}\leavevmode
\includegraphics[width=.7\textwidth]{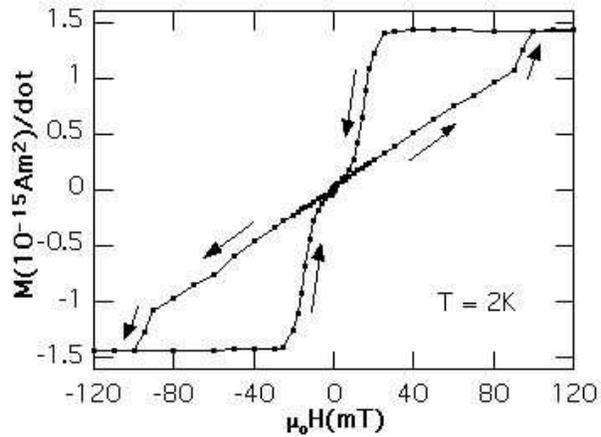}
\caption{Hysteresis loops of the magnetic moment of the array
of 1.8 $\times$10$^7$ Co particles (300 nm $\times$ 200 nm $\times$30 
nm).
The in-plane field is applied along the long axis of the particles.}
\label{fig_Co_array}
\end{center}
\end{figure}

After studying the magnetization reversal of
individual particles, the question arises of which manner the
properties of a macroscopic sample are based on one-particle
properties. In order to answer this question, we fabricated a sample
consisting of 1.8 $\times$10$^7$ nearly 
identical elliptic Co particles of about
the same dimensions and material as the individual particle
studied above. These particles are placed on a Si substrate
with a spacing of 2 $\mu$m. Because of this large spacing, dipole
interactions between particles are negligible. Fig.~\ref{fig_Co_array}
shows the hysteresis loop of the array of Co particles when
the field is applied parallel to the long axis of the particle.
This hysteresis loop shows the same characteristics as the
hysteresis loop of one particle (Fig.~\ref{fig_Co_domain}), {\it i.e.}, 
nucleation and
annihilation of domain walls. 
Because of switching field distributions
mainly due to surface defects and a slight distribution
of particle sizes, shapes, etc., the domain wall nucleation
and annihilation are no longer discontinuous
although still irreversible. They take
place along continuous curves with a width of about 10 mT.

% Bla-bla conclusion~\cite{Hehn96}

%%%%%%%%%%%%%%%%%%%%%%%%%%%%%%%%
\title[Influence of temperature on the magnetization reversal]
{Influence of temperature on the magnetization reversal}
\label{chap_reversal_clas}
\markboth{Influence of temperature on the magnetization reversal}
{Influence of temperature on the magnetization reversal}
%%%%%%%%%%%%%%%%%%%%%%%%%%%%%%%%

The thermal fluctuations of the magnetic moment 
of a single-domain ferromagnetic 
particle and its decay towards thermal equilibrium 
were introduced by N\'eel~\cite{Neel49a,Neel49b} 
and further developed by Bean and Livingston~\cite{Bean55,Bean59}, and 
Brown~\cite{Brown59,Brown63a,Brown63b}.
The simplest case is an assembly of independent particles
having no magnetic anisotropy.
In the absence of an applied magnetic field, the magnetic 
moments are randomly oriented. 
The situation is similar to paramagnetic atoms 
where the temperature dependence of the magnetic susceptibility 
follows a Curie behavior, and the field dependence 
of magnetization is described by a Brillouin function. 
The only difference is that the magnetic moments 
of the particles are much larger than those of the paramagnetic atoms.
Therefore, the quantum mechanical Brillouin function can be replaced 
by the classical limit for larger magnetic 
moments, namely the Langevin function.
This theory is called superparamagnetism. 
The situation changes however as soon as 
magnetic anisotropy is present which 
establishes one or more preferred orientations of the 
particle's magnetization (Sect.~\ref{chap_reversal_T_0}).
In the following, we present an overview over
the simplest model describing 
thermally activated magnetization reversal
of single isolated nanoparticles
which is called the N\'eel--Brown Model.
After a brief review of the model (Sect.~\ref{Neel_Brown}),
we present experimental methods to study
the thermally activated magnetization 
reversal (Sect.~\ref{methods_NB}). Finally, we
discuss some applications of the N\'eel--Brown 
Model (Sect.~\ref{Exp_NB}).

%%%%%%%%%%%%%%%%%%%%%%%%%%%%%%%%
\section{N\'eel--Brown model of thermally 
activated magnetization reversal}
\label{Neel_Brown}
%%%%%%%%%%%%%%%%%%%%%%%%%%%%%%%%

In N\'eel and Brown's model of thermally 
activated magnetization reversal, a 
single domain magnetic particle has two 
equivalent ground states of opposite 
magnetization separated by an energy barrier 
which is due to shape and crystalline 
anisotropy. The system can escape from one 
state to the other by thermal activation over 
the barrier. Just as in the Stoner--Wohlfarth 
model, they assumed uniform magnetization 
and uniaxial anisotropy in order to derive a 
single relaxation time. N\'eel supposed further 
that the energy barrier between the two equilibrium 
states is large in comparison to the 
thermal energy $k_{\rm B}T$ which justified a 
discrete orientation 
approximation~\cite{Neel49a,Neel49b}. 
Brown criticized 
N\'eel's model because the system is not 
explicitly treated as a gyromagnetic 
one~\cite{Brown59,Brown63a,Brown63b}. Brown 
considered the magnetization 
vector in a particle to wiggle 
around an energy minimum, then jump to the vicinity of 
the other minimum, then wiggle around 
there before jumping again. He supposed that the orientation 
of the magnetic moment may 
be described by a Gilbert equation with a random 
field term which is assumed to be white 
noise. On the basis of these assumptions, Brown 
was able to derive a Fokker--Planck 
equation for the distribution of magnetization 
orientations. Brown did not solve his 
differential equation. Instead he tried some analytic 
approximations and an asymptotic 
expansion for the case of the field parallel or 
perpendicular to the easy axis of 
magnetization. More recently, 
Coffey et al.~\cite{Coffey95,Coffey98a} 
found by numerical methods an exact solution 
of Brown's differential equation for uniaxial 
anisotropy and an arbitrary applied field 
direction. They also derived an asymptotic general 
solution for the case of large energy barriers 
in comparison to the thermal energy 
$k_{\rm B}T$. 
This asymptotic solution is of particular interest 
for single particle measurements and 
is reviewed in the following.

For a general asymmetric bistable
energy potential $E = E(\vec{m},\vec{H})$ 
(Eq.~\ref{eq_E}) with the orientation of magnetization 
$\vec{m} =  \vec{M}/M_{\rm s}$
($M_{\rm s}$ is the spontaneous magnetization), 
$\vec{H}$ is the applied field, and with minima 
at $\vec{n}_1$ and $\vec{n}_2$ separated 
by a potential barrier 
containing a saddle point at $\vec{n}_0$ 
(with the $\vec{n}_i$ coplanar), 
and in the case of $\beta (E_0 - E_i) >> 1$ 
where $\beta = 1/k_{\rm B}T$, 
and $E_i = E(\vec{n}_i,\vec{H})$, 
Coffey et al. showed that the longest relaxation time 
\footnote{The inverse of the longest relaxation time is determined by 
the smallest non-vanishing eigenvalue of the appropriate 
Fokker--Planck equation~\cite{Coffey95,Coffey98a}.
All other eigenvalues can be neglected in the considered asymptotic 
limit of $\beta (E_0 - E_i) >> 1$.}
is given by 
the following equation which is valid in the intermediate 
to high damping limit (IHD) defined 
by $\alpha \beta (E_0 - E_i) > 1$~\cite{Coffey98c}:
\begin{equation}
\tau^{-1} = \frac {\Omega_0}{2\pi \omega_0} \left\lbrack
	\omega_1 e^{-\beta(E_0 - E_1)} +
	\omega_2 e^{-\beta(E_0 - E_2)}
	\right\rbrack \;,
\label{eq_tau_IHD}
\end{equation}
where $\omega_0$ and $\Omega_0$ are the saddle and damped saddle 
angular frequencies:
\begin{equation}
\omega_0 = \frac {\gamma}{M_{\rm s}} \sqrt{-c^{(0)}_1 c^{(0)}_2}\;,
\label{eq_omega_0}
\end{equation}
\begin{equation}
\Omega_0 = 	\frac {\gamma}{M_{\rm s}} 
		\frac {\alpha}{1 + \alpha^2} 
	\left\lbrack
	- c^{(0)}_1 - c^{(0)}_2 +
	\sqrt{(c^{(0)}_2 - c^{(0)}_1)^2 - 4 \alpha^{-2} c^{(0)}_1 c^{(0)}_2} 
	\right\rbrack \;.
\label{eq_Omega_0}
\end{equation}

$\omega_1$ and $\omega_2$ are the well angular frequencies:
\begin{equation}
\omega_i = \frac {\gamma}{M_{\rm s}} \sqrt{c^{(i)}_1 c^{(i)}_2}\;,
\label{eq_omega_i}
\end{equation}
with $i = 1$ and 2. 
$c^{(j)}_1$ and $c^{(j)}_2$ ($j = 0, 1, 2$) are the coefficients in 
the truncated Taylor series of the potential at well and saddle points, 
{\it i.e.} the curvatures of the potential at well and saddle points. 
$\gamma$ is the gyromagnetic ratio, 
$\alpha = \nu \gamma M_{\rm s}$ is the dimensionless damping factor and 
$\nu$ is the friction in Gilbert's equation (ohmic damping).

Whereas in the low damping limit (LD), 
defined by $\alpha \beta (E_0 - E_i) < 1$, the longest 
relaxation time is given by~\cite{Klik90a,Klik90b}:
\begin{equation}
\tau^{-1} = \frac {\alpha}{2\pi} \left\lbrack
	\omega_1 \beta(E_0 - E_1) e^{-\beta(E_0 - E_1)} +
	\omega_2 \beta(E_0 - E_2) e^{-\beta(E_0 - E_2)}
	\right\rbrack \;,
\label{eq_tau_LD}
\end{equation}
In this case, the energy dissipated in one cycle of motion in the well is very 
small in comparison to the thermal energy $k_{\rm B}T$. 

Experimentally, relaxation is observed only if $\tau$ 
is of the order of magnitude of
the measuring time of the experiment. This implies for all known 
single particle measurement techniques that $\beta (E_0 - E_i) >> 1$, 
{\it i.e.} the asymptotic solutions (\ref{eq_tau_IHD}) and (\ref{eq_tau_LD}) 
are always a very good approximation to the exact solution of 
Brown's Fokker--Planck equation~\cite{Coffey98b}. 
Due to an applied field,  $\beta (E_0 - E_1) >> \beta (E_0 - E_2)$ 
(taking $E_2$ as the metastable minimum) might be true.
Then the first exponential in equations 
(\ref{eq_tau_IHD}) and (\ref{eq_tau_LD}) can be neglected.

Concerning the possible values of $\alpha$, we remark that little information 
is available. Typical values should be between 0.01 and 5~\cite{Dormann97}, 
meaning that in practice $\alpha \beta (E_0 - E_i)$ 
can be $>>1$, $<<1$, or $\approx 1$. 
Thus the distinction between Eqs. (\ref{eq_tau_IHD}) and 
(\ref{eq_tau_LD}) becomes important.

Finally, we note that $c^{(j)}_1$ and $c^{(j)}_2$ ($j = 0, 1, 2$) 
can be found experimentally by measuring the critical surface of 
the switching field and applying the
calculation of Thiaville (Sect.~\ref{St_W_general})~\cite{Thiaville00}.

%%%%%%%%%%%%%%%%%%%%%%%%%%%%%%%%
\section{Experimental methods for the study of the N\'eel--Brown model} 
\label{methods_NB}
%of Thermally Activated Magnetization Reversal}
%%%%%%%%%%%%%%%%%%%%%%%%%%%%%%%%

As discussed in the previous section, in the N\'eel--Brown model 
of thermally activated magnetization reversal, a single domain 
magnetic particle has two equivalent ground states of opposite 
magnetization separated by an energy barrier due to for instant 
shape and crystalline anisotropy. 
The system can escape from one state to the other either by 
thermal activation over the barrier at high temperatures, 
or by quantum tunneling at low temperatures (Sect.~\ref{chap_MQT}). 
At sufficiently low temperatures and at zero field, 
the energy barrier between the two states of opposite 
magnetization is much too high to observe an escape process. 
However, the barrier can be lowered by applying a magnetic field 
in the opposite direction to that of the particle's magnetization. 
When the applied field is close enough to the switching field 
at zero temperature $H_{sw}^0$, thermal fluctuations 
are sufficient to allow the system to overcome the barrier, 
and the magnetization is reversed. 

In the following, we discuss three different experimental methods 
for studying this stochastic escape process which are called 
waiting time, switching field and telegraph noise measurements.

\begin{figure}
\begin{center}\leavevmode
\includegraphics[width=.8\textwidth]{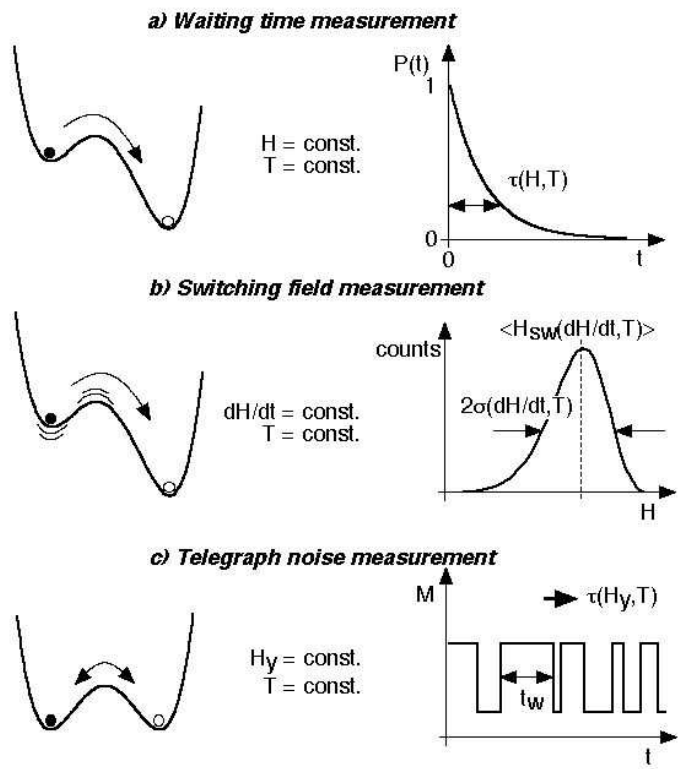}
\caption{Schema of three methods for studying the escape from a potential 
well: waiting time and telegraph noise measurements give direct access to the 
switching time probability $P(t)$, whereas switching field measurements 
yield histograms of switching fields}
\label{fig3_mes_techniques}
\end{center}
\end{figure}

%%%%%%%%%%%%%%%%%%%%%%%%%%%%%%%%
\subsection{Waiting time measurements}
\label{waiting_time}
%%%%%%%%%%%%%%%%%%%%%%%%%%%%%%%%

The waiting time method consists in measuring the
probability that the magnetization has not 
switched after a certain time. In the case
of an assembly of identical and isolated
particles, it corresponds to measurements
of the relaxation of magnetization.
However, in most particle assemblies,
broad distributions of switching fields
lead to logarithmic decay of magnetization,
and the switching probability is hidden behind
the unknown distributions functions~\cite{Dormann97}.
For individual particle studies, 
waiting time measurements give direct access to the 
switching probability (Fig.~\ref{fig3_mes_techniques}). 
At a given temperature, the magnetic field $H$ is increased 
to a waiting field $H_{\rm w}$ near the switching 
field $H_{\rm sw}^0$. Next, the elapsed time 
until the magnetization switches is 
measured. This process is repeated several 
hundred times, yielding a waiting time 
histogram. The integral of this histogram 
and proper normalization yields 
the probability that the magnetization has not 
switched after a time $t$.
This probability is measured at different waiting 
fields $H_{\rm w}$ and temperatures
in order to explore several barrier heights
and thermal activation energies.

According to the  N\'eel--Brown model, the probability 
that the magnetization has not 
switched after a time $t$ is given by:
\begin{equation}
	P(t) = e^{-t/\tau} \;,
\label{eq_P_t}
\end{equation}
and $\tau$ (inverse of the switching rate) can be expressed by an 
Arrhenius law of the form:
\begin{equation}
	\tau^{-1}(\varepsilon) = B \varepsilon^{a+b-1} e^{-A \varepsilon^a}  \;,
\label{eq_tau}
\end{equation}
where $\varepsilon = \left(1 - H/H_{\rm sw}^0\right)$ and
$A$, $B$, $a$ and $b$ depends on damping, temperature, 
energy barrier height (Eqs.~\ref{eq_E_1a} to~\ref{eq_Pfeiffer}, 
and~\ref{eq_tau_IHD} to~\ref{eq_tau_LD}), 
curvatures at well and saddle points, 
and reversal mechanism (thermal or quantum) 
(cf. Tab.~1 of Ref.~\cite{Garg95}). 
For simplicity, experimentalists have 
often supposed a constant pre-exponential factor $\tau_0^{-1}$
instead of $B \varepsilon^{a+b-1}$.

The adjustment of equation (\ref{eq_P_t}) to the measured 
switching probabilities yields a set of mean 
waiting times $\tau^{-1}(H_{\rm w}, T)$. 
In order to adjust the N\'eel--Brown model 
to this set of data, 
we propose the following relation which can be found 
by inserting 
$\varepsilon = \left(1 - H_{\rm w}/H_{\rm sw}^0\right)$ 
into Eq. (\ref{eq_tau}):
\begin{equation}
H_{\rm w} = H_{\rm sw}^0 \left(
	1 - \left\lbrack
	\frac {1}{A} \ln\left(\tau B \varepsilon^{a + b - 1}\right)
	\right\rbrack^{1/a}
\right) \;,
\label{eq_H_w}
\end{equation}
When plotting the $H_{\rm w}$ values as a function of 
$\left\lbrack T \ln\left(\tau B \varepsilon^{a + b - 1}
\right) \right\rbrack^{1/a}$, 
all points should gather on a straight line 
(master curve) by choosing the 
proper value for the constants $B$, $a$ and $b$
($a$ and $b$ should be given by 
Eqs.~\ref{eq_E_1a} to~\ref{eq_Pfeiffer}, 
and~\ref{eq_tau_IHD} to~\ref{eq_tau_LD}).
$A$ can be obtained from the slope of the master.

The number of exploitable decades for $\tau$ values is limited for 
waiting time measurements: short-time (milliseconds) experiments 
are limited by the inductance of the field coils
\footnote{A solution to this problem might be a superposition
of a constant applied field and a small pulse field.}
and long-time (minutes) 
studies by the stability of the experimental setup. 
Furthermore, the total acquisition time for a set 
of $\tau^{-1}(H_{\rm w}, T)$ is rather long (weeks). 
Thus a more convenient method is 
needed for single particle measurements which 
is the switching field method.

%%%%%%%%%%%%%%%%%%%%%%%%%%%%%%%%
\subsection{Switching field measurements}
\label{switching_field}
%%%%%%%%%%%%%%%%%%%%%%%%%%%%%%%%

For single particle studies, it is often 
more convenient to study magnetization reversal by ramping the applied field at a 
given rate and measuring the field value as soon as the particle magnetization switches. 
Next, the field ramp is reversed and the process repeated. 
After several hundred cycles, switching field histograms are established, 
yielding the mean switching field $\langle H_{\rm sw} \rangle$ and 
the width $\sigma_{\rm sw}$ (rms. deviation). 
Both mean values are measured as a function of 
the field sweeping rate and temperature (Fig.~\ref{fig3_mes_techniques}).

From the point of view of thermally activated magnetization reversal,
switching field measurements are equivalent to waiting time measurements 
as the time scale for the sweeping rate is typically more than 8 orders of 
magnitude greater than the time scale of the exponential prefactor 
which is in general around $10^{-10}$ s. 
We can therefore apply the N\'eel--Brown model described above.
The mathematical transformation from a switching time probability 
(Eqs.~\ref{eq_P_t}~--~\ref{eq_tau})
to a switching field probability was first given by 
Kurkij$\ddot{a}$rvi~\cite{Kurkijarvi72} 
for the critical current in SQUIDs. 
A more general calculation was evaluated by Garg~\cite{Garg95}. 
In many cases, the mean switching field 
$\langle H_{\rm sw} \rangle$ can be approximated 
by the first two terms of the development of Garg~\cite{Garg95}:
\begin{equation}
\langle H_{\rm sw}(T,v) \rangle \approx H_{\rm sw}^0 \left(
	1 - \left\lbrack
	\frac {1}{A} 
	\ln\left(\frac {H_{\rm sw}^0 B}{v a A^{1-b/a}}\right)
	\right\rbrack^{1/a}
\right) \;,
\label{eq_H_sw}
\end{equation}
where the field sweeping rate is given by $v = dH/dt$. 
The width of the switching field distribution $\sigma_{\rm sw}$ can be 
approximated by the first term of Garg's development:
\begin{equation}
\sigma_{\rm sw} \approx H_{\rm sw}^0 \frac{\pi}{\sqrt{6} a} 
	\left(\frac{1}{A}\right)^{1/a} 
	\left\lbrack
	\ln\left(\frac {H_{\rm sw}^0 B}{v a A^{1-b/a}}\right)
	\right\rbrack^{(1-a)/a} \;.
\label{eq_sigma}
\end{equation}

In the case of a constant pre-exponential factor $\tau_0^{-1}$, 
the calculation of $\langle H_{\rm sw} \rangle$ and $\sigma_{\rm sw}$ is more simple 
and is given by equations (4) and (5) in Ref.~\cite{WW_PRL97_Co}, respectively.

Similar to the waiting time measurements, a scaling of the model to a set 
of $\langle H_{\rm sw}(T,v) \rangle$ values can be done by plotting the 
$\langle H_{\rm sw}(T,v) \rangle$ values as a function of 
$\left\lbrack T \ln\left(\frac {H_{\rm sw}^0 B}{v a A^{1-b/a}}\right) 
\right\rbrack^{1/a}$.
All points should gather on a straight line 
by choosing the proper value 
for the constants ($a$ and $b$ should be given by 
Eqs.~\ref{eq_E_1a} to~\ref{eq_Pfeiffer}, 
and~\ref{eq_tau_IHD} to~\ref{eq_tau_LD}).

The entire switching field distribution $P(H)$ can be calculated iteratively by the 
following equation ~\cite{Kurkijarvi72}: 
\begin{equation}
P(H) = \tau^{-1}(H) v^{-1}  
	\left\lbrack
	1 - \int_0^H {P(H')dH'}
	\right\rbrack \;.
\label{eq_P_H}
\end{equation}

%%%%%%%%%%%%%%%%%%%%%%%%%%%%%%%%
\subsection{Telegraph noise measurements}
\label{telegraph_noise}
%%%%%%%%%%%%%%%%%%%%%%%%%%%%%%%%

In order to study the superparamagnetic state 
\footnote{At zero applied field, a single domain magnetic 
particle has two equivalent ground 
states of opposite magnetization separated 
by an energy barrier. When the thermal energy 
$k_{\rm B}T$ is sufficiently high, the total magnetic 
moment of the particle can 
fluctuate thermally, like a single spin in a 
paramagnetic material.  
Such magnetic behavior of an assembly 
of independent single domain 
particles is called superparamagnetism
~\cite{Dormann97,Bean55,Bean59}.} 
of a single particle, it is simply 
necessary to measure the particle's 
magnetization as a function of time. 
We call this  telegraph noise measurement as 
stochastic fluctuations between two states are expected. 
According to the N\'eel--Brown model, 
the mean time $\tau$ spent in one state of 
magnetization is given by an Arrhenius 
law of the form of Eq.~\ref{eq_tau}. 
As $\tau$ increases exponentially 
with decreasing temperature, it is 
very unlikely that an escape 
process will be observed at low 
temperature. However, applying a constant field in 
direction of a hard axis (hard plane) of 
magnetization reduces the height of the 
energy barrier (Fig.~\ref{fig3_mes_techniques}). 
When the energy barrier is sufficiently small, 
the particle's magnetization can fluctuate 
between two orientations which are close to a 
hard axis (hard plane) of magnetization. 
The time spent in each state follows an 
exponential switching probability law as 
given by Eqs.~\ref{eq_P_t} and \ref{eq_tau} 
with $a$ = 2 (Eq.~\ref{eq_E_3}). 
\footnote{Note that for a slightly 
asymmetric energy potential, one switching probability 
can be so long that two--level 
fluctuation becomes practically unobservable.}

%%%%%%%%%%%%%%%%%%%%%%%%%%%%%%%%
\section{Experimental evidence for the N\'eel--Brown model}
\label{Exp_NB}
%%%%%%%%%%%%%%%%%%%%%%%%%%%%%%%%

The N\'eel--Brown model is widely used in magnetism, 
particularly in order to describe the 
time dependence of the magnetization of 
collections of particles, thin films and bulk 
materials. However until recently, all the reported 
measurements, performed on 
individual particles, were not consistent with the 
N\'eel--Brown theory. This disagreement 
was attributed to the fact that real samples contain 
defects, ends and surfaces which could 
play an important, if not dominant, role in the 
physics of magnetization reversal. It was 
suggested that the dynamics of reversal occurs 
via a complex path in configuration space, 
and that a new theoretical approach is required 
to provide a correct description of 
thermally activated magnetization reversal even 
in single-domain ferromagnetic particles 
~\cite{Ledermann94,WW_JMMM95b}. Similar conclusions 
were drawn from numerical 
simulations of the magnetization reversal
~\cite{Richards96,Gonzalez96,Garcia96,Hinzke98,Boerner97}.

A few years later, micro-SQUID measurements on 
individual Co nanoparticles showed for the first 
time a very good agreement 
with the N\'eel--Brown model by using waiting time, 
switching field and telegraph noise 
measurements~\cite{WW_PRL97_Co,WW_PRL97_BaFeO,Jamet01a}. 
It was also found that sample defects, 
especially sample oxidation, play a 
crucial role in the physics of magnetization reversal.

In the following subsections, we review some typical results 
concerning nanoparticles (Sect.~\ref{nanopart}), 
clusters (Sect.~\ref{cluster_T}) and 
wires (Sect.~\ref{Ni_wire_T}). In 
Sect.~\ref{deviations}, 
we point out the main deviations from 
the N\'eel--Brown model which 
are due to defects.

%%%%%%%%%%%%%%%%%%%%%%%%%%%%%%%%
\subsection{Application to nanoparticles}
\label{nanopart}
%%%%%%%%%%%%%%%%%%%%%%%%%%%%%%%%

\begin{figure}[t]
\begin{center}\leavevmode
\includegraphics[width=.8\textwidth]{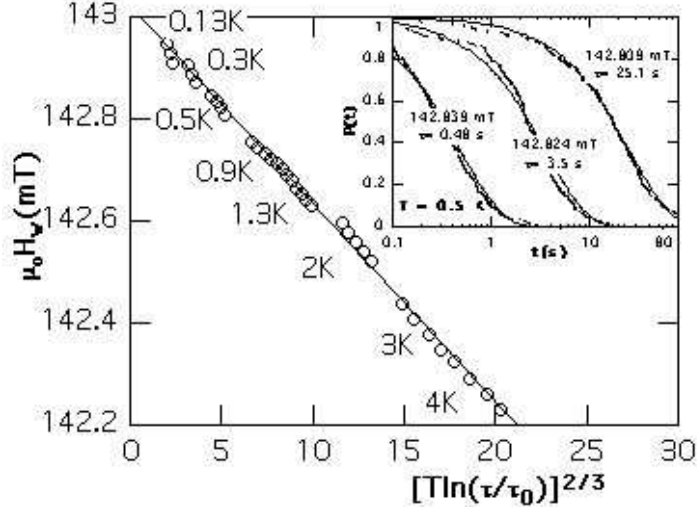}
\caption{Scaling plot of the mean switching 
time $\tau(H_{\rm w},T)$ for several 
waiting fields $H_{\rm w}$ and temperatures 
(0.1~s $< \tau(H_{\rm w},T) <$ 60~s) 
for a Co nanoparticle. 
The scaling yields $\tau_0 \approx 3 \times10^{-9}$s. 
{\it Inset:}~examples of the probability 
of not-switching of magnetization as a function of time 
for different applied fields and at 0.5~K. 
{\it Full lines} are data fits with an exponential 
function: $P(t)Ê=Êe^{-t/\tau}$}
\label{P_t_scaling}
\end{center}
\end{figure}

One of the important predictions of 
the N\'eel--Brown model concerns the 
exponential not-switching probability $P(t)$ 
(Eq.~\ref{eq_P_t}) which can be measured directly 
via waiting time measurements (Sect.~\ref{waiting_time}): 
at a given temperature, the 
magnetic field is increased to a waiting field $H_{\rm w}$ which 
is close to the switching field. Then, the elapsed time is 
measured until the magnetization switches. This process is 
repeated several hundred times, in order to obtain a waiting 
time histogram. The integral of this histogram gives the 
not-switching probability $P(t)$ which is measured at several 
temperatures $T$ and waiting fields $H_{\rm w}$. 
The inset of Fig.~\ref{P_t_scaling} displays typical 
measurements of $P(t)$ performed on a Co nanoparticle. 
All measurements show that $P(t)$ is given by an 
exponential function described by a single relaxation time $\tau$.

The validity of Eqs. \ref{eq_E_1a} and \ref{eq_tau}) 
is tested by plotting the waiting field $H_{\rm w}$ as a 
function of $[T ln(\tau/\tau_0)]^{2/3}$. 
\footnote{$a = 3/2$ because the field was applied at 
about $20^{\circ}$ from the easy axis of 
magnetization (Eq.~\ref{eq_E_1a}).}
If the N\'eel--Brown model applies, all 
points should collapse onto one straight line (master curve) by choosing 
the proper values for $\tau_0$. Fig.~\ref{P_t_scaling} shows that the data set 
$\tau(H_{\rm w},T)$ falls on a master curve provided that
$\tau_0 \approx 3 \times10^{-9}$s. The slope and intercept yield 
the values $E_0$ = 214\,000~K and $H_{\rm sw}^0$ = 143.05~mT. 
The energy barrier $E_0$ can be approximately converted to a 
thermally ``activated volume" by using 
$V = E_0/(\mu_0 M_{\rm S} H_{\rm sw}^0) \approx$ (25~nm)$^3$ 
which is very close to the particle volume estimated by SEM. 
This agreement is another confirmation of a magnetization 
reversal by uniform rotation. The result of the waiting time 
measurements are confirmed by switching field and 
telegraph noise measurements
~\cite{WW_PRL97_Co,WW_PRL97_BaFeO}. 
The field and temperature dependence of the 
exponential prefactor $\tau_0$ 
is taken into account in~\cite{Coffey98a}. 

%%%%%%%%%%%%%%%%%%%%%%%%%%%%%%%%
\subsection{Application to Co clusters}
\label{cluster_T}
%%%%%%%%%%%%%%%%%%%%%%%%%%%%%%%%

\begin{figure}
\begin{center}\leavevmode
\includegraphics[width=.7\textwidth]{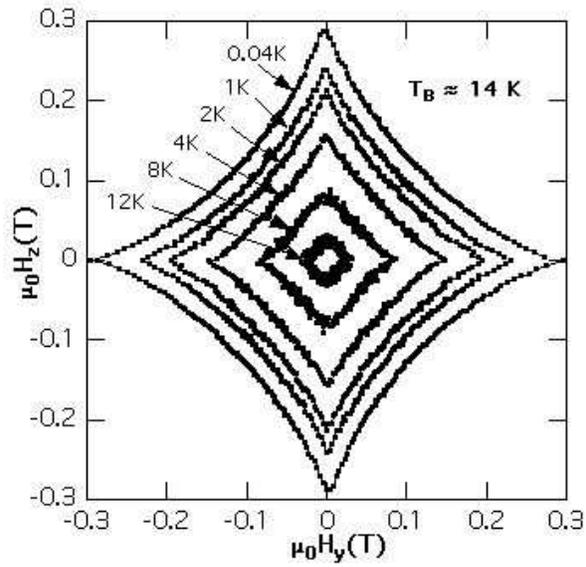}
\caption{Temperature dependence of the switching field 
of a 3 nm Co cluster, measured in the
plane defined by the easy and medium hard axes 
($H_{y}-H_{z}$ plane in Fig.~\ref{ast_Co_cluster}).
The data were recorded using the blind mode method (Sect.~\ref{blind})
with a waiting time of the applied field of $\Delta t$ = 0.1s.
The scattering of the data is due to stochastic and in 
good agreement with Eq.~\ref{eq_sigma}.}
\label{ast_T}
\end{center}
\end{figure}

Fig.~\ref{ast_T} presents the angular dependence of the switching field of 
a 3 nm Co cluster measured at different temperatures. At 0.03 K, the 
measurement is very close to the standard Stoner--Wohlfarth astroid 
(Fig.~\ref{ast_ST_W}). For higher temperatures the switching field becomes 
smaller and smaller. It reaches the origin at about 14 K yielding the 
blocking temperature $T_{\rm B}$ = 14 K of the cluster magnetization. 
$T_{\rm B}$ is defined as the temperature
for which the waiting time $\Delta t$ becomes equal to
the relaxation time $\tau$ of the particle's magnetization at
$\vec{H}$ = $\vec{0}$. $T_{\rm B}$ can be used to estimate the
total number $N_{tot}$ of magnetic Co atoms in the cluster. 
Using an Arrhenius-like law (Eq.~\ref{eq_tau}) which
can be written as:
 $\Delta t = \tau = \tau_0 exp(K_{at}N_{tot}/k_{\rm B}T_{\rm B})$,
where $\tau_0^{-1}$ is the attempt frequency typically
between $10^{10}$ to $10^{11}$ Hz~\cite{Respaud98}, 
K$_{at}$ is an
effective anisotropy energy per atom 
and k$_{B}$ is the Boltzmann constant. 
Using the expression of the
switching field at $T$ = 0 K and for $\theta = 0$:
$\mu_{0}H_{sw} = 2K_{at}/\mu_{at}= 0.3 T$ (Fig.~\ref{ast_T}),
the atomic moment $\mu_{at}=1.7\mu_{B}$,
$\Delta t$ = 0.01 s,
$\tau_0 = 10^{-10}$ s, and $T_{\rm B}$ = 14 K,
we deduce $N_{tot} \approx 1500$, which
corresponds very well to a 3 nm Co cluster (Fig.~\ref{cluster_scheme}).

\begin{figure}
\begin{center}\leavevmode
\includegraphics[width=.7\textwidth]{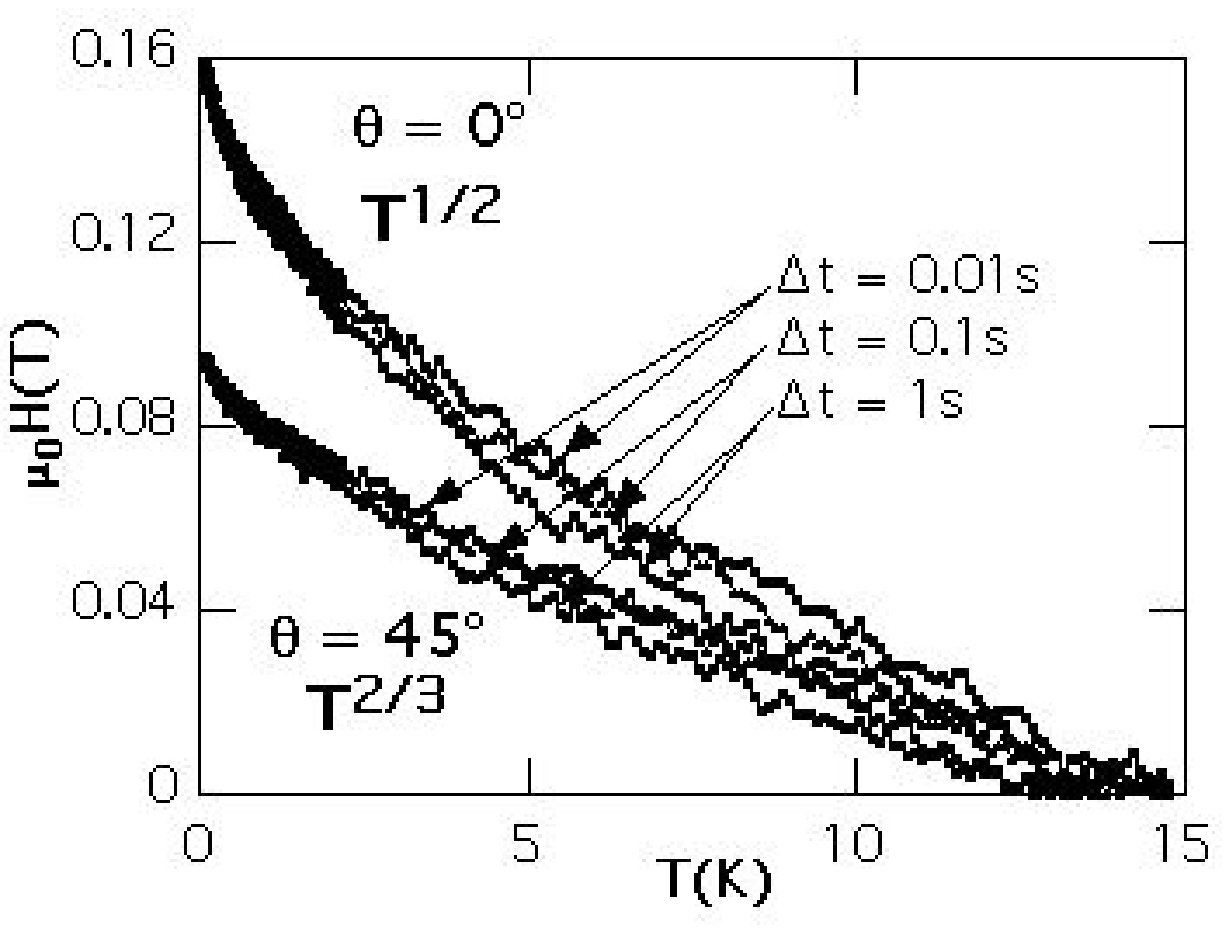}
\caption{Temperature dependence of the switching field 
of a 3 nm Co cluster, measured at $0^{\circ}$ and $45^{\circ}$. 
The data were recorded using the blind mode method (Sect.~\ref{blind})
with different waiting time  $\Delta t$ of the applied field.
The scattering of the data is due to stochastics and is in 
good agreement with Eq.~\ref{eq_sigma}.}
\label{Hsw_T}
\end{center}
\end{figure}

\begin{figure}
\begin{center}\leavevmode
\includegraphics[width=.7\textwidth]{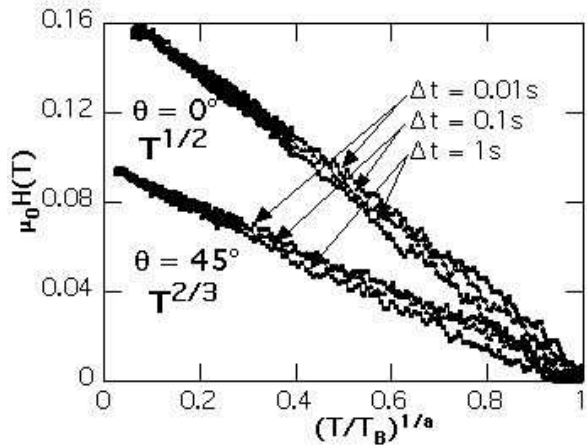}
\caption{Temperature dependence of the switching field 
of a 3 nm Co cluster as in Fig.~\ref{Hsw_T} but
plotted as a function of $(T/T_{\rm B})^{1/a}$ with $T_{\rm B}$ = 14 K
and $a$ = 2 or 3/2 for $\theta = 0^{\circ}$ and $45^{\circ}$,
respectively, and for three waiting times $\Delta t$.}
\label{Hsw_T_a}
\end{center}
\end{figure}

Fig.~\ref{Hsw_T} presents a detailed measurement of 
the temperature dependence 
of the switching field at $0^{\circ}$ and $45^{\circ}$
and for three waiting times $\Delta t$. This measurement 
allows us to check the predictions of the field dependence of 
the barrier height.
Eq.~\ref{eq_H_sw} predicts that the mean switching field should be 
proportional to $T^{1/a}$ where $a$ depends on the direction
of the applied field (Eqs.~\ref{eq_E_1a} to~\ref{eq_Pfeiffer}):
$a = 2$ for $\theta = 0^{\circ}$ and $90^{\circ}$,
and $a = 3/2$ for all other angles 
which are not too close to $\theta = 0^{\circ}$ and $90^{\circ}$.
We found a good agreement with this model (Fig.~\ref{Hsw_T_a}).

%%%%%%%%%%%%%%%%%%%%%%%%%%%%%%%%
\subsection{Application to Ni wires}
\label{Ni_wire_T}
%%%%%%%%%%%%%%%%%%%%%%%%%%%%%%%%

Electrodeposited wires (with diameters ranging from 40 
to 100~nm and lengths up to 5000~nm, see Fig.~\ref{SQUID_SEM}) 
were studied~\cite{WW_PRL96_Ni,WW_PRB97_Ni} 
using the micro-SQUID technique (Sect.~\ref{curling_exp}).
For diameter values under 50 nm, the switching probability as a 
function of time could be described by a single 
exponential function (Eq.~\ref{eq_P_t}). 
The mean waiting time $\tau$ followed an Arrhenius law 
(Eq.~\ref{eq_tau})
as proposed by the N\'eel--Brown model. Temperature and field sweeping
rate dependence of the mean switching field could be described 
by the model of Kurkij$\ddot{a}$rvi 
(Sect.~\ref{switching_field}) which is based
on thermally assisted magnetization reversal over a 
simple potential barrier. This measurements allowed us to 
estimate an activation volume which was two orders of magnitude 
smaller than the wire volume. 
This confirmed the idea of the reversal of the
magnetization caused by a nucleation of a reversed 
fraction of the cylinder, rapidly propagating along the
whole sample. This result was also in good agreement with 
a micromagnetic model of Braun~\cite{Braun99}.

A pinning of the propagation of the magnetization 
reversal occurred for a few samples, where
several jumps were observed in the hysteresis curves. 
The pinning of a domain wall was probably
due to structure defects. 
The dynamic reversal properties of depinning were quite
different from those of nucleation of a domain wall. 
For example, the probability 
of depinning as a function of
time did not follow a single exponential law. 
A similar effect was also observed in 
single submicron Co particles
having one domain wall~\cite{WW_PRB96}, 
showing a domain wall annihilation
process (Sect.~\ref{nucleation}).

%%%%%%%%%%%%%%%%%%%%%%%%%%%%%%%%
\subsection{Deviations form the N\'eel--Brown model}
\label{deviations}
%%%%%%%%%%%%%%%%%%%%%%%%%%%%%%%%

Anomalous magnetic properties of oxidized or ferrimagnetic
nanoparticles have been reported previously by several
authors~\cite{Berkowitz75,Richardson91}.
These properties are, e.g., the lack of saturation in
high fields and shifted hysteresis loops after cooling in the
presence of a magnetic field. These behaviors have been attributed
to uncompensated surface spins of the particles and
surface spin disorder~\cite{Kodama96,Kodama99}.

Concerning our single-particle studies,
we systematically observed aging effects which we attribute
to an oxidation of the surface of the sample, forming
antiferromagnetic CoO or Ni0~\cite{WW_JMMM95b,WW_PRB97_Ni}. 
We found that the antiferromagnetic
coupling between the core of the particle or wire 
and its oxidized surface
changed the dynamic reversal properties.
For instant, we repeated the measurements 
of the magnetization reversal
of a Ni wire two days after fabrication, six weeks
after, and finally after three months~\cite{WW_PRB97_Ni}. 
Between these measurements,
the wire stayed in a dry box. The quasi-static
micro-SQUID measurements did reveal only small
changes. The saturation magnetization measured after six
weeks was unchanged and was reduced by one to two percent
after three months. The angular dependence of the
switching field changed also only slightly.
The dynamic measurements showed the aging
effects more clearly, as evidenced by
\begin{itemize}
\item
a non exponential probability of not switching;
\item
an increase of the width of the switching field distributions;
\item
a decrease of the activation energy.
\end{itemize}
We measured a similar behavior on lithographic fabricated
Co particles with an oxidized border~\cite{WW_JMMM95b}.

\begin{figure}
\begin{center}\leavevmode
\includegraphics[width=.7\textwidth]{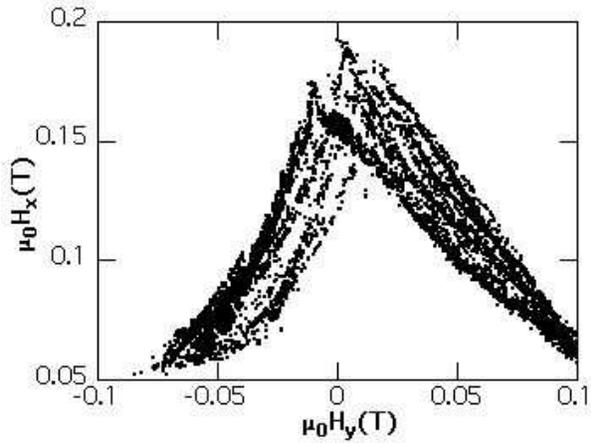}
\caption{Angular dependence of switching fields of a 3 nm Fe cluster 
having (probably) a slightly oxidized surface. Each point corresponds to
on of the 10000 switching field measurement. The huge variations of 
the switching field might be due to exchange bias of frustrated spin 
configurations. However, quantum effects like those described in
Sect.~\ref{chap_MQT} are not completely excluded.}
\label{Fe_bizarre}
\end{center}
\end{figure}

\begin{figure}
\begin{center}\leavevmode
\includegraphics[width=.7\textwidth]{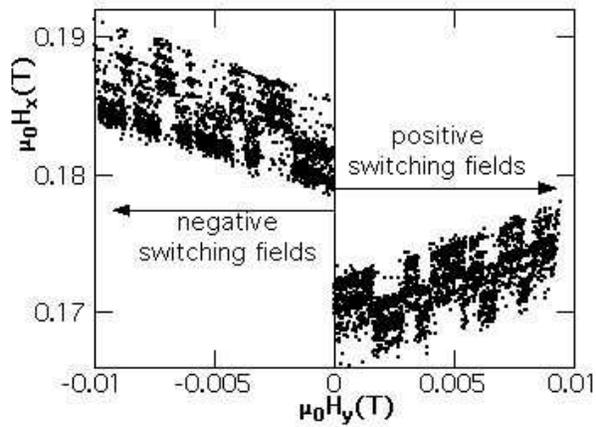}
\caption{Detail of the angular dependence of switching 
fields of a 3 nm Fe cluster having (probably)  
a slightly oxidized surface.
Each point corresponds to one of the 3000
switching field measurement.
Stochastic fluctuation between 
different switching field 
distributions are observed. The \lq mean' 
hysteresis loop is shifted
to negative fields.}
\label{Fe_bizarre_tele}
\end{center}
\end{figure}

Fig.~\ref{Fe_bizarre} presents the angular dependence 
of switching fields of a 3 nm Fe cluster 
having a slightly oxidized surface. Huge variation of the switching 
fields can be observed which might be due to exchange bias of frustrated 
spin configurations at the surface 
of the cluster (Fig.~\ref{Fe_bizarre_tele}).

We propose that the magnetization reversal of a ferromagnetic
particle with an antiferromagnetic surface layer is
mainly governed by two mechanisms which are both due to
spin frustration at the interface between the ferromagnetic
core and the antiferromagnetic surface layer(s).
The first mechanism may come from the spin frustration
differing slightly from one cycle to another, thus producing
a varying energy landscape. These energy variations are less important
at high temperatures when the thermal energy
($k_{\rm B} T$) is much larger. 
However at lower temperature, the magnetization reversal becomes
sensitive to the energy variations. During the hysteresis
loop the system chooses randomly a path through the energy
landscape which leads to broad switching field distributions. 
A second mechanism may become dominant at high temperatures:
the magnetization reversal may be governed by a
relaxation of the spin frustration, hence by a relaxation of the
energy barrier. This relaxation is thermally activated, {\it i.e.},
slower at lower temperatures.

%%%%%%%%%%%%%%%%%%%%%%%%%%%%%%%%
\title[Magnetization reversal by quantum tunneling]
{Magnetization reversal by quantum tunneling}
\label{chap_MQT}
%%%%%%%%%%%%%%%%%%%%%%%%%%%%%%%%

Studying the boundary between classical and 
quantum physics has become a very attractive 
field of research which is known as \lq mesoscopic' 
physics (Fig.~\ref{fig_scale}). 
New and fascinating mesoscopic effects can occur 
when characteristic system dimensions are smaller than the 
length over which the quantum wave function 
of a physical quantity remains sensitive to 
phase changes. Quantum interference effects in
mesoscopic systems have, until now, 
involved phase interference between paths of particles 
moving in real space as in SQUIDs 
or mesoscopic rings. For magnetic systems, similar 
effects have been proposed for 
spins moving in spin space, such as magnetization 
tunneling out of a metastable potential 
well, or coherent tunneling between classically 
degenerate directions of magnetization 
~\cite{Leggett87,QTM94}. 

We have seen in the previous sections that the 
intrinsic quantum character of the magnetic 
moment can be neglected for nanoparticles with 
dimensions of the order of the domain 
wall width $\delta$ and the exchange length $\lambda$, 
{\it i.e.}, particles with a collective spin 
of $S = 10^5$ or larger. However, recent measurements 
on molecular clusters with a 
collective spin of $S = 10$ suggest that quantum 
phenomena might be observed at larger 
system sizes with $S >> 1$. Indeed, it has been predicted 
that macroscopic quantum tunneling of 
magnetization can be observed in magnetic 
systems with low dissipation. 
In this case, it is 
the tunneling of the magnetization vector of a 
single-domain particle through its 
anisotropy energy barrier or the tunneling 
of a domain wall through its pinning energy. 
These phenomena have been studied theoretically 
and experimentally~\cite{QTM94}.

The following sections review the most important 
results concerning the observed 
quantum phenomena in molecular clusters which 
are mesoscopic model systems
to test quantum tunneling theories and the effects 
of the environmental decoherence.
Their understanding requires a knowledge of many
physical phenomena, and are therefore particularly
interesting for fundamental studies. 
We then focuses on Magnetic Quantum Tunneling (MQT) 
studied in individual nanoparticles 
or nanowires. We concentrate on the necessary experimental 
conditions for MQT and review some experimental results
which suggest that quantum effects might even
be important in nanoparticles with $S = 10^5$ or larger.

%%%%%%%%%%%%%%%%%%%%%%%%%%%%%%%%
\section{Quantum tunneling of magnetization 
in molecular clusters}
\label{mol_cluster}
%%%%%%%%%%%%%%%%%%%%%%%%%%%%%%%%

\begin{figure}[t]
\begin{center}\leavevmode
\includegraphics[width=.7\textwidth]{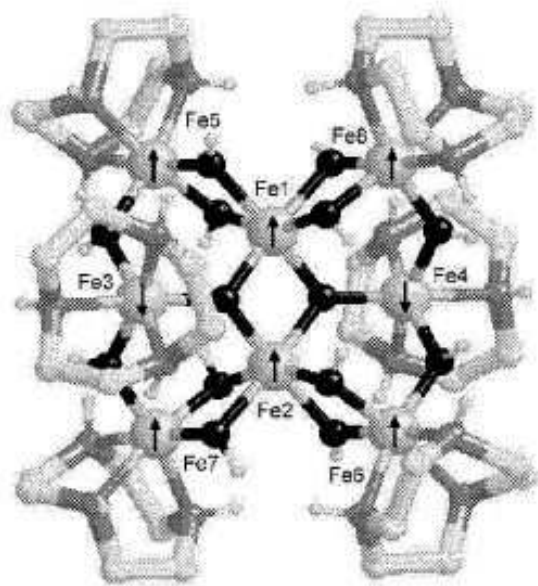}
\caption{Schematic view of the magnetic core of the 
Fe$_8$ cluster. The oxygen atoms are black, 
the nitrogen gray, and carbon atoms are white. 
For the sake of clarity only the hydrogen atoms that
are exchanged with deuterium are shown as small 
spheres (Sect.~\ref{hyperfine_Fe8}). The
arrows represent the spin structure of the ground 
state $S = 10$ as
experimentally determined through 
polarized neutron diffraction
experiments~\cite{Pontillon99}.
The exact orientation of easy, medium and hard axis of
magnetization (Fig. \ref{sphere}) can be found in
~\cite{Barra00}.}
\label{fig_Fe8}
\end{center}
\end{figure}

Magnetic molecular clusters are the final point in 
the series of smaller and smaller units from bulk matter to atoms.
Up to now, they have been the most promising candidates for 
observing quantum phenomena since they have a well 
defined structure with well characterized 
spin ground state and magnetic anisotropy. 
These molecules can be regularly assembled in 
large crystals where all molecules often have 
the same orientation. Hence, macroscopic 
measurements can give direct access to 
single molecule properties. 
The most prominent 
examples are a dodecanuclear mixed-valence 
manganese-oxo cluster with acetate ligands, 
short Mn$_{12}$ acetate~\cite{Sessoli93}, and an 
octanuclear iron(III) oxo-hydroxo cluster of 
formula [Fe$_8$O$_2$(OH)$_{12}$(tacn)$_6$]$^{8+}$ 
where tacn is a macrocyclic ligand,
short Fe$_8$ (Fig.~\ref{fig_Fe8})~\cite{Barra96}. 
Both systems have a spin ground 
state of $S = 10$, and an Ising-type 
magneto-crystalline anisotropy, which stabilizes the 
spin states with $m = \pm10$ and generates 
an energy barrier for the reversal of the 
magnetization of about 67~K for Mn$_{12}$ 
acetate and 25~K for Fe$_8$.

Thermally-activated quantum tunneling of the 
magnetization has first been evidenced in both 
systems \cite{Novak95,Friedman96,Thomas96,Sangregorio97}. 
Theoretical discussion of this assumes that 
thermal processes (principally phonons) promote the 
molecules up to high levels with small 
quantum numbers $|m|$, 
not far below the top of the energy barrier, 
and the molecules then tunnel inelastically 
to the other side. Thus the transition is almost 
entirely accomplished via thermal transitions 
and the characteristic relaxation time is 
strongly temperature dependent. 
An alternative explication was also presented \cite{Garg98}.
For Fe$_8$ however the relaxation time becomes 
temperature independent below 
0.36 K \cite{Sangregorio97,Ohm98a} 
showing that a pure tunneling mechanism between 
the only populated ground states 
$m = \pm S = \pm 10$ is responsible 
for the relaxation of the magnetization. 
On the other hand in the Mn$_{12}$ acetate 
system one sees temperature independent relaxation 
only for strong applied fields
and below about 0.6 K \cite{Perenboom98,Kent00}.
During the last years, several new molecular magnets
were presented (see for instance~\cite{Caneschi99,
Aubin98,Price99,Yoo_Jae00}) 
which show also tunneling at low temperatures.

The following subsections review 
the most appealing results concerning 
the Fe$_8$ system with can be seen as 
an {\it ideal}  \lq model molecule' to study
quantum phenomena in magnetic nanostructures.
We stress that the tunneling in large
spins is remarkable because it does not
show up at lowest orders of perturbation
theory.

All measurements on Fe$_8$ were performed using an array of 
micro-SQUIDs (Sect.~\ref{SQUID_array}). 
The high sensitivity of this 
magnetometer allows us to study single Fe$_8$ crystals 
\cite{Wieghardt84} of sizes of the order of 10 to 500 $\mu$m.
For ac--susceptibility measurements and 
magnetization measurements at $T > 6$ K, 
we used a home built Hall probe 
magnetometer~\cite{Kent94,WW_PhysB00c}. 
It works in the temperature range between 0.03~K and 100~K, 
for frequencies between 1~Hz and 100~kHz.

After discussing the magnetic anisotropy of Fe$_8$, we present 
the observed quantum phenomena. 
The discussions of the following sections
neglect environmental decoherence effects for the sake of simplicity. 
In Sects.~\ref{decoherence}, we focus
on effects of the environment (dipolar coupling, nuclear spins, 
and temperature)
onto the tunneling.
This review should help to set up a complete theory 
which describes {\it real} magnetic quantum systems. 

%%%%%%%%%%%%%%%%%%%%%%%%%%%%%%%%
\subsection{Magnetic anisotropy in Fe$_8$}
\label{anisotropy_Fe8} 
%%%%%%%%%%%%%%%%%%%%%%%%%%%%%%%%

The  octanuclear iron(III) oxo-hydroxo cluster of formula 
[Fe$_8$O$_2$(OH)$_{12}$(tacn)$_6$]$^{8+}$ where tacn is a 
macrocyclic ligand, short Fe$_8$ (Fig.~\ref{fig_Fe8}) 
was first synthesized by
Wieghardt {\it et al.} in 1984~\cite{Wieghardt84}. 
Four central iron (III) ions with $S=5/2$ are bridged by two oxo groups. 
The other four iron (III) ions are bridged 
by hydroxo groups to the central iron
ions in an almost planar arrangement, as shown in
Fig.~\ref{fig_Fe8}. The clusters have approximate $D_2$ symmetry but
crystallize in the triclinic system~\cite{Barra00}. 

Fe$_8$ has a $S=10$ ground state which
originates from antiferromagnetic interactions that
do not give complete compensation of the magnetic
moment~\cite{Delfs93}. 
Spin--orbital moments can be neglected as the magnetic
ions are in an \lq orbital singlet' as a result
of Hund's rules.
The spin structure of the ground
state schematized by the arrows in Fig.~\ref{fig_Fe8} has been
recently confirmed by a single-crystal polarized
neutron investigation that provided a magnetization 
density map of the cluster~\cite{Pontillon99}. 

\begin{figure}
\begin{center}
\includegraphics[width=.7\textwidth]{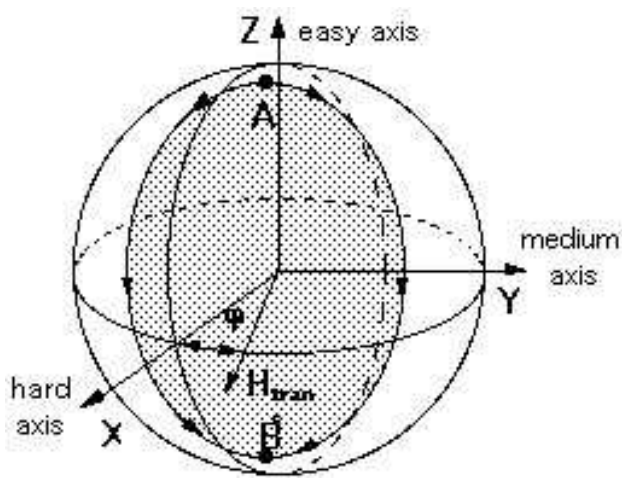}
\caption{Unit sphere showing degenerate minima 
{\bf A} and {\bf B} which are joined by two tunnel 
paths (heavy lines). The hard, medium, 
and easy axes are taken in $x$--, $y$-- and 
$z$--direction, respectively. 
The constant transverse field $H_{trans}$ for tunnel splitting
measurements is applied in 
the $xy$--plane at an azimuth angle 
$\varphi$. At zero applied field $\vec{H} = 0$, the giant spin 
reversal results from the interference of two 
quantum spin paths of opposite direction 
in the easy anisotropy $yz$--plane.
For transverse fields in direction of the hard axis,
the two quantum spin paths are in a plane which is parallel to the 
$yz$--plane, as indicated in the figure.
By using Stokes theorem it has been shown \cite{Garg93} that 
the path integrals can be converted in an area integral, 
giving that destructive interference, that is a 
quench of the tunneling rate, occurs whenever 
the shaded area is $k \pi / S$, where $k$ is an odd integer.
The interference effects disappear quickly when the transverse field 
has a component in the $y$--direction because the tunneling is then 
dominated by only one quantum spin path.}
\label{sphere}
\end{center}
\end{figure}

\begin{figure}
\begin{center}
\includegraphics[width=.7\textwidth]{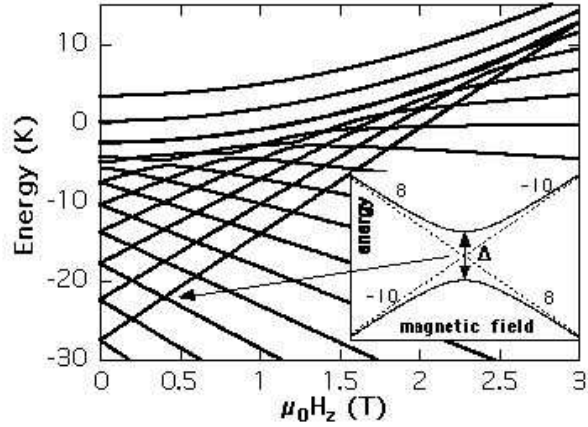}
\caption{Zeeman diagram of the 21 levels of 
the $S = 10$ manifold of Fe$_8$ as a 
function of the field applied along the easy axis (Eq.~\ref{eq_H_biax}). 
From bottom to top, the levels are labeled 
with quantum numbers $m = \pm10, \pm9, ...0$. 
The levels cross 
at fields given by $\mu_0 H_z \approx n \times$ 0.22 T, with $n = 1, 2, 3 ...$. 
The {\it inset} displays 
the detail at a level crossing where the 
transverse terms (terms containing $S_x$ 
or/and $S_y$ spin operators) turn the crossing 
into an avoided level crossing. 
The greater the tunnel splitting $\Delta$, the higher the tunnel rate.}
\label{fig_level}
\end{center}
\end{figure}

The simplest model describing the spin system of Fe$_8$ molecular clusters 
(called the giant spin model) has the following Hamiltonian~\cite{Barra96}:
\begin{equation}
	H = -D S_z^2 + E \left(S_x^2 - S_y^2\right) 
	+ g \mu_{\rm B} \mu_0 \vec{S}\cdot\vec{H} \;.
\label{eq_H_biax}
\end{equation}
$S_x$, $S_y$, and $S_z$ are the three 
components of the spin operator, 
$D$  and $E$ are the anisotropy constants 
which were determined via HF-EPR 
($D/k_{\rm B} \approx$ 0.275 K and 
$E/k_{\rm B} \approx$ 0.046 K~\cite{Barra96}), 
and the last term of the Hamiltonian describes the Zeeman 
energy associated with an applied field $\vec{H}$. 
This Hamiltonian defines a hard, medium, 
and easy axes of magnetization in $x$, $y$ and $z$ directions, 
respectively (Fig.~\ref{sphere}). 
It has an energy level 
spectrum with $(2S+1) = 21$ values which, 
to a first approximation, can be labeled by the 
quantum numbers $m = -10, -9, ..., 10$
choosing the $z$--axis as quantization axis. The energy spectrum, 
shown in Fig.~\ref{fig_level}, can be obtained by using 
standard diagonalisation techniques of the $[21 \times 21]$ 
matrix describing the spin 
Hamiltonian $S = 10$. At $\vec{H} = 0$, the levels $m = \pm10$ 
have the lowest energy. 
When a field $H_z$ is applied, the energy levels with 
$m < -2$ increase, while those 
with $m > 2$ decrease (Fig.~\ref{fig_level}). Therefore, 
energy levels of positive and negative 
quantum numbers cross at certain fields $H_z$. 
It turns out that for Fe$_8$ the levels cross 
at fields given by $\mu_0 H_z \approx n \times$ 0.22~T, 
with $n = 1, 2, 3 ...$. The inset of Fig.~\ref{fig_level} 
displays the details at a level crossing where transverse terms 
containing $S_x$ or $S_y$ spin operators turn the 
crossing into an \lq avoided level crossing'. 
The spin $S$ is \lq in resonance' between two 
states when the local longitudinal field is close 
to an avoided level crossing. 
The energy gap, the so-called 
\lq tunnel spitting' $\Delta$, can be tuned by 
an applied field in the $xy -$ plane 
(Fig.~\ref{sphere}) via the $S_xH_x$ and $S_yH_y$ 
Zeeman terms (Sect.~\ref{oscillations}).

\begin{figure}
\begin{center}
\includegraphics[width=.7\textwidth]{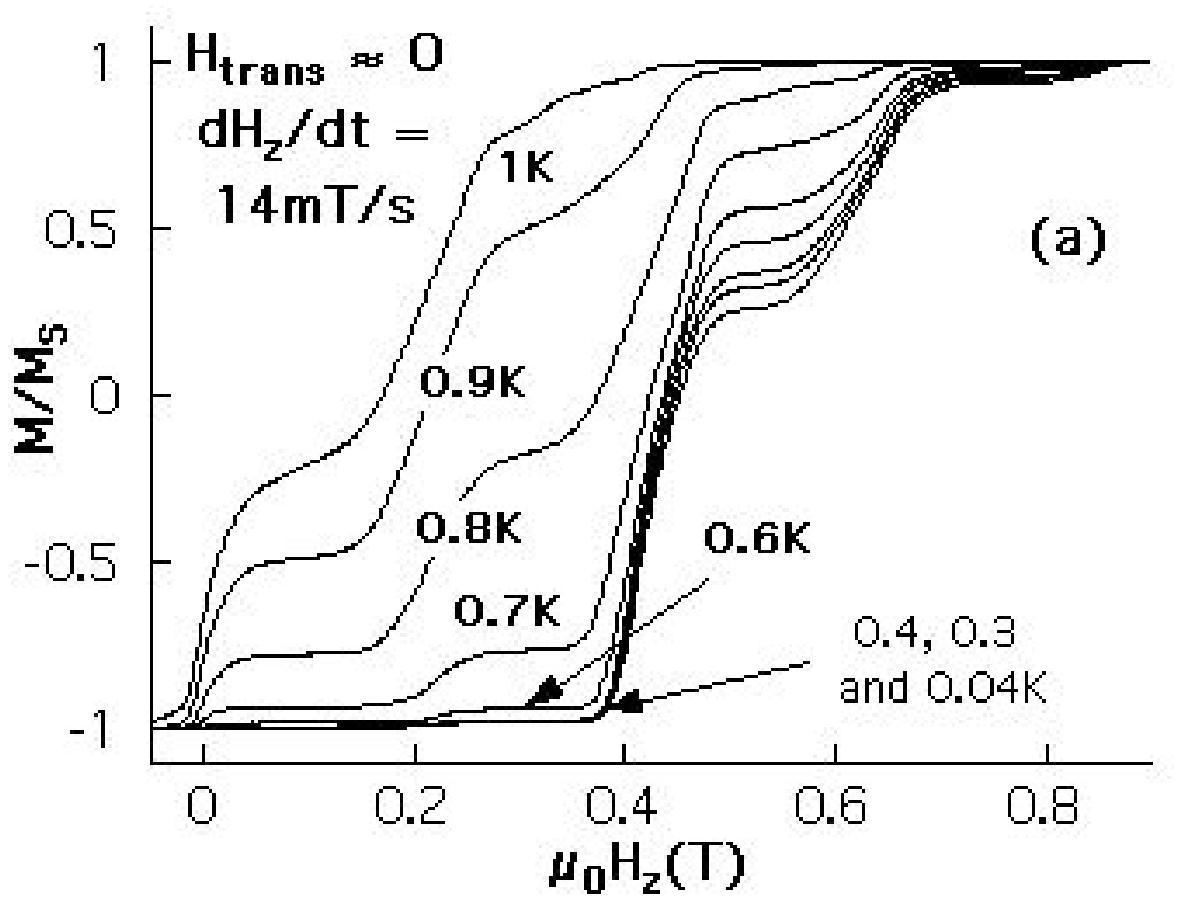}
\includegraphics[width=.7\textwidth]{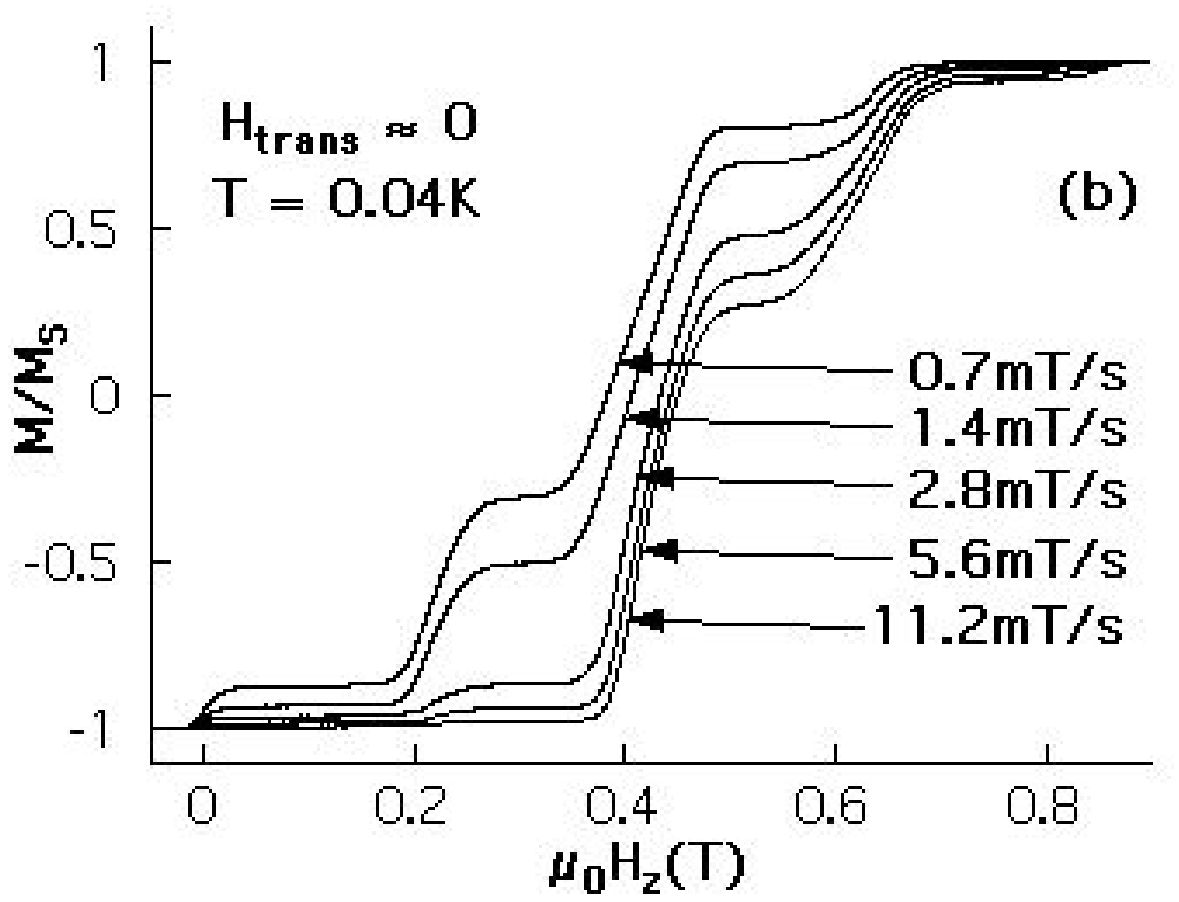}
\caption{Hysteresis loops of a single crystal 
of Fe$_8$ molecular clusters: 
(a) at different temperatures and a constant 
sweeping rate $dH_z/dt$ = 0.014 T/s, and
(b) at 0.04 K and different field sweeping rates. 
The loops display a series of steps, 
separated by plateaus. As the temperature is lowered, 
there is a decrease in 
the transition rate due to reduced thermal 
assisted tunneling. The hysteresis 
loops become temperature independent 
below 0.35~K, demonstrating 
quantum tunneling at the lowest energy levels. 
The resonance widths at small fields $H_z$ 
of about 0.05~T are mainly
due to dipolar fields between the molecular 
clusters~\cite{Ohm98a,WW_PRL99}.}
\label{hyst_Fe8}
\end{center}
\end{figure}

The effect of these avoided level crossings can be seen in hysteresis loop 
measurements (Fig.~\ref{hyst_Fe8}). When the applied field is near an 
avoided level crossing, the magnetization relaxes faster, yielding steps 
separated by plateaus. As the temperature is lowered, there is a decrease 
in the transition rate due to reduced thermal-assisted tunneling. 
A similar behavior was observed in Mn$_{12}$ acetate clusters 
\cite{Novak95,Friedman96,Thomas96} where equally 
separated steps were observed at 
$H_z \approx n \times 0.45$~T. The main difference 
between both clusters is that 
the hysteresis loops of Fe$_8$ become 
temperature independent below 0.36 K
whereas measurements on Mn$_{12}$ acetate 
indicate a temperature
independence only for strong applied fields
and below 0.6K~\cite{Perenboom98,Kent00}. 

\begin{figure}
\begin{center}
\includegraphics[width=.7\textwidth]{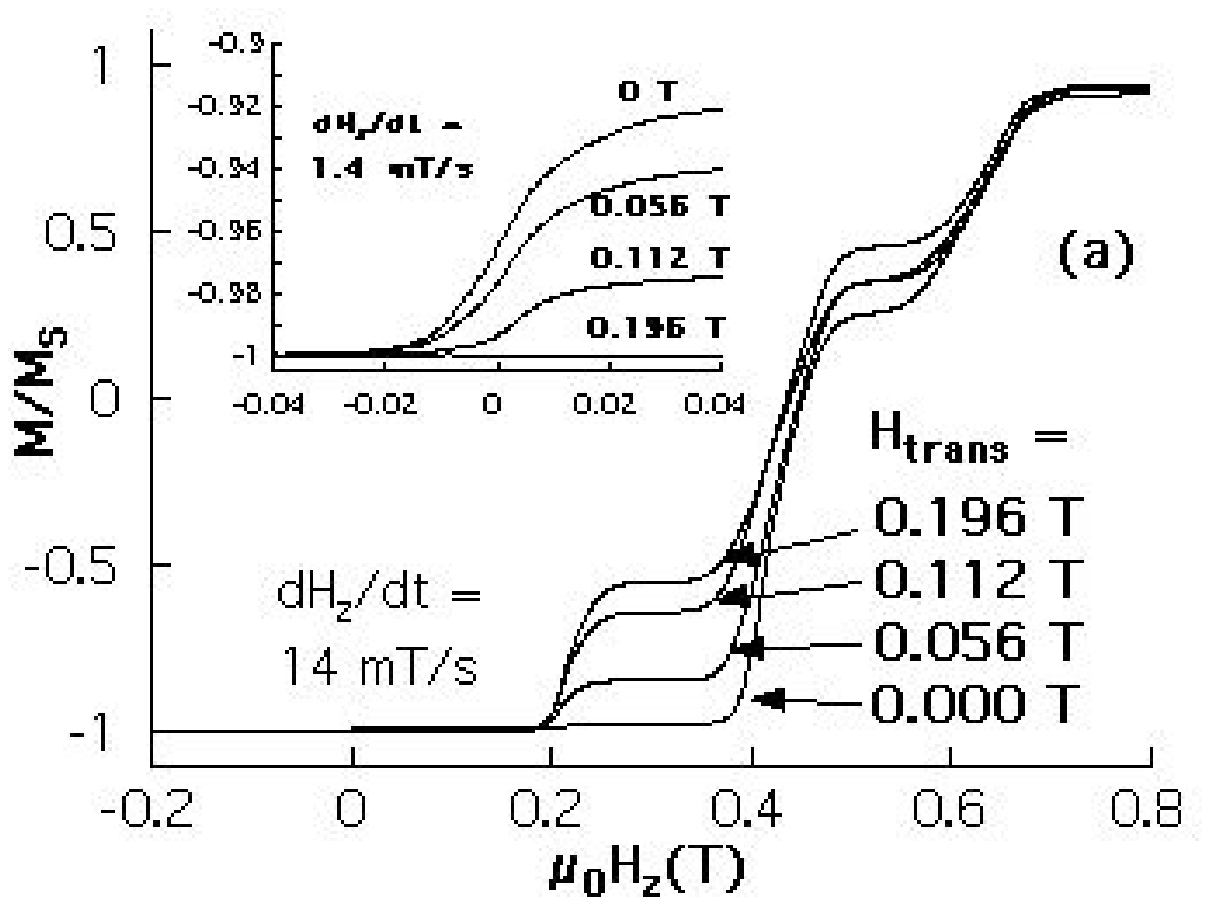}
\includegraphics[width=.7\textwidth]{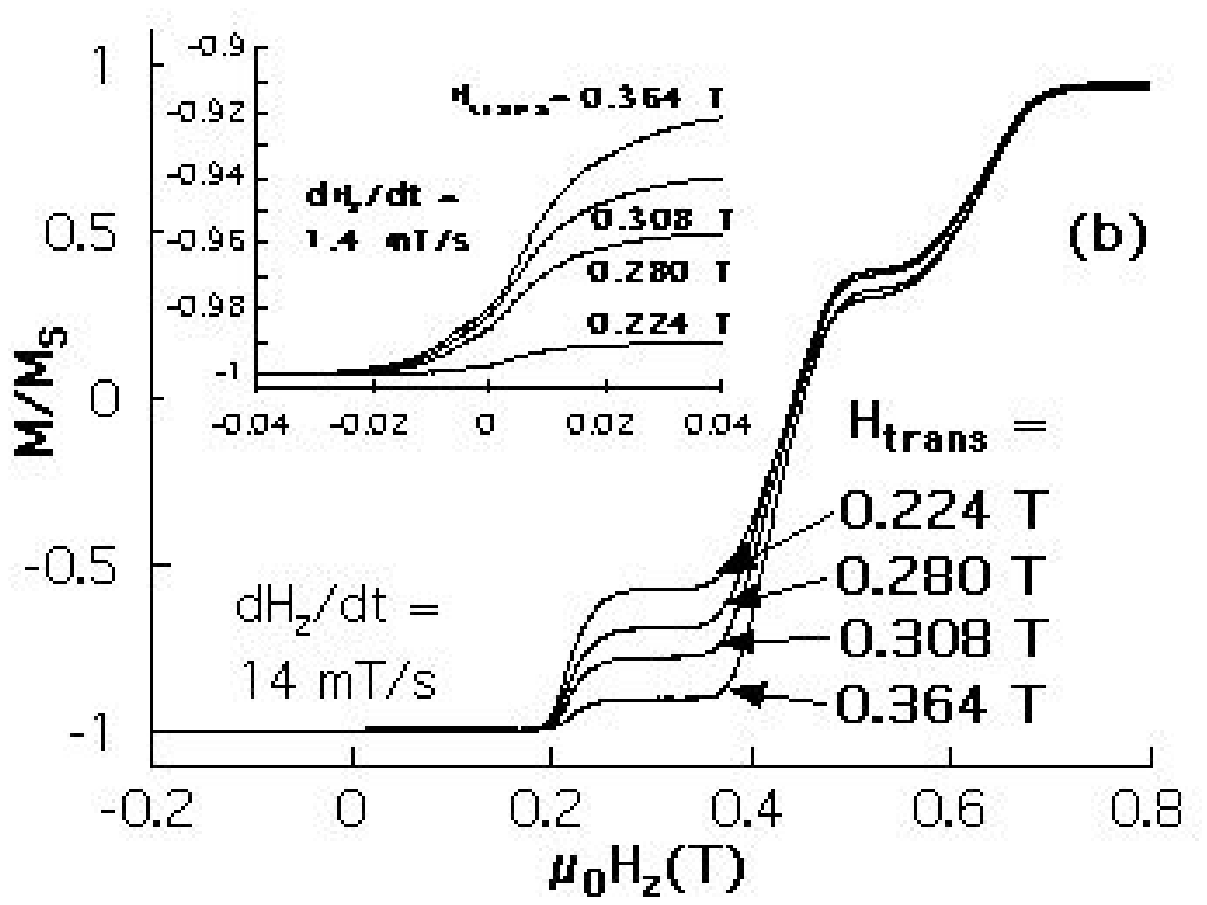}
\caption{Hysteresis loops measured along
$H_z$ in the presence of a 
constant transverse field at 0.04 K. 
Insets: enlargement around the field $H_z = 0$. 
Notice that the sweeping rate is ten times slower 
for the measurements in the insets than that of the main figures.}
\label{fig_hyst_Hx}
\end{center}
\end{figure}

Another important difference is that the step 
heights ({\it i.e.} the relaxation rates) change periodically 
when a constant transverse field is applied 
(Fig.~\ref{fig_hyst_Hx}). It is 
the purpose of the next subsections to 
present a detailed study of this behavior.

%%%%%%%%%%%%%%%%%%%%%%%%%%%%%%%%
\subsection{Landau--Zener tunneling in Fe$_8$}
\label{LZ} 
%%%%%%%%%%%%%%%%%%%%%%%%%%%%%%%%

\begin{figure}
\begin{center}\leavevmode
\includegraphics[width=.7\textwidth]{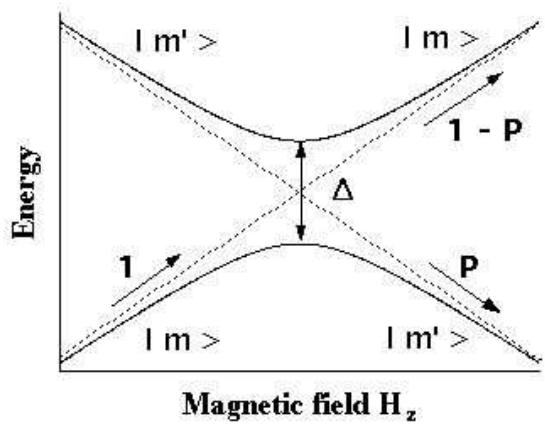}
\caption{Detail of the energy level diagram near an avoided level crossing.
$m$ and $m'$ are the quantum numbers of the energy level.
$P_{m,m'}$ is the Landau--Zener tunnel probability when sweeping the applied 
field from the left to the right over the anti-crossing.
The greater the gap $\Delta$ and the slower the sweeping rate, 
the higher is the tunnel rate (Eq.~\ref{eq_LZ}).}
\label{LZ_crossing}
\end{center}
\end{figure}

The non-adiabatic transition between the two states 
in a two-level system has first been discussed
by Landau, Zener and St$\ddot{u}$ckelberg
~\cite{Landau32,Zener32,Stuckelberg32}. 
The original work by Zener concentrates on
the electronic states of a bi-atomic molecule, 
while Landau and St$\ddot{u}$ckelberg
considered two atoms which undergo a scattering process.
Their solution of the time dependent Schr$\ddot{o}$dinger equation
of a two-level system could be applied to many physical systems
and it became an important tool for studying tunneling transitions.
The Landau--Zener model has also been applied to spin tunneling 
in nanoparticles and clusters
~\cite{Miyashita95,Miyashita96,Rose98,Rose99,Thorwart00,Leuenberger00}.
The tunneling probability $P$ when sweeping the 
longitudinal field $H_z$ at a constant rate 
over an avoided energy level crossing 
(Fig.~\ref{LZ_crossing}) is given by:
\begin{equation}
P_{m,m'} = 1 - {\rm exp}\left\lbrack-
     \frac {\pi \Delta_{m,m'}^2}
     {2 \hbar g \mu_{\rm B} |m - m'| \mu_0 dH_z/dt}\right\rbrack
\label{eq_LZ}
\end{equation}
Here, $m$ and $m'$ are the quantum numbers of the avoided level crossing, 
$dH_z/dt$ is the constant field sweeping rates, 
$g~\approx~2$, $\mu_{\rm B}$ the Bohr magneton, and 
$\hbar$ is Planck's constant. 

With the  Landau--Zener model in mind, we can now start to
understand qualitatively the hysteresis loops (Fig.~\ref{hyst_Fe8}b).
Let us start at a large negative magnetic field $H_z$. At very low
temperature, all molecules are in the $m = -10$ ground state.
When the applied field $H_z$ is ramped down to zero, all
molecules will stay in the $m = -10$ ground state. 
When ramping the field over the $\Delta_{-10,10}$--region
at $H_z \approx 0$, 
there is a Landau--Zener tunnel probability 
$P_{-10,10}$ to tunnel from the
$m = -10$ to the $m = 10$ state. 
$P_{-10,10}$ depends
on the sweeping rate (Eq.~\ref{eq_LZ}), {\it i.e.} the slower the 
sweeping rate, the larger $P_{-10,10}$. This is clearly demonstrated
in the hysteresis loop measurements showing larger steps
for slower sweeping rates (Fig.~\ref{hyst_Fe8}b). 
When the field $H_z$ is now further
increased, there is a remaining fraction of molecules in the
$m = -10$ state which became a metastable state. The next
chance to escape from this state is when the  field reaches
the $\Delta_{-10,9}$--region. There is a Landau--Zener
tunnel probability $P_{-10,9}$ to tunnel from the
$m = -10$ to the $m = 9$ state. As $m = 9$ is an excited state,
the molecules in this state desexcite quickly to 
the $m = 10$ state by emitting a phonon.
An analogous procedure happens when the  applied field reaches 
the $\Delta_{-10,10-n}$--regions ($n$ = 2, 3, \ldots) 
until all molecules are
in the $m = 10$ ground state, {\it i.e.} the 
magnetization of all molecules
is reversed. As phonon emission can 
only change the molecule state 
by $\Delta m$ = 1 or 2, there is a phonon 
cascade for higher applied fields. 
\footnote{Phonon induced transitions with $|\Delta m| > 2$
are very small \cite{Villain97,Fort98,Leuenberger00a}.}

In order to apply quantitatively the Landau--Zener 
formula (Eq.~\ref{eq_LZ}), we first saturated the 
crystal of Fe$_8$ clusters in a field of $H_z = -1.4$~T, 
yielding an initial magnetization $M_{\rm in} = -M_{\rm s}$. 
\footnote{In order to avoid heating problems for 
measurements of $\Delta$ for $n > 1$, we started in a thermally 
annealed sample with $M_{\rm in} = 0.95 M_{\rm s}$ 
instead of $M_{\rm in} = - M_{\rm s}$ or $M_{\rm in} = 0$.} 
Then, we swept the applied field at a constant rate 
over one of the resonance transitions and 
measured the fraction of molecules which 
reversed their spin. This procedure yields the 
tunneling rate $P_{-10,10-n}$ and thus the tunnel 
splitting $\Delta_{-10,10-n}$ (Eq.~\ref{eq_LZ})
with $n$ = 0, 1, 2, \ldots.

For very small tunneling probabilities $P_{-10,10-n}$, we did multiple sweeps 
over the resonance transition. 
The magnetization $M$ after $N$ 
sweeps is given by:
\begin{equation}
	M(N) \approx M_{\rm eq} + (M_{\rm in} - M_{\rm eq})
	              e^{-k P_{-10,10-n} N} =
				  M_{\rm eq} + (M_{\rm in} - M_{\rm eq}) e^{-\Gamma t}
\label{eq_M_N}
\end{equation}
Here $M_{\rm in}$ is the initial magnetization, 
$M_{\rm eq}(H_z)$ is the equilibrium magnetization,
$N~= \frac{1}{A}\frac{dH_z}{dt} t$ is the number 
of sweeps over the level crossing, 
$\Gamma~= k P_{-10,10-n} \frac{1}{A}\frac{dH_z}{dt}$ 
is the overall Landau--Zener transition rate, 
$k$ = 2 for $n$ = 0 and  $k$ = 1 for $n$ = 1, 2, \ldots,
and $A$ is the amplitude of the ramp-field.
\footnote{We supposed here that the forth and 
back sweeps give the same tunnel 
probability. This is a good approximation for $P << 1$ 
where next nearest neighbor (molecule) effects can be neglected.}
We have therefore a simple tool to obtain the tunnel splitting by 
measuring $P_{-10,10-n}$, or $M(N)$ for $P_{-10,10-n} << 1$.

\begin{figure}
\begin{center}
\includegraphics[width=.7\textwidth]{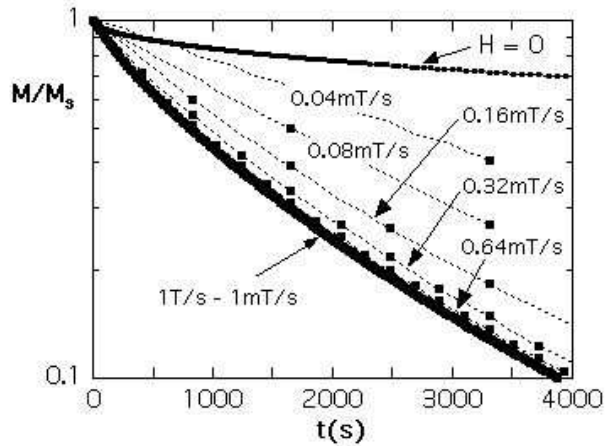}
\caption{Scaling plot for the Landau--Zener method 
showing the predicted field sweeping 
rate dependence for 1~T/s to 1~mT/s. 
Each point indicates the magnetization after a 
field sweep over the $m = \pm10$ resonance. 
The dotted lines are guides for the eyes.
For comparison, the figure displays also a 
relaxation curve at a constant field $\vec{H} = 0$
(Fig.~\ref{fig_hyst_Hx})
which shows much slower relaxation~\cite{WW_PRL99}.}
\label{LZ_test_1}
\end{center}
\end{figure}

\begin{figure}
\begin{center}
\includegraphics[width=.7\textwidth]{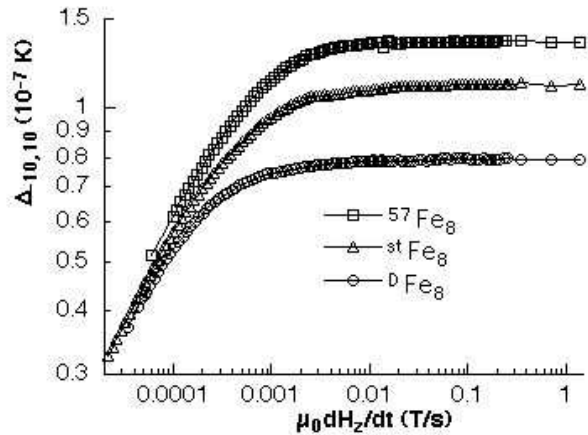}
\caption{Scaling plot for the Landau--Zener 
method showing the predicted field sweeping 
rate dependence for 1~T/s to 1~mT/s similar to Fig.~\ref{LZ_test_1} 
but for three isotopically substituted Fe$_8$ samples.
Further details are presented in Sect.~\ref{hyperfine_Fe8}
}
\label{LZ_test_st_57_D}
\end{center}
\end{figure}

We first checked the predicted Landau--Zener 
sweeping field dependence of the tunneling rate. 
This can be done, for example, by plotting the 
relaxation of magnetization as a function of 
$t = N \frac{A}{dH_z/dt}$. The Landau 
Zener model predicts that all measurements should 
fall on one curve which was indeed the 
case for sweeping rates between 1 and 0.001~T/s (Fig.~\ref{LZ_test_1}) 
for the $m = \pm10$ transition. The deviations 
at lower sweeping rates, are mainly due to the 
{\it hole-digging mechanism}~\cite{WW_PRL99} 
which slows down the relaxation (see Sect.~\ref{digging}).
\footnote{Roughly speaking, at very low field sweeping rates
internal fields change faster than the external field.}
In the ideal case, we should find an exponential 
curve (Eq.~\ref{eq_M_N}). However, we found clear
deviations from the exponential curve 
(Fig.~\ref{LZ_test_st_57_D}). 
which might be 
due to molecules with different amounts 
of nuclear spins. For example, two 
percent of natural iron has a nuclear spin, 
{\it i.e.} about 10 percent of 
Fe$_8$ has at least one nuclear spin on the iron.
This interpretation is supported by measurements on isotopically 
substituted Fe$_8$ samples (Fig.~\ref{LZ_test_st_57_D}). 

\begin{figure}
\begin{center}
\includegraphics[width=.7\textwidth]{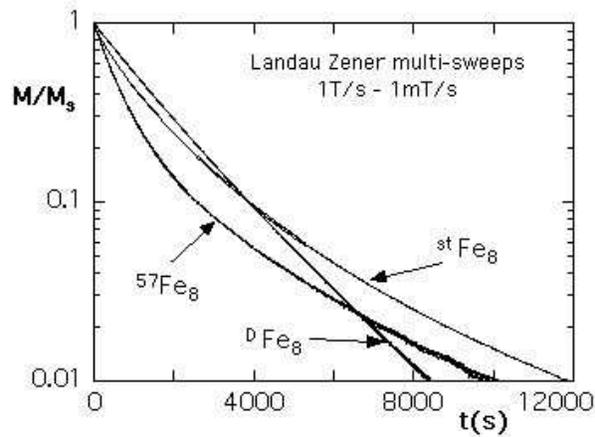}
\caption{Field sweeping rate dependence of the tunnel splitting 
$\Delta_{-10,10}$ 
measured by a Landau--Zener method for three Fe$_8$ samples,
for $H_x$ = 0. 
The Landau--Zener method
works in the region of high sweeping rates 
where $\Delta_{-10,10}$ is sweeping rate independent. 
Note that the differences of $\Delta_{-10,10}$ between 
the three isotopically substituted samples 
are rather small in comparison 
to the oscillations in Fig.~\ref{Delta_Hx}.}
\label{LZ_test_2}
\end{center}
\end{figure}

Another way of checking the Landau--Zener 
sweeping field dependence of the tunneling rate is presented
in Fig.~\ref{LZ_test_2} showing a sweeping rate independent
$\Delta_{-10,10}$ between 1 and 0.001~T/s. 
\footnote{Recent measurements confirmed the good agreement up to 30 T/s.}
The measurements on isotopically substituted Fe$_8$ samples show a 
small dependence of $\Delta_{m,m'}$ on the 
hyperfine coupling (Fig.~\ref{LZ_test_2}). 
Such an effect has been predicted for a constant applied 
field by Tupitsyn {\it et al.} \cite{Tupitsyn97},
and for a ramped field by Rose \cite{Rose99}.
Further details are presented in Sect.~\ref{hyperfine_Fe8}.

We also compared the tunneling rates found 
by the Landau--Zener method with those 
found using a square-root decay method 
which was proposed by Prokof'ev and 
Stamp~\cite{Prokofev98}, and 
found a good agreement~\cite{WW_PRL99,WW_PRL00} (Sect.~\ref{Prok_Stamp}).

Our measurements showed for the first time that the Landau--Zener 
method is particularly adapted for 
molecular clusters because it works even in 
the presence of dipolar fields 
which spread the resonance transition provided 
that the field sweeping rate is not too 
small. Furthermore, our measurements show 
a small but clear influence of the 
hyperfine coupling which should be included in a generalized
Landau--Zener model~\cite{Rose99}.

%
%%%%%%%%%%%%%%%%%%%%%%%%%%%%%%%%
\subsection{Oscillations of tunnel splitting}
\label{oscillations} 
\begin{figure}
\begin{center}
\includegraphics[width=.7\textwidth]{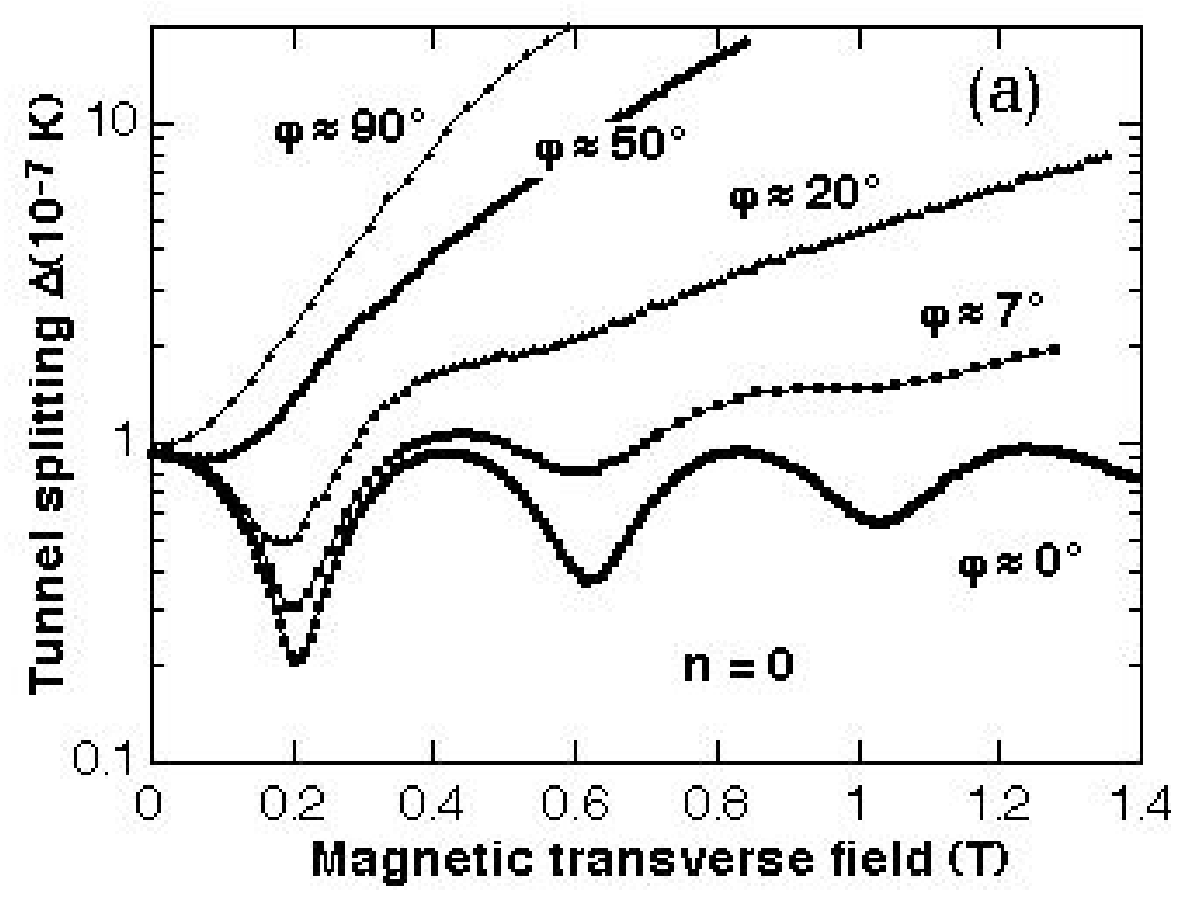}
\includegraphics[width=.7\textwidth]{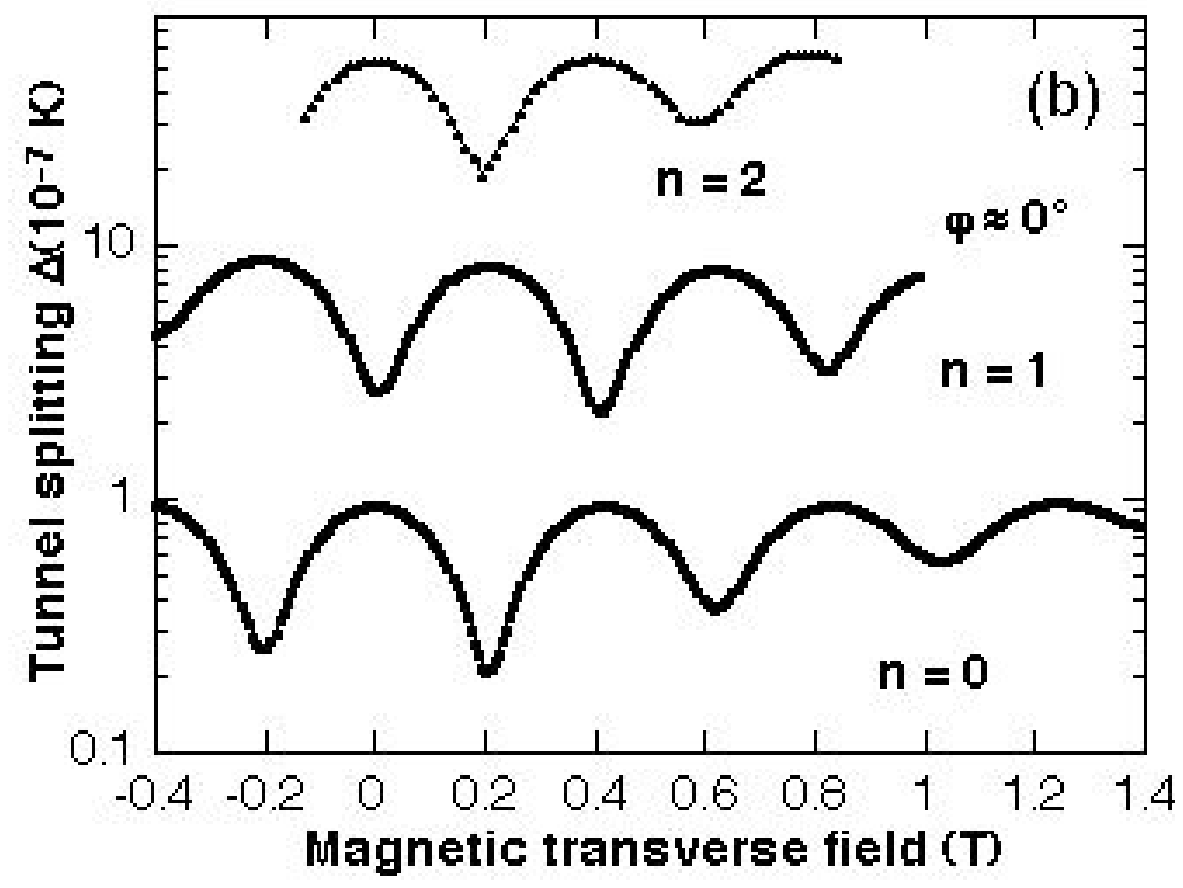}
\caption{Measured tunnel splitting $\Delta$ as a 
function of transverse field for (a) several azimuth angles 
$\varphi$ at $m = \pm 10$, and (b) $\varphi 
\approx 0^{\circ}$, and for quantum transition between 
$m = -10$ and $(10 - n)$. Note the parity 
effect which is analogous to the suppression of 
tunneling predicted for half integer spins. 
It should also be mentioned that 
internal dipolar and hyperfine fields hinder a 
quench of $\Delta$ which is predicted for an 
isolated spin (Figs.~\ref{Delta_Hx_theo} and \ref{Delta_Hx_Min}).}
\label{Delta_Hx}
\end{center}
\end{figure}

%%%%%%%%%%%%%%%%%%%%%%%%%%%%%%%%
%
An applied field in the $xy-$plane can tune
the tunnel splittings $\Delta_{m,m'}$ via the $S_x$ and $S_y$ 
spin operators of the Zeeman terms which do not commute with
the spin Hamiltonian. 
This effect can be demonstrated by using the Landau--Zener method 
(Sect.~\ref{LZ}).
Fig.~\ref{Delta_Hx} presents a detailed study 
of the tunnel splitting $\Delta_{\pm10}$  
at the tunnel transition between $m = \pm10$, 
as a function of transverse fields applied at different angles $\varphi$, 
defined as the azimuth angle between the anisotropy hard axis and the 
transverse field (Fig.~\ref{sphere}).
For small angles $\varphi$ the tunneling rate 
oscillates with a period of 
ca. 0.4~T, whereas no oscillations showed 
up for large angles $\varphi$~\cite{WW_Science99}. 
In the latter case, a much stronger increase of $\Delta_{\pm10}$ with 
transverse field is observed. The transverse 
field dependence of the tunneling rate for 
different resonance conditions between 
the state $m = -10$ and $(10 - n)$ can be 
observed by sweeping the longitudinal field around 
$\mu_0H_z = n \times 0.22$~T with $n$~= 0, 1, 2,~... 
The corresponding tunnel splittings $\Delta_{-10,10-n}$ 
oscillate with almost the same 
period of ca. 0.4~T (Fig.~\ref{Delta_Hx}). 
In addition, comparing quantum transitions between 
$m = -10$ and $(10 - n)$, with $n$ even or odd, 
revealed a parity (or symmetry) effect which is 
analogous to the Kramers' suppression of 
tunneling predicted for half integer 
spins~\cite{Loss92,Delft92}. This behavior has been observed 
for $n~=~0$ to~4.
\footnote{The tunneling rate were too fast for $n > 4$.}
A similar strong 
dependence on the azimuth angle $\varphi$ 
was observed for all the resonances.

%%%%%%%%%%%%%%%%%%%%%%%%%%%%%%%%
\subsubsection{Semi-classical descriptions}
\label{Berry} 
%%%%%%%%%%%%%%%%%%%%%%%%%%%%%%%%
%
Before showing that the above results can be derived
by an exact numerical calculation using the 
quantum operator formalism, it is useful
to discuss semi-classical models. The original
prediction of oscillation of the tunnel splitting
was done by using the path integral formalism~\cite{Feynman70}.
Here~\cite{Garg93}, the oscillations 
are explained by constructive 
or destructive interference of quantum 
spin phases (Berry phases) of two tunnel paths
(instanton trajectories) (Fig.~\ref{sphere}).
Since our experiments were reported, 
the Wentzel--Kramers--Brillouin theory has been
used independently by Garg \cite{Garg99b} and
Villain and Fort \cite{Villain00}.
The surprise is that although these 
models~\cite{Garg93,Garg99b,Villain00} are 
derived semiclassically, and should
have higher order corrections in $1/S$, they appear 
to be exact as written! This has first been
noted in Refs.~\cite{Garg99b,Villain00}
and then proven in Ref.~\cite{Garg00b}
Some extensions or alternative explications of 
Garg's result can be found in 
Refs.~\cite{Barnes99,Liang00,Yoo00,Hu00}.

The period of oscillation 
is given by \cite{Garg93}:
\begin{equation}
\Delta H = \frac {2 k_{\rm B}}{g \mu_{\rm B}} \sqrt{2 E (E + D)}
\label{eq_Garg}
\end{equation}
where $D$ and $E$ are defined in Eq.~\ref{eq_H_biax}.
We find a period of oscillation of $\Delta H$~= 0.26~T for $D$~= 0.275~K  
and $E$~= 0.046~K as in Ref.~\cite{Barra96}. 
This is somewhat smaller than the experimental value of ca. 0.4 T.
We believe that this is due to higher order terms of the spin 
Hamiltonian which are neglected in Garg's calculation. These terms can
easily be included in the operator formalism
as show in next subsection.

%%%%%%%%%%%%%%%%%%%%%%%%%%%%%%%%
\subsubsection{Exact numerical diagonalisation}
\label{numeric} 
%%%%%%%%%%%%%%%%%%%%%%%%%%%%%%%%
In order to 
quantitatively reproduce the observed periodicity 
we included forth order terms in the 
spin Hamiltonian (Eq.~\ref{eq_H_biax}) as recently employed in 
the simulation of inelastic neutron 
scattering measurements \cite{Caciuffo98,Caciuffo00} and performed a 
diagonalisation of the $[21 \times 21]$ 
matrix describing the $S = 10$ system. 
For the calculation of the tunnel splitting 
we used $D = 0.289$ K, $E = 0.055$ K (Eq.~\ref{eq_H_biax}) and
the forth order terms as defined 
in~\cite{Caciuffo98} with 
$B^0_4 = 0.72 \times$ 10$^{-6}$ K, 
$B^2_4 = 1.01 \times$ 10$^{-5}$ K, 
$B^4_4 = -0.43 \times$ 10$^{-4}$ K, 
which are close to the values obtained by EPR 
measurements in ~\cite{Barra00} and neutron 
scattering measurements \cite{Caciuffo00}.

\begin{figure}
\begin{center}
\includegraphics[width=.7\textwidth]{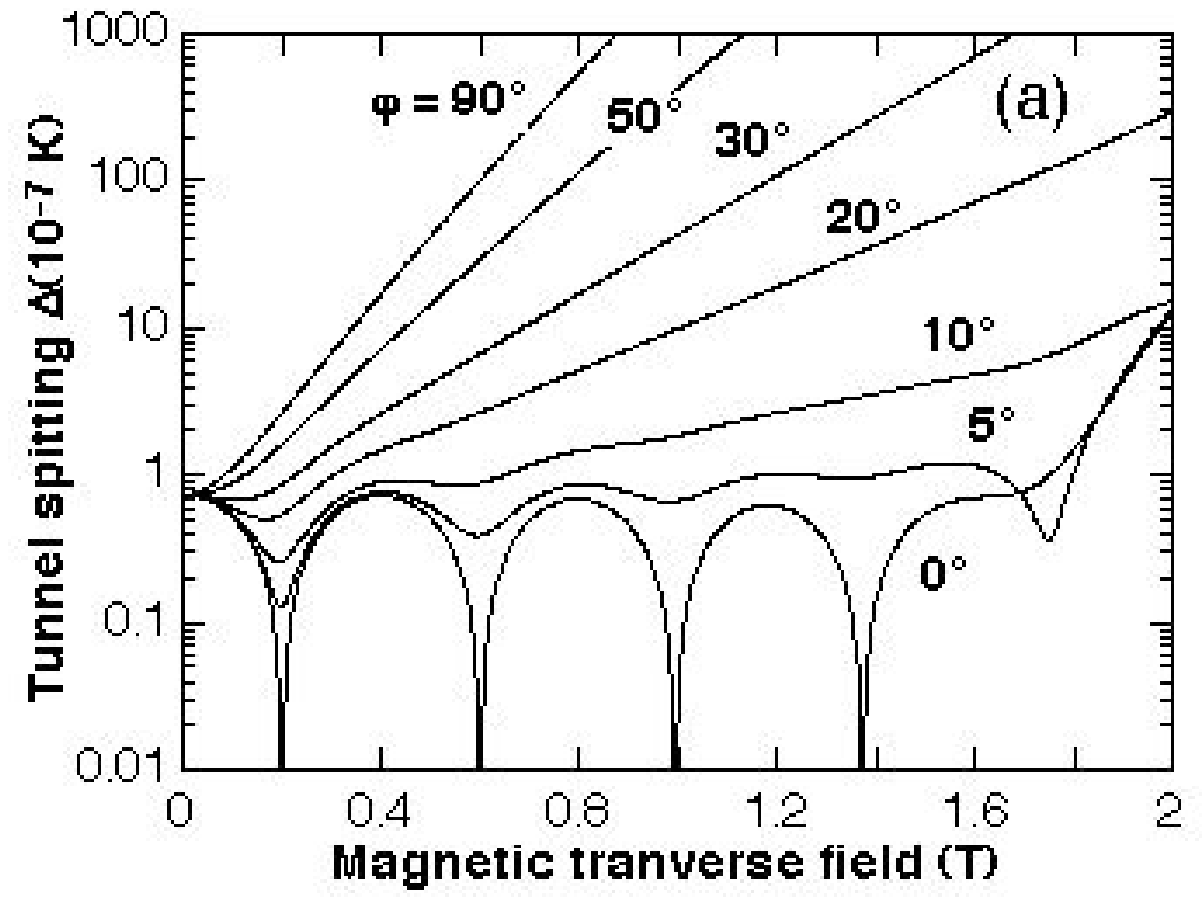}
\includegraphics[width=.7\textwidth]{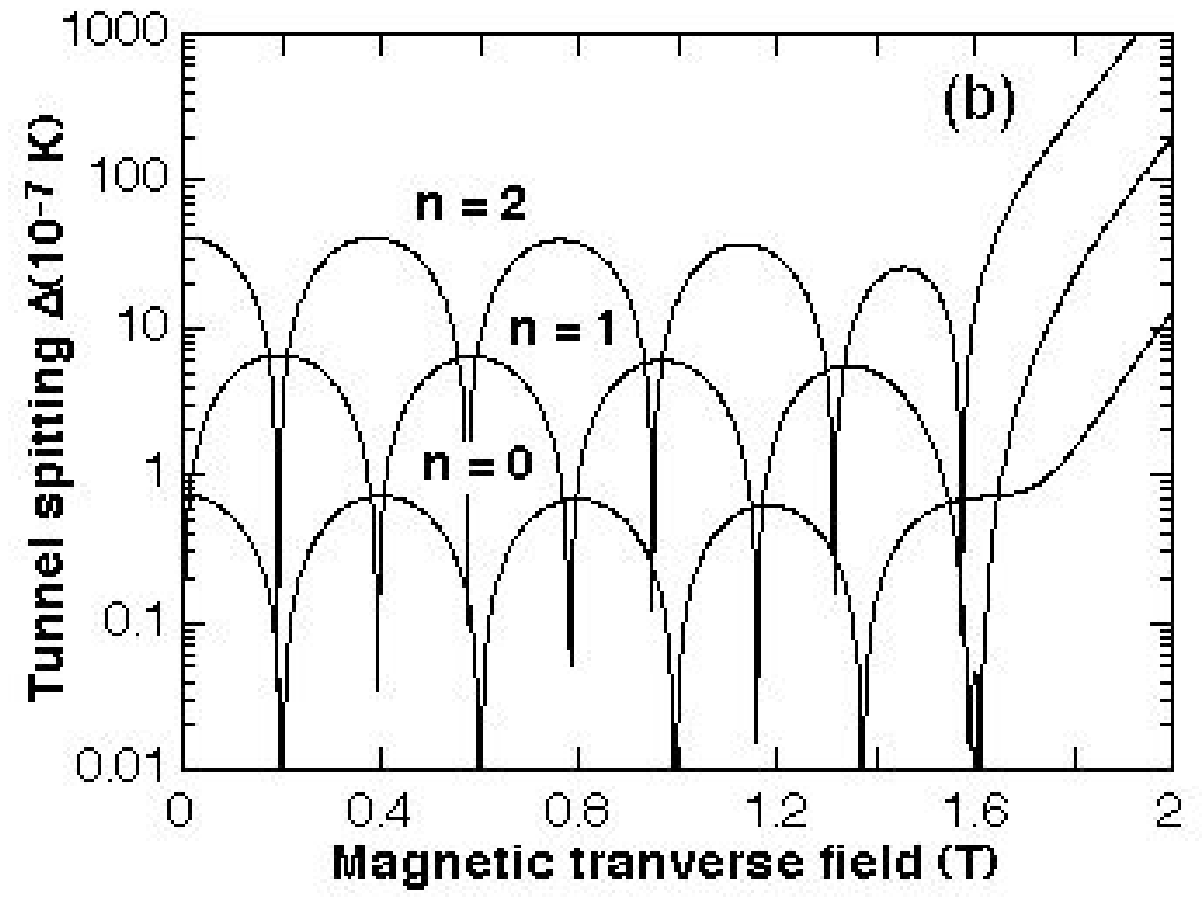}
\caption{Calculated tunnel splitting $\Delta$ 
as a function of transverse field for (a)
quantum transition between $m = \pm10$ at several
azimuth angles $\varphi$, and (b)
quantum transition between $m = -10$ and $(10 - n)$ 
at $\varphi = 0^{\circ}$ (Sect.~\ref{numeric}).
The forth order terms 
suppress the oscillations of $\Delta$ for large 
transverse fields $|H_x|$.}
\label{Delta_Hx_theo}
\end{center}
\end{figure}

The calculated tunnel splittings for the states involved 
in the tunneling process 
at the resonances $n$ = 0, 1, and 2 are 
reported in Fig. \ref{Delta_Hx_theo}, 
showing the oscillations as well as 
the parity effect for odd 
resonances. 
The calculated tunneling splitting is however  
ca. 1.5 times smaller  than the observed one. 
This small discrepancy could be reduced by 
introducing higher order terms. We believe that this
is not relevant because the above model neglects for example 
the influence of nuclear spins which seems to increase the
measured (effective) tunnel splittings 
(Fig.~\ref{LZ_test_2} and Sect.~\ref{hyperfine_Fe8}).

Our choice of the forth order terms 
suppresses the oscillations of large transverse 
fields (Fig.~\ref{Delta_Hx_theo}). This region could 
not be studied in the current set-up. 
Future measurements should 
focus on the higher field region in order 
to find a better effective Hamiltonian.

%%%%%%%%%%%%%%%%%%%%%%%%%%%%%%%%
\section{Environmental decoherence effects in molecular clusters}
\label{decoherence} 
%%%%%%%%%%%%%%%%%%%%%%%%%%%%%%%%

At temperatures below 0.36~K, Fe$_8$ molecular clusters display a 
clear crossover from thermally activated relaxation to a temperature 
independent quantum regime, with a pronounced resonance structure of 
the relaxation time as a function of the external field 
(Sect.~\ref{anisotropy_Fe8}).
It was surprising however that the observed 
relaxation of the magnetization in the quantum regime was found to be 
non-exponential and the resonance width orders of magnitude too large 
\cite{Sangregorio97,Ohm98a}. The key to understand this seemingly 
anomalous behavior involves the hyperfine 
fields as well as the evolving distribution of the weak 
dipole fields of the nanomagnets themselves \cite{Prokofev98}.
Both effects showed to be the main source of decoherence
at very low temperature. At higher temperatures, phonons
are another source of decoherence.

In the following sections, we focus on the 
low temperature and low field limits, 
where phonon-mediated relaxation is astronomically long and can be 
neglected. In this limit, the $m~= \pm S$ spin 
states are coupled due to the tunneling 
splitting $\Delta_{\pm S}$ which is about 
10$^{-7}$~K for Fe$_8$ (Sect.~\ref{oscillations})
and 10$^{-11}$~K for Mn$_{12}$~\cite{AlSaqer00} with $S$ = 10. 
In order to tunnel between these 
states, the longitudinal magnetic energy 
bias $\xi=g\mu_{\rm B}SH_{\rm local}$ due to the local 
magnetic field $H_{\rm local}$ on a molecule must be smaller than $\Delta_{\pm S}$
implying a local field smaller than $10^{-8}$~T for Fe$_8$ 
clusters. Since the typical intermolecular dipole fields are of the 
order of 0.05~T, it seems at first that almost all molecules should be 
blocked from tunneling by a very large energy bias. Prokof'ev and Stamp 
have proposed a solution to this dilemma by proposing that fast dynamic 
nuclear fluctuations broaden the resonance, and the gradual adjustment 
of the dipole fields in the sample caused by the tunneling, brings 
other molecules into resonance and allows continuous relaxation 
\cite{Prokofev98}. Some interesting predictions are briefly reviewed 
in the following section.

%%%%%%%%%%%%%%%%%%%%%%%%%%%%%%%%
\subsection{Prokof'ev--Stamp theory}
\label{Prok_Stamp} 
%%%%%%%%%%%%%%%%%%%%%%%%%%%%%%%%

Prokof'ev and Stamp were the first who realized that 
there are localized couplings of
environmental modes with mesoscopic systems which cannot be modeled 
with an \lq oscillator bath' model~\cite{Feynman63} 
describing delocalized environmental modes such as electrons, phonons, 
photons, etc. They found that these localized modes such as 
nuclear and paramagnetic spins are often strong and described them 
with a spin bath model~\cite{Prokofev96}.
We do not review this theory
\footnote{For a review, see~\cite{Prokofev00}.}
but focus on one particular application 
which is interesting for molecular clusters~\cite{Prokofev98}.
Prokof'ev and Stamp showed that at a 
given longitudinal applied field $H_z$, the magnetization of a crystal
of molecular clusters should relax 
at short times with a square-root time dependence
which is due to a gradual modification 
of the dipole fields in the sample caused by the tunneling:
\begin{equation}
	M(H_z,t)=M_{\rm in}+(M_{\rm eq}(H_z)-M_{\rm in})
	       \sqrt{\Gamma_{\rm sqrt}(H_z)t} 
\label{M_sqrt}
\end{equation}
Here $M_{\rm in}$ is the initial magnetization at time $t$~= 0 
({\it i.e.} after a rapid field change), 
and $M_{\rm eq}(H_z)$ is the equilibrium magnetization at $H_z$. 
The rate function $\Gamma_{\rm sqrt}(H_z)$ is 
proportional to the normalized distribution $P(H_z)$ of molecules 
which are in resonance at $H_z$:
\begin{equation}
	\Gamma_{\rm sqrt}(H_z) = 
	        c \frac{\xi_0}{E_D}
	          \frac{\Delta_{\pm S}^2}{4 \hbar} P(H_z) 
\label{gamma_sqrt}
\end{equation}
where $\xi_0$ is the line width coming 
from the nuclear spins, 
$E_D$ is the Gaussian half-width of $P(H_z)$, 
and $c$ is a constant of the order 
of unity which depends on the sample shape. 
If these simple relations 
are exact, then measurements of the 
short time relaxation as a 
function of the applied field $H_z$ gives directly 
the distribution $P(H_z)$, and allows one to measure the 
tunnel splitting $\Delta_{\pm S}$ which is described
in the next section.

%%%%%%%%%%%%%%%%%%%%%%%%%%%%%%%%
\subsection{Hole digging method to study 
dipolar distributions and hyperfine couplings}
\label{digging} 
%%%%%%%%%%%%%%%%%%%%%%%%%%%%%%%%

\begin{figure}
\begin{center}
\includegraphics[width=.7\textwidth]{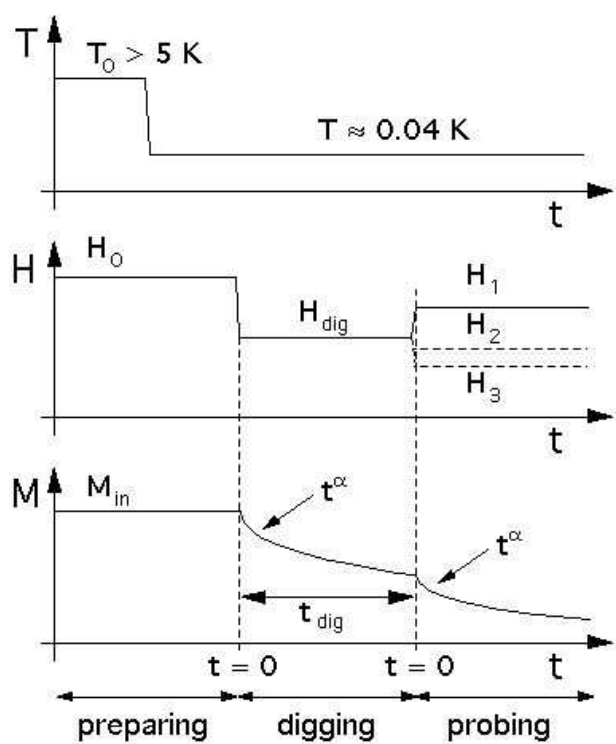}
\caption{Schema of the hole digging method presenting
the time dependence of temperature, applied field and magnetization of
the sample.}
\label{fig_dig}
\end{center}
\end{figure}

Motivated by the Prokof'ev--Stamp theory~\cite{Prokofev98},
we developed a new technique, which we call the
{\it hole digging method} that can be 
used to observe the time evolution of
molecular states in crystals of molecular clusters.
It allowed us to measure the statistical 
distribution of magnetic bias
fields in the Fe$_8$ system that 
arise from the weak dipole fields of 
the clusters themselves. A
hole can be \lq dug' into the distribution 
by depleting the available spins at
a given applied field.
Our method is based on the simple idea that after 
a rapid field change, the resulting short time relaxation 
of the magnetization is directly related to the
number of molecules which are in resonance at the given applied field.
Prokof'ev and Stamp have suggested that the short time relaxation should
follow a $\sqrt{t}-$relaxation law (Eq.~\ref{M_sqrt}). 
However, the hole digging method should
work with any short time relaxation
law, for example a power law:
\begin{equation}
	M(H_z,t)=M_{\rm in}+(M_{\rm eq}(H_z)-M_{\rm in})
	       (\Gamma_{\rm short}(H_z)t)^\alpha 
\label{M_power}
\end{equation}
where $\Gamma_{\rm short}$ is a characteristic short time
relaxation rate which is directly related to the
number of molecules which are in resonance at 
the applied field $H_z$, and $0 < \alpha <  1$ in most cases.
$\alpha$ = 0.5 in the Prokof'ev--Stamp theory (Eq.~\ref{M_sqrt})
and $\Gamma_{\rm sqrt}$ is directly proportional 
to $P(H_z)$ (Eq.~\ref{gamma_sqrt}).
The {\it hole digging method} can be divided into three steps 
(Fig.~\ref{fig_dig}):

\begin{enumerate}
\item
{\bf Preparing the initial state:} 
A well defined initial magnetization 
state of the crystal of molecular 
clusters can be achieved by rapidly 
cooling the sample from high 
down to low temperatures in a constant applied 
field $H_z^0$. For zero applied 
field ($H_z$ = 0) or rather 
large applied fields ($H_z >$ 1 T), one 
yields the demagnetized or saturated 
magnetization state of the 
entire crystal, respectively. 
One can also quench the sample 
in a small field of few milliteslas
yielding any possible initial magnetization $M_{\rm in}$. 
When the quench is fast 
($<$ 1 s), the sample's magnetization does 
not have time to relax, either by 
thermal or quantum transitions. 
This procedure yields a frozen thermal 
equilibrium distribution.
Whereas for slow cooling rates, 
the molecule spin states in the crystal might tend to
certain dipolar ordered ground state.  

\item
{\bf Modifying the initial state -- hole digging:}  
After preparing the initial state, a field $H_{\rm dig}$ is applied 
during a time $t_{\rm dig}$, called \lq digging field and digging time', 
respectively. During the digging time and depending on $H_{\rm dig}$, 
a fraction of the molecular spins tunnel (back and/or forth), {\it i.e.} 
they reverse the direction of magnetization. 
\footnote{The field sweeping
rate to apply $H_{\rm dig}$ should be fast enough to minimize the change
of the initial state during the field sweep.}

\item
{\bf Probing the final state:} 
Finally, a field $H_z^{\rm probe}$ 
is applied (Fig.~\ref{fig_dig}) 
to measure the 
short time relaxation from which 
one yields $\Gamma_{\rm short}$ 
(Eq.~\ref{M_power}) which is related to the number of spins 
which are still free for tunneling after step (2).
\end{enumerate}

The entire procedure is then repeated many times but at other 
fields $H_z^{\rm probe}$ yielding 
$\Gamma_{\rm short}(H_z,H_{\rm dig},t_{\rm dig})$
which is related to the distribution of spins 
$P(H_z,H_{\rm dig},t_{\rm dig})$
which are still free for tunneling after the hole digging.
For $t_{\rm dig}$ = 0, this method maps out the initial distribution.

%%%%%%%%%%%%%%%%%%%%%%%%%%%%%%%%
\subsection{Intermolecular dipole interaction in Fe$_8$}
\label{dipolar_Fe8} 
%%%%%%%%%%%%%%%%%%%%%%%%%%%%%%%%

\begin{figure}
\begin{center}
\includegraphics[width=.7\textwidth]{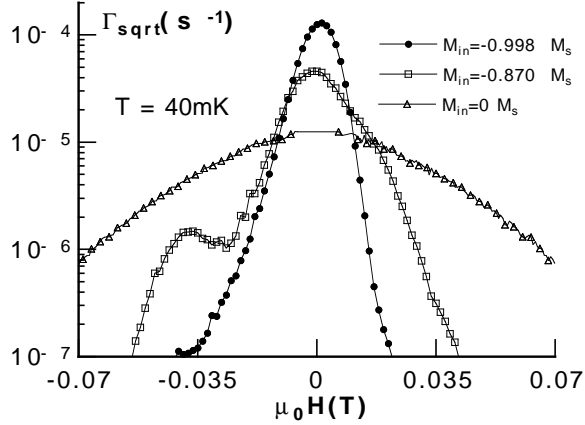}
\caption{Field dependence of the short time square-root relaxation 
rates $\Gamma_{\rm sqrt}(H_z)$ for three different values of the 
initial magnetization $M_{\rm in}$. According to Eq.~(\ref{gamma_sqrt}), the 
curves are proportional to the distribution $P(H_z)$ of magnetic 
energy bias due to local dipole field distributions in the 
sample. Note the logarithmic scale for $\Gamma_{\rm sqrt}$. The 
peaked distribution labeled $M_{\rm in}=-0.998 M_{\rm s}$ was obtained by 
saturating the sample, whereas the other distributions were obtained 
by thermal annealing. $M_{\rm in}=-0.870 M_{\rm s}$ is distorted 
by nearest neighbor lattice effects.}
\label{dist_dipole}
\end{center}
\end{figure}

We applied the hole digging method to several sample
of molecular clusters and quantum spin glasses. 
The most detailed study has been
done on the Fe$_8$ system. We found the 
predicted $\sqrt{t}$ relaxation (Eq.~(\ref{M_sqrt})) in 
experiments on fully saturated Fe$_8$ crystals \cite{Ohm98a,Ohm98b} 
and on non-saturated samples \cite{WW_PRL99}. 
Fig.~\ref{dist_dipole} displays a detailed study of the dipolar 
distributions revealing a remarkable structure 
which is due to next nearest neighbor effects~\cite{WW_PRL99}.
\footnote{The peak at $Ð0.04$ T as well as the shoulder at 
0.02 T and 0.04 T are originated by the clusters 
which have one nearest neighbor cluster 
with reversed magnetization: 
The peak at $Ð0.04$ T corresponds to the reversal 
of the neighboring cluster along the {\bf a} crystallographic 
axis, which almost coincide with the easy axis of magnetization, 
while the shoulder at 0.02 T and 0.04 T are due 
to the clusters along {\bf b} and {\bf c}.}
These results are in good agreement with 
simulations~\cite{Ohm98c,Cuccoli99}.

For a saturated initial state, the Prokof'ev--Stamp theory 
allows one to estimate 
the tunnel splitting $\Delta_{\pm S}$. 
Using Eqs. (3), (9) and (12) of \cite{Prokofev98},
and integration, we find $\int \Gamma_{\rm sqrt}d\xi = 
c \frac{\xi_0}{E_D} \frac{\Delta_{\pm S}^2}{4 \hbar}$, 
where $c$ is a constant of the order 
of unity which depends on the sample shape. 
With $E_D$ = 15 mT, 
$\xi_0$ = 0.8 mT, $c$ = 1 and $\Gamma_{\rm sqrt}$ 
from Fig.~\ref{dist_dipole} in \cite{WW_PRL99}, 
we find $\Delta_{\pm 10} = 1.2 \times 10^{-7}$ K 
which is close to the result of $\Delta_{\pm 10} = 
1.0 \times 10^{-7}$ K obtained 
by using a Landau--Zener method (Sect.~\ref{LZ}) \cite{WW_Science99}.

\begin{figure}
\begin{center}
\includegraphics[width=.7\textwidth]{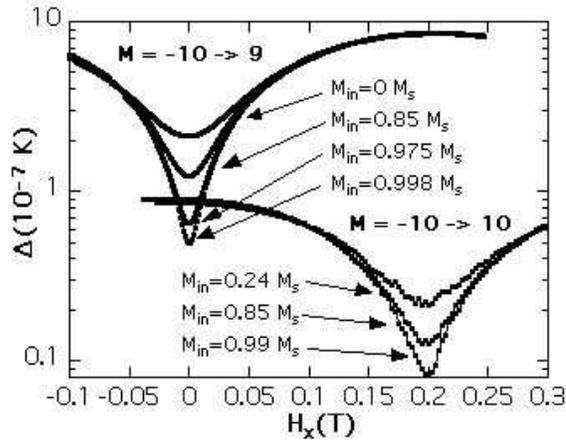}
\caption{Detailed measurement of the tunnel 
splitting $\Delta$ around a 
topological quench for the quantum transition between 
$m = -10$ and $(10 - n)$ at $\varphi = 0^{\circ}$. 
Note the strong dependence on the initial magnetization $M_{\rm in}$ 
which demonstrates the transverse dipolar interaction between 
Fe$_8$ molecular clusters \cite{WW_PRL99}.}
\label{Delta_Hx_Min}
\end{center}
\end{figure}

Whereas the hole digging method probes the longitudinal
dipolar distribution ($H_z$ direction), 
the Landau--Zener method can be used to probes the transverse
dipolar distribution by measuring the tunnel splittings 
$\Delta$ around a topological quench.
Fig.~\ref{Delta_Hx_Min} displays such a study 
for the quantum transition between $m = \pm10$, 
and $m = -10$ and $9$. 
Particular effort were made to align well the transverse 
field in direction of the hard axis. 
The initial magnetizations $0 \leq M_{\rm in} \leq M_{\rm s}$ were 
prepared by rapidly quenching 
the sample from 2~K in the present of a longitudinal 
applied field $H_z$. The quench 
takes approximately one second and thus the sample does not 
have time to relax, either by 
thermal activation or by quantum transitions, so that 
the high temperature \lq thermal 
equilibrium' spin distribution is effectively frozen in. 
For $H_z~>$~1~T, one gets an 
almost saturated magnetization state. 

The measurements of $\Delta(M_{\rm in})$ show a strong 
dependence of the minimal tunnel 
splittings on the initial magnetization (Fig.~\ref{Delta_Hx_Min}) 
which demonstrates the transverse dipolar 
interaction between Fe$_8$ molecular clusters being 
largest of $M_{\rm in} = 0$ similar 
to the longitudinal dipolar interaction (Fig.~\ref{dist_dipole}).

%%%%%%%%%%%%%%%%%%%%%%%%%%%%%%%%
\subsection{Hyperfine interaction in Fe$_8$ and Mn$_{12}$}
\label{hyperfine_Fe8} 
%%%%%%%%%%%%%%%%%%%%%%%%%%%%%%%%

The strong influence of nuclear spins on 
resonant quantum tunneling in the molecular
cluster Fe$_8$ was demonstrated for the first time~\cite{WW_PRL00} 
by comparing the relaxation rate of the
standard Fe$_8$ sample with two isotopic modified samples:
(i) $^{56}$Fe is replaced by $^{57}$Fe, and
(ii) a fraction of $^1$H is replaced by $^2$H.
By using the hole digging method, 
we measured an intrinsic broadening 
which is driven by the hyperfine fields. 
Our measurements are in good agreement 
with numerical hyperfine calculations~\cite{WW_PRL00,Rose99}.
For $T >$ 1.5 K, the influence of nuclear 
spins on the relaxation rate is
less important, suggesting that spin--phonon 
coupling dominates the relaxation rate.

Concerning Mn$_{12}$  we did \emph{not} find that the
relaxation follows the $\sqrt{t}-$ relaxation law
at low temperatures~\cite{WW_EPL99}. 
It is well known that the situation in this sample is
more complicated due to the fact that 
there are several coexisting species of
Mn$_{12}$ in any crystal, each with different relaxation times.
In Ref.~\cite{WW_EPL99} we were able to isolate 
one faster relaxing species.
The relaxation could be {\it approximately} fit to
the $\sqrt{t}-$ relaxation law, but in fact is better fit to a power
law $t^\alpha$ with $0.3 < \alpha < 0.5$ (depending on the applied field). 
We applied the hole digging method to this species, and found evidence for
intrinsic line broadening below 0.3K which we suggest comes from nuclear
spins in analogy with Fe$_8$. 
We also measured the relaxation of Mn$_{12}$ 
at higher temperature (0.04 - 5 K) and small fields ($<$ 0.1 T),
and found no evidence
for a short time $\sqrt{t}-$ relaxation. 

%%%%%%%%%%%%%%%%%%%%%%%%%%%%%%%%
\subsection{Temperature dependence of the 
Landau--Zener tunneling probability}
\label{LZ_T} 
%%%%%%%%%%%%%%%%%%%%%%%%%%%%%%%%

In this section, we present studies of the temperature 
dependence of the Landau--Zener tunneling probability $P$
yielding a deeper insight into the spin dynamics of the Fe$_8$ cluster. 
By comparing the three isotopic samples (Sect.~\ref{hyperfine_Fe8}) 
we demonstrate the influence
of nuclear spins on the tunneling mechanism 
and in particular on the
lifetime of the first excited states.
Our measurements show the need of 
a generalized Landau--Zener transition rate theory 
taking into account environmental 
effects such as hyperfine and spin--phonon coupling 
\footnote{Spin--phonon interactions mainly originate from the 
perturbation of the crystal field by lattice vibration,
which produce both a fluctuating local strain
and a fluctuating local rotation~\cite{Abragam70,Fort98,Leuenberger00a}.
It is sufficient to retain 
the lowest order terms which are quadratic with respect
to spin operators. The resulting spin--phonon Hamiltonian
contains (i) terms which commute with $S_z$ and do not
contribute to the relaxation; (ii) terms proportional
to $S_zS_+$ and $S_zS_-$; (iii) terms proportional
to $S_+^2$ and $S_-^2$. Thus, the spin--phonon 
interaction has matrix elements between states
with quantum numbers $m$ and 
$m'$ if $|m-m'|=$ 1 or 2~\cite{Fort98}.}.

All measurement so far were done in the pure quantum regime 
($T < 0.36$ K) where transition via 
excited spin levels can be neglected. 
We discuss now the temperature region 
of small thermal activation ($T <$ 1 K) 
where we should consider transition 
via excited spin levels as well~\cite{Fort98,Leuenberger00}. 

In order to measure the temperature dependence of the tunneling probability, 
we used the Landau--Zener method 
as described in Sect.~\ref{LZ} with a phenomenological 
modification of the tunneling probability $P$ (for a negative saturated 
magnetization):
\begin{equation}
P = n_{-10}P_{-10,10} + P_{th}
\label{eq_LZ_T}
\end{equation}
where $P_{-10,10}$ is given by Eq.~\ref{eq_LZ} , $n_{-10}$ is the 
Boltzmann population of the $m = -10$ spin level, and 
$P_{th}$ is the overall tunneling probability via excited spin levels.
$n_{-10} \approx 1$ for the considered temperature $T < 1$ K 
and a negative saturated magnetization of the sample.

\begin{figure}
\begin{center}
\includegraphics[width=.7\textwidth]{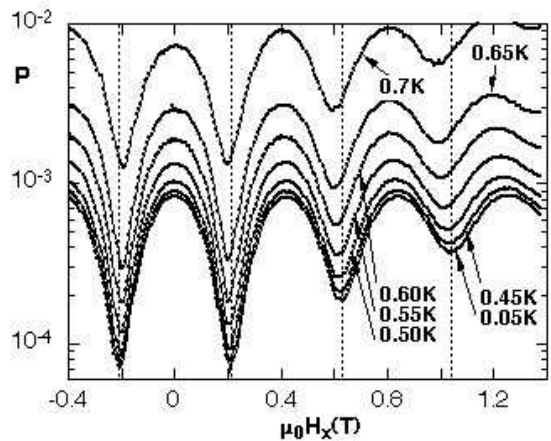}
\caption{Transverse field dependence of the 
tunneling probability $P$ at several 
temperatures, and the ground state 
tunneling probability $P_{-10,10}$ measured 
at $T = 0.05$ K and for $^{st}$Fe$_8$. 
The field sweeping rate was 0.14 T/s. 
The dotted lines indicate the minima of $P_{-10,10}$.}
\label{P_Hx_T}
\end{center}
\end{figure}

\begin{figure}
\begin{center}
\includegraphics[width=.7\textwidth]{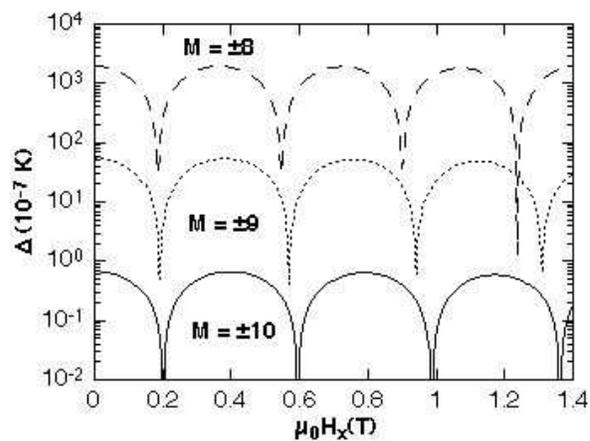}
\caption{Calculated tunnel splitting $\Delta_{m,m'}$  
as a function of the transverse field $H_x$ for 
quantum transition between $m = \pm10, \pm9$ and $\pm8$ 
(Sect.~\ref{numeric}).}
\label{Delta_10_9_8}
\end{center}
\end{figure}

\begin{figure}
\begin{center}
\includegraphics[width=.7\textwidth]{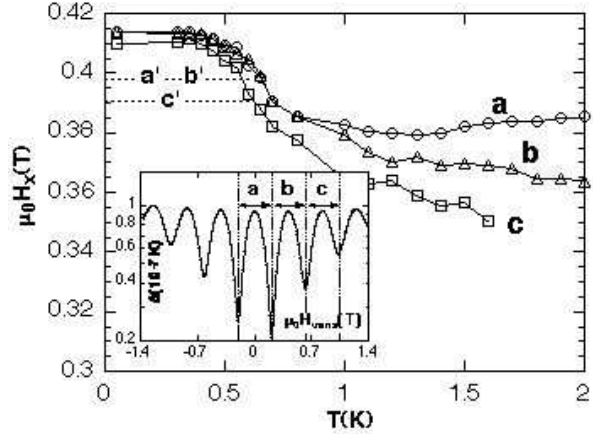}
\caption{Temperature dependence of the field between minima of 
the tunneling probability $P$ in Fig.~\ref{P_Hx_T}. 
a, b, and c are defined in the inset. 
The dotted line labeled with a', b', and c' where take from 
$P_{th}$ of Fig.~\ref{P_th_Hx_T}; see also \cite{WW_PhysB00c}.}
\label{period_T}
\end{center}
\end{figure}

Fig.~\ref{P_Hx_T} displays the measured tunneling probability $P$ 
for $^{st}$Fe$_8$ as a 
function of a transverse field $H_x$ and for several temperatures. 
The oscillation of $P$ are seen for all temperatures but the periods 
of oscillations decreases for increasing temperature 
(Fig.~\ref{period_T}). This behavior can be explained by the giant 
spin model (Eq.~\ref{eq_H_biax}) with forth order 
transverse terms (Sect.~\ref{numeric}). Indeed, the tunnel 
splittings of excited spin levels oscillate 
as a function of $H_x$ with 
decreasing periods (Fig.~\ref{Delta_10_9_8}).

\begin{figure}
\begin{center}
\includegraphics[width=.7\textwidth]{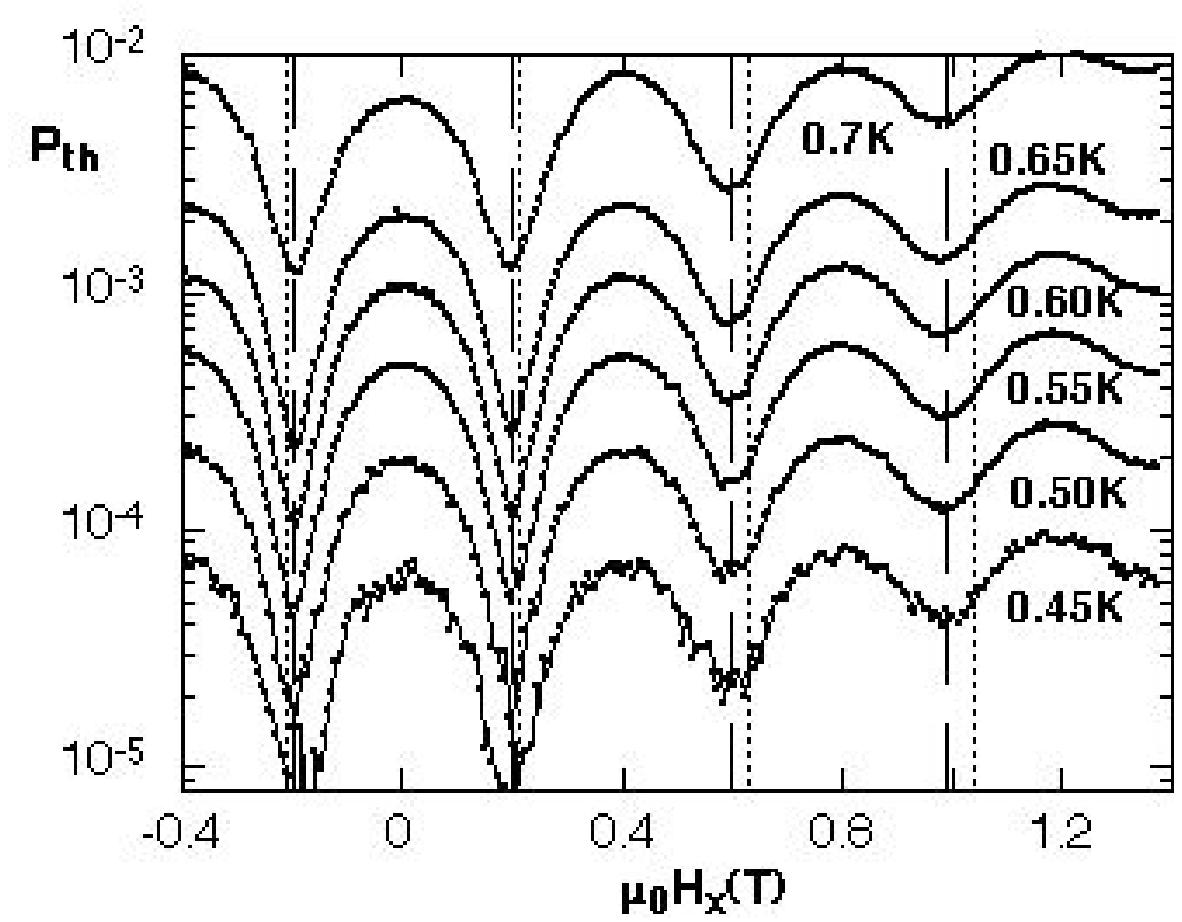}
\caption{Transverse field dependence of $P_{th}$ 
which is the difference 
between the measured tunnel probability $P$ and 
the ground state tunnel probability $n_{-10}P_{-10,10}$ 
measured at T = 0.05K (see Fig.~\ref{P_Hx_T}. 
The field sweeping rate was 0.14 T/s. The long dotted 
lines indicate the minima of $P_{th}$ whereas the short 
dotted lines indicate the minima of $P_{-10,10}$.}
\label{P_th_Hx_T}
\end{center}
\end{figure}

Fig.~\ref{P_th_Hx_T} presents the tunneling 
probability via excited spin levels
$P_{th} = P - n_{-10}P_{-10,10}$. 
Surprisingly, the periods of 
$P_{th}$ are temperature independent 
in the region $T~<$~0.7~K. 
This suggests that only transitions via 
excited levels $m = \pm9$ are 
important in this temperature regime. 
This statement is confirmed by 
the following estimation~\cite{WW_EPL00}.

Using Eq.~\ref{eq_LZ}, typical field sweeping rates of 0.1 T/s, and 
tunnel splittings from Fig.~\ref{Delta_10_9_8}, 
one easily finds that the Landau--Zener 
tunneling probability of excited levels are $P_{-m,m} \approx 1$ for 
$m < 10$ and $\vec{H} \approx 0$. This means that the relaxation 
rates via excited levels are 
mainly governed by the lifetime of the excited levels and the time 
$\tau_{res,m}$ during which these levels are in resonance. 
The later can be estimated by 
\begin{equation}
\tau_{res,m} = \frac{\Delta_{-m,m}}{g \mu_{\rm B} m \mu_0 dH_z/dt}.
\label{eq_tau_res}
\end{equation}

\begin{figure}
\begin{center}
\includegraphics[width=.7\textwidth]{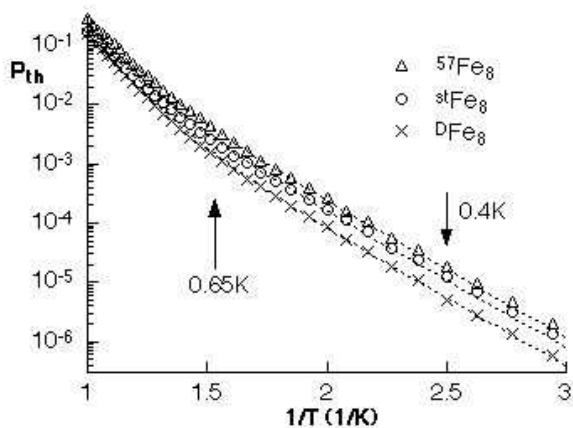}
\caption{Temperature dependencies of $P_{th}$ for $H_x$ = 0 
for three Fe$_8$ samples. The field sweeping rate was 0.14 T/s. 
The dotted lines are fits of the data using Eq.~\ref{eq_P_th_9}.}
\label{P_th_T}
\end{center}
\end{figure}

The probability for a spin to pass into the excited level $m$ 
can be estimated by $\tau_m^{-1}e^{-E_{10,m}/k_{\rm B}T}$, where $E_{10,m}$ 
is the energy gap between the levels $10$ and $m$, and $\tau_m$ is 
the lifetime of the excited level $m$. One gets:
\begin{equation}
P_{th} \approx \sum_{m = 9, 8} \frac{\tau_{res,m}}{\tau_m}e^{-E_{10,m}/k_{\rm B}T} 
\approx \sum_{m = 9, 8} \frac{\Delta_{-m,m}}{\tau_m g \mu_{\rm B} 
m \mu_0 dH_z/dt}e^{-E_{10,m}/k_{\rm B}T}.
\label{eq_P_th_9}
\end{equation}
Note that this estimation neglects higher excited levels with $|m| < 8$.
\footnote{The probability of phonon induced transitions with $|\Delta m| > 2$
are very small \cite{Villain97,Fort98,Leuenberger00}. 
Also the Boltzmann factor $e^{-E_{10,m}/k_{\rm B}T}$
is very small for $m < 8$ and $T < 1$ K.}
Fig.~\ref{P_th_T} displays the measured $P_{th}$ for the 
three isotopic Fe$_8$ samples. 
For 0.4 K $< T <$ 1 K we fitted Eq.~\ref{eq_P_th_9} 
to the data leaving {\it only}
the level lifetimes $\tau_9$ and $\tau_8$ as adjustable parameters. All other
parameters are calculated using the parameters in Sect.~\ref{numeric}. 
We obtain $\tau_9 = 1.0, 0.5,$ and $0.3 \times10^{-6}$s, and 
$\tau_8 = 0.7, 0.5,$ and $0.4 \times10^{-7}$s 
for $^{D}$Fe$_8$, $^{st}$Fe$_8$, and $^{57}$Fe$_8$, respectively. 
These results evidence that only the first excited 
level have to be considered for 0.4 K $< T <$ 0.7 K. 
Indeed, the second term of the summation in 
Eq.~\ref{eq_P_th_9} is negligible in this temperature interval.
It is interesting to note that this finding 
is in contrast to hysteresis loop 
measurements on Mn$_{12}$ \cite{Kent00,Kent00b} 
which suggested an abrupt 
transition between thermal assisted and 
pure quantum tunneling \cite{Garanin99}.
Furthermore, our result shows clearly the influence of
nuclear spins which seem to decrease the level lifetimes $\tau_m$,
{\it i.e.} to increase dissipative effects.

The nuclear magnetic moment and not the mass of the nuclei 
seems to have the major effect on the dynamics of the
magnetization. In fact the mass is increased in both isotopically modified
samples whereas the effect on the relaxation rate is opposite. 
On the other hand ac susceptibility measurements at
$T~>$~1.5~K showed no clear difference between
the three samples suggesting that
above this temperature, where the relaxation is predominately due to
spin--phonon coupling~\cite{Fort98,Leuenberger00a}, the role of the nuclear
spins is less important. Although the increased mass of the isotopes
changes the spin--phonon coupling, this effect seems to be small.

We can also exclude that the change of mass for the three isotopic samples
has induced a significant change in the magnetic anisotropy of the
clusters. In fact the measurements below $T <$ 0.35~K,
where spin--phonon coupling is negligible, have shown that
(i) relative positions of the resonances as a function of the
longitudinal field $H_z$ are unchanged~\footnote{
We observed a small shift of the resonances of the 
order of magnitude of 1~mT, positive for $^{57}$Fe$_8$ and negative for
$^{\rm D}$Fe$_8$ ($M_{\rm init}  = -M_{\rm s}$). 
This can also be attributed to the modified 
hyperfine fields. However, a quantitative measurement is complicated 
by the fact that it is impossible to have two crystal with exactly 
the same shape, {\it i.e.} the same internal fields.
}, and (ii) all three
samples have the same period of oscillation of $\Delta$ as a function of
the transverse field $H_x$~\cite{WW_Science99}, a period which is
very sensitive to any change of the anisotropy constants.

%%%%%%%%%%%%%%%%%%%%%%%%%%%%%%%%
\subsection{Conclusion on molecular magnets}
\label{conclusion_mol} 
%%%%%%%%%%%%%%%%%%%%%%%%%%%%%%%%
%
In conclusion, we presented detailed measurements which 
demonstrated that molecular magnets 
offer an unique opportunity to explore the 
quantum dynamics of a large but finite spin.
\footnote{Molecules with small spin have also
been studied. For example, time resolved magnetization 
measurements were 
performed on a spin $1/2$ molecular complex,
so-called V$_{15}$~\cite{ChiorescuJAP00}. 
Despite the absence of a barrier, 
magnetic hysteresis is observed over a time scale of
several seconds. A detailed analysis in terms of a 
dissipative two-level model has been given, in which fluctuations
and splittings are of the same energy. Spin--phonon 
coupling leads to long relaxation times and to
a particular \lq butterfly' hysteresis 
loop~\cite{ChiorescuPRL00,Dobrovitski00}.}
We focused our discussion on the Fe$_8$ molecular magnet
because it is the first system where studies in the pure
quantum regime were possible. The tunneling in this system
is remarkable because it does not show up at lowest 
order of perturbation theory.

What remains still debated is the possibility of observing 
quantum coherence between states of opposite magnetization. 
Dipole-dipole and hyperfine interactions are source of decoherence. 
In other words when a spin has tunneled through the barrier, 
it experiences a huge modification of its environment 
(hyperfine and dipolar) which prohibits the back tunneling. 
Prokof'ev and Stamp suggested  three possible strategies 
to suppress the decoherence~\cite{Prokofev95}. 
(i) Choosing a system were the 
NMR frequencies far exceed the tunnel frequencies making 
any coupling impossible. (ii) Isotopically purify the sample 
to remove all nuclear spins. (iii) Applying a transverse field 
to increase the tunnel rate to frequencies much larger than 
hyperfine field fluctuations. All three strategies are difficult 
to realize. However, some authors tried to realize the last 
one by performing EPR experiments in the presence of a large 
transverse field~\cite{Barco00}. 
Absorption of radio-frequency electromagnetic 
fields were observed which might be due to induced transitions 
near the tunnel splitting. However, no experiments showed 
the oscillatory behavior in the time domain which might be 
evidenced by a spin echo type of experiment. 

Concerning the perspectives of the field of single molecule magnets 
we expect that chemistry is going to play a major role through 
the synthesis of novel larger spin clusters with strong anisotropy. 
We want to stress
that there are already many other molecular magnets
(see for instance~\cite{Caneschi99,Aubin98,Price99,Yoo_Jae00}) 
which are possible model systems.
We believe that more sophisticated 
theories are needed which describe the
dephasing effects of the environment 
onto the quantum system.
These studies are important for sing the quantum character
of molecular clusters for applications like
\lq Quantum Computers'. 
The first implementation of Grover's algorithm with 
molecular magnets has been proposed~\cite{Leuenberger01}

%%%%%%%%%%%%%%%%%%%%%%%%%%%%%%%%
\section{Quantum tunneling of magnetization 
in individual single-domain nanoparticles}
%%%%%%%%%%%%%%%%%%%%%%%%%%%%%%%%

The following sections focuses on 
Magnetic Quantum Tunneling (MQT) 
studied in individual nanoparticles 
or nanowires where 
the complications due to distributions of particle size, 
shape, and so on are avoided. The 
experimental evidence for MQT in a single-domain particle 
or in assemblies of particles is 
still a controversial subject. We shall therefore 
concentrate on the necessary experimental 
conditions for MQT and review some experimental results
which suggest that quantum effects might even
be important in nanoparticles with $S = 10^5$ or larger.
We start by reviewing some 
important predictions concerning MQT in a single-domain particle.

%%%%%%%%%%%%%%%%%%%%%%%%%%%%%%%%
\subsection{Magnetic Quantum Tunneling in nanoparticles}
%%%%%%%%%%%%%%%%%%%%%%%%%%%%%%%%

On the theoretical side, it has been shown 
that in small magnetic particles, a 
large number of spins coupled by strong exchange interaction, 
can tunnel through the 
energy barrier created by magnetic anisotropy. 
It has been proposed that there is a 
characteristic crossover temperature $T_{\rm c}$ 
below which the 
escape of the magnetization from a metastable state 
is dominated by quantum barrier 
transitions, rather than by thermal over barrier 
activation. Above $T_{\rm c}$ the escape rate is 
given by thermal over barrier activation 
(Sect.~\ref{chap_reversal_clas}).

In order to compare experiment with theory, predictions 
of the crossover temperature 
$T_{\rm c}$ and the escape rate $\Gamma_{\rm QT}$ in 
the quantum regime are relevant. Both 
variables should be expressed as a function of 
parameters which can be changed 
experimentally. Typical parameters are the number of 
spins $S$, effective anisotropy 
constants, applied field strength and direction, 
coupling to the environment (dissipation) etc.
Many theoretical papers have been published during the last few 
years~\cite{QTM94}. We discuss here a result specially 
adapted for single particle 
measurements, which concerns the field dependence of 
the crossover temperature $T_{\rm c}$.

\begin{figure}[t]
\begin{center}\leavevmode
\includegraphics[width=0.6\linewidth]{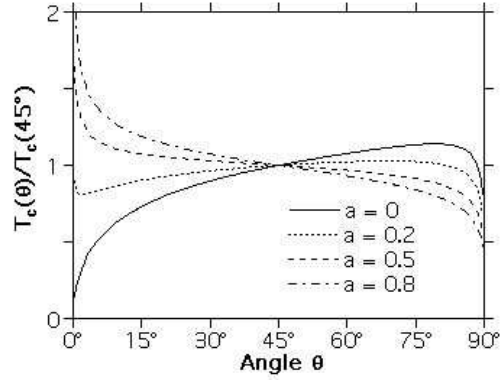}
\caption{Normalized crossover temperature $T_{\rm c}$ as given 
by (\ref{eq_Tc_theo}), and for 
several values of the ratio $a = H_{\perp}/H_{\parallel}$}
\label{fig_Tc_theo}
\end{center}
\end{figure}

The crossover temperature $T_{\rm c}$ can be defined as 
the temperature where the quantum 
switching rate equals the thermal one. 
The case of a magnetic particle, as a 
function of the applied field direction, has been considered 
by several authors~\cite{Zaslavskii90,Miguel96,Kim97}. 
We have chosen the result for a particle with biaxial 
anisotropy as the 
effective anisotropy of most particles can be 
approximately described by strong uniaxial 
and weak transverse anisotropy. 
The result due to Kim can be written in the following form~\cite{Kim97}: 

\begin{eqnarray}
	T_{\rm c}(\theta) & \sim & \mu_0 H_{\parallel} \varepsilon^{1/4}
	\sqrt{1+a\left(1+\mid \cos\theta \mid^{2/3}\right)}
	\frac {\mid \cos\theta \mid^{1/6}}{1+\mid \cos\theta \mid^{2/3}} \;,
\label{eq_Tc_theo}
\end{eqnarray}
where $\mu_0 H_{\parallel} = K_{\parallel}/M_{\rm S}$ and 
$\mu_0 H_{\perp} = K_{\perp}/M_{\rm s}$ are the parallel and transverse 
anisotropy fields given in Tesla, $K_{\parallel}$ and $K_{\perp}$ 
are the parallel and transverse 
anisotropy constants of the biaxial anisotropy, 
$\theta$ is the angle between the easy axis of 
magnetization and the direction of the applied field, 
and $\varepsilon = (1 - H/H_{sw}^0)$. 
Equation (\ref{eq_Tc_theo}) is valid for 
any ratio $a = H_{\perp}/H_{\parallel}$. 
The proportionality coefficient of (\ref{eq_Tc_theo}) 
is of the order of unity ($T_{\rm c}$ is in units 
of Kelvin) and depends on the approach used for calculation~\cite{Kim97}. 
Equation (\ref{eq_Tc_theo}) is plotted in Fig.~\ref{fig_Tc_theo} 
for several values of the ratio $a$. It is valid in the 
range $\sqrt{\varepsilon} < \theta < \pi/2 - \sqrt{\varepsilon}$. 

The most interesting feature which may be drawn from 
(\ref{eq_Tc_theo}) is that the 
crossover temperature is tunable using the external field 
strength and direction (Fig.~\ref{fig_Tc_theo}) because 
the tunneling probability is increased by the 
transverse component of the applied field. 
Although at high transverse fields, $T_{\rm c}$ decreases again due to a 
broadening of the anisotropy barrier. 
Therefore, quantum tunneling experiments should 
always include studies of angular dependencies. 
When the effective magnetic anisotropy 
of the particle is known, MQT theories give clear 
predictions with no fitting parameters.
MQT could also be studied as a function of the 
effective magnetic anisotropy. In 
practice, it is well known for single particle 
measurements that each particle is somewhat 
different. Therefore, the effective magnetic 
anisotropy has to be determined for each 
particle (Sect.~\ref{St_W_general}).

Finally, it is important to note that most 
of the MQT theories neglect damping 
mechanisms. We discussed the case of ohmic 
damping in Sect.~\ref{Neel_Brown}~\cite{Coffey98a} which is 
the simplest form of damping. More complicated 
damping mechanisms 
might play an important role. 
We expect more theoretical work on this in future.

%%%%%%%%%%%%%%%%%%%%%%%%%%%%%%%%
\subsection{Magnetization measurements of individual single-domain 
nanoparticles and wires at very low temperatures}
%%%%%%%%%%%%%%%%%%%%%%%%%%%%%%%%

In order to avoid the complications due to distributions of 
particle size, shape, and so on, some groups have tried to study the 
temperature and field dependence of magnetization reversal of 
individual magnetic particles or wires. 
Most of the recent studies were done using 
Magnetic Force Microscopy at room temperature. 
Low temperature investigations were 
mainly performed via resistance measurements (Sect.~\ref{mes_techniques}).

The first magnetization measurements of individual 
single-domain nanoparticles at low 
temperature (0.1~K $< T <$ 6~K) were presented by 
Wernsdorfer {\it et al.}~\cite{WW_JMMM95a}. 
The detector (a Nb micro-bridge-DC-SQUID, 
see Sect.~\ref{micro_SQUID}) 
and the particles studied (ellipses with axes 
between 50 and 1000~nm and 
thickness between 5 and 50~nm) were fabricated 
using electron beam lithography. 
Electrodeposited wires (with diameters ranging from 
40 to 100~nm and lengths up to 5000 nm) 
were also studied~\cite{WW_PRL96_Ni,WW_PRB97_Ni}. 
Waiting time and switching 
field measurements (Sect.~\ref{methods_NB}) showed that the magnetization 
reversal of these particles and wires 
results from a single thermally activated domain wall nucleation, 
followed by a fast wall 
propagation reversing the particle's magnetization. 
For nanocrystalline Co particles of 
about 50~nm and below 1~K, a flattening of the 
temperature dependence of the mean 
switching field was observed which could not be 
explained by thermal activation. These 
results were discussed in the context of MQT. 
However, the width of the 
switching field distribution and the probability 
of switching are in disagreement with such 
a model because nucleation is very sensitive to 
factors like surface defects, surface 
oxidation and perhaps nuclear spins. The fine 
structure of pre-reversal magnetization 
states is then governed by a multivalley energy 
landscape (in a few cases distinct 
magnetization reversal paths were effectively 
observed~\cite{WW_JMMM95b}) and the 
dynamics of reversal occurs via a complex path in configuration space. 

Coppinger {\it et al.}~\cite{Coppinger95} used telegraph noise 
spectroscopy to investigate 
two-level fluctuations (TLF) observed in the conductance 
of a sample containing self-assembled ErAs quantum wires 
and dots in a semi-insulating GaAs matrix. They showed 
that the TLF could be related to two possible magnetic 
states of a ErAs cluster and that the 
energy difference between the two states was a linear 
function of the magnetic field. They 
deduced that the ErAs cluster should contain a few tens 
of Er atoms. At temperatures 
between 0.35~K and 1~K, the associated switching rates 
of the TLF were thermally activated, 
whilst below 0.35 K the switching rate became 
temperature independent. Tunneling of 
the magnetization was proposed in order to explain 
the observed behavior. 

Some open questions remain: 
what is the object which is really probed by TLF? 
If this is a single ErAs particle, as assumed by the authors, 
the switching probability  should be an 
exponential function of time. The pre-exponential factor 
$\tau_0^{-1}$ (sometimes called attempt frequency) was 
found to lie between $10^3$ and $10^6$~s$^{-1}$ whereas 
expected values are between $10^9$ and $10^{12}$~s$^{-1}$. 
Why must one apply fields of about 2~T in order to 
measure two-level fluctuations which should be 
expected near zero field? What is the influence of the 
measurement technique on the sample? 

By measuring the electrical resistance of 
isolated Ni wires with diameters between 20 and 
40~nm, Hong and Giordano studied the motion of 
magnetic domain walls~\cite{Hong95}. 
Because of surface roughness and oxidation, 
the domain walls of a single wire are trapped 
at pinning centers. The pinning barrier 
decreases with an increase in the magnetic field. 
When the barrier is sufficiently small, 
thermally activated escape of the wall occurs. This 
is a stochastic process which can be characterized 
by a switching (depinning) field 
distribution. A flattening of the temperature 
dependence of the mean switching field and a 
saturation of the width of the switching field 
distribution (rms. deviation $\sigma$) were 
observed below about 5~K. The authors proposed that 
a domain wall escapes from its pinning site 
by thermal activation at high temperatures and by quantum tunneling 
below $T_{\rm c} \sim$ 5~K.

These measurements pose several questions: 
what is the origin of the pinning center 
which may be related to surface roughness, impurities, 
oxidation and so on? The sweeping rate 
dependence of the depinning field, as well as the 
depinning probability, could not be 
measured even in the thermally activated regime. 
Therefore, it was not possible to check 
the validity of the N\'eel--Brown model
~\cite{Neel49a,Neel49b,Brown59,Brown63a,Brown63b} 
or to compare measured and predicted 
rms. deviations $\sigma$. Finally, a crossover 
temperature $T_{\rm c}$ of about 5~K is three 
orders of magnitude 
higher than $T_{\rm c}$ predicted by current theories.

Later, Wernsdorfer et al. published results obtained 
on nanoparticles synthesized by 
arc discharge, with dimensions between 
10 and 30~nm~\cite{WW_PRL97_Co}. These particles 
were single crystalline, and the surface roughness 
was about two atomic layers. Their 
measurements showed for the first time that the 
magnetization reversal of a ferromagnetic 
nanoparticle of good quality can be described by 
thermal activation over a single-energy 
barrier as proposed by N\'eel and 
Brown~\cite{Neel49a,Neel49b,Brown59,Brown63a,Brown63b} 
(see Sect.~\ref{Exp_NB}). 
The activation volume, which is the 
volume of magnetization overcoming the barrier, 
was very close to the particle volume, predicted for 
magnetization reversal by uniform rotation. No quantum effects 
were found down to 0.2~K. This was not surprising because the 
predicted cross-over temperature is $T_{\rm c} \sim$ 0.02~K. 
The results of Wernsdorfer et al. constitute the preconditions for the 
experimental observation of MQT of magnetization on a single particle.

\begin{figure}[t]
\begin{center}\leavevmode
\includegraphics[width=0.6\linewidth]{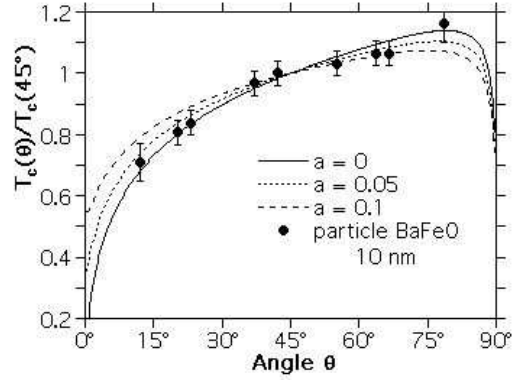}
\caption{Angular dependence of the crossover 
temperature $T_{\rm c}$ for a 10~nm BaFe$_{10.4}$Co$_{0.8}$Ti$_{0.8}$O$_{19}$ 
particle with $S \approx 10^5$. The lines are given by (\ref{eq_Tc_theo}) 
for different values of the ratio $a = H_{\perp}/H_{\parallel}$. The experimental data are 
normalized by $T_{\rm c}(45^{\circ})$ = 0.31~K}
\label{fig_Tc_exp}
\end{center}
\end{figure}

Just as the results obtained with Co 
nanoparticles~\cite{WW_PRL97_Co}, a quantitative 
agreement with the N\'eel--Brown model of 
magnetization reversal was found on 
BaFe$_{12-2x}$Co$_x$Ti$_x$O$_{19}$ 
nanoparticles($0 < x < 1$)~\cite{WW_PRL97_BaFeO}, which we will call
BaFeO, in the size range of 10 -- 20~nm. 
However, strong deviations from this model were 
evidenced for the smallest particles containing 
about $10^5 \mu_{\rm B}$ and for temperatures 
below 0.4~K. These deviations are in good agreement 
with the theory of macroscopic 
quantum tunneling of magnetization. 
Indeed, the measured angular dependence of $T_{\rm c}(\theta)$ is in 
excellent agreement with the 
prediction given by (\ref{eq_Tc_theo}) (Fig.~\ref{fig_Tc_exp}). 
The normalization value $T_{\rm c}(45^{\circ})$~= 0.31~K compares well with the 
theoretical value of about 0.2~K.

Although the above measurements are in good agreement 
with MQT theory, we should 
not forget that MQT is based on several strong 
assumptions. Among them, there is the 
assumption of a giant spin, {\it i.e.}, all magnetic 
moments in the particle are rigidly coupled 
together by strong exchange interaction. 
This approximation might be good in the 
temperature range where thermal activation is dominant 
but is it not yet clear if this 
can be made for very low energy barriers. 
Future measurements might give us the answer.

%%%%%%%%%%%%%%%%%%%%%%%%%%%%%%%%
\subsection{Quantization of the magnetization}
%%%%%%%%%%%%%%%%%%%%%%%%%%%%%%%%

In order to give a definite proof that MQT 
can occur in a magnetic nanoparticle 
we propose to surge for the energy level 
quantization of its collective spin state. 
This was recently evidenced in molecular 
cluster like Fe$_8$ having a collective 
spin state $S = 10$ (Sect.~\ref{mol_cluster}).
In the case of the BaFeO particles with $S \approx 
10^5$~\cite{WW_PRL97_BaFeO}, 
the field separation associated with level quantization is 
rather small: ${\rm \Delta}H \approx H_{\rm a}/ 2 S 
\approx$ 0.002~mT where $H_{\rm a}$ is the anisotropy field. 
However, for a 3 nm Co cluster with $S \approx 10^3$
the field separation ${\rm \Delta}H = H_{\rm a}/ 2 S \sim$ 0.2~mT
might be large enough to be measurable.

\begin{figure}
\begin{center}
\includegraphics[width=.7\textwidth]{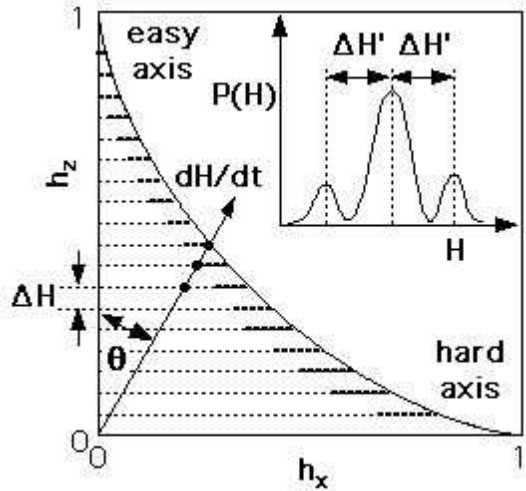}
\caption{Schematic view of the resonance fields of a giant spin $S$.
The continuous line is the classical switching fields of Stoner--Wohlfarth
(Sect.~\ref{St_W}). The inset presents schematically a switching field 
histogram with ${\rm \Delta}H' \approx \frac{H_{\rm a}}{2S}\frac{1}{\cos{\theta}}$ }
\label{ast_St_W_QT}
\end{center}
\end{figure}

Fig.~\ref{ast_St_W_QT} displays schematically the field values of 
resonances between quantum states of $S$. When the applied field is 
ramped in a certain direction, the resonance might occur for fields 
$H_{\rm res} \approx n \times \frac{H_{\rm a}}{2S}\frac{1}{\cos{\theta}}$,
with $n$ = 1, 2, 3, \ldots. $\theta$ is the angle between the applied 
field and the easy axis of magnetization. For large spins $S$, 
tunneling might be observable only for fields which are close to the 
classical switching field (Fig.~\ref{ast_St_W_QT}).
The resonance fields could be evidenced by measuring switching field 
distributions (inset of Fig.~\ref{ast_St_W_QT}) 
as a function of the angle $\theta$. 

\begin{figure}
\begin{center}
\includegraphics[width=.7\textwidth]{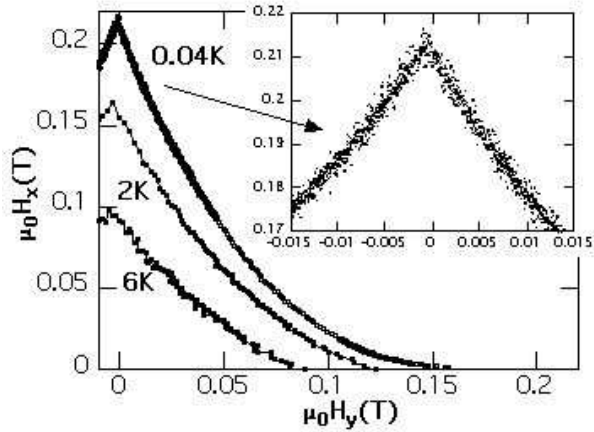}
\caption{Angular dependence of the switching field of a 3 nm Fe 
cluster with $S \approx$ 800. The inset presents detailed 
measurements near the easy axis of magnetization.}
\label{ast_QT_Fe}
\end{center}
\end{figure}

Such a study is presented in Fig.~\ref{ast_QT_Fe} for a 3 nm Fe 
cluster with $S \approx$ 800. The estimated field separation 
${\rm \Delta}H = H_{\rm a}/ 2 S$ is about 0.1~mT
whereas the width of the switching field distribution is about
ten times larger. We observed sometimes
a small periodic fine structure which is close to the expected 
${\rm \Delta}H$. However this fine structure always disappeared when 
averaging over more measurements.
A possible origin might be hyperfine couplings which broaden the 
energy levels (Sect.~\ref{hyperfine_Fe8}) 
leading to a complete overlap of adjacent energy levels. 
It is also important to mention that the switching field distributions 
were temperature independent for 0.04~K $< T <$ 0.2~K.

In some cases, we observed huge variations of 
the switching field (Figs.~\ref{Fe_bizarre} and \ref{Fe_bizarre_tele})
which might be due to exchange bias of frustrated spin 
configurations. However, quantum effects are not completely excluded.

Future measurements should focus 
on the level quantization of collective spin states 
of $S = 10^2$.

%%%%%%%%%%%%%%%%%%%%%%%%%%%%%%%%
\title[Summary and conclusion]
{Summary and conclusion}
\label{conclusion}
%%%%%%%%%%%%%%%%%%%%%%%%%%%%%%%%

Nanometer-sized magnetic particles have generated continuous 
interest as the study of their properties has proved 
to be scientifically and technologically very challenging.
In the last few years, new fabrication techniques 
have led to the possibility of making 
small objects with the required structural and chemical 
qualities. In order to study these 
objects new local measuring techniques were developed 
such as magnetic force microscopy, 
magnetometry based on micro-Hall probes 
or micro-SQUIDs. This led to a new 
understanding of the magnetic behavior of nanoparticles. 

In this article we reviewed the most important 
theories and experimental results 
concerning the magnetization reversal of 
single-domain particles and clusters. 
Special emphasis is laid 
on single particle measurements avoiding complications 
due to distributions of particle 
size, shape etc. Measurements on particle assemblies 
has been reviewed in~\cite{Dormann97}. 
We mainly discuss the low temperature 
regime in order to avoid spin excitations.

Sect.~\ref{chap_mes} reviews briefly the commonly 
used measuring techniques. Among them, 
electrical transport measurements, Hall probes and 
micro-SQUID techniques seem to be the 
most convenient techniques for low temperature measurements. 

Sect.~\ref{chap_reversal_T_0} discusses the 
mechanisms of magnetization reversal in single domain 
particles at zero Kelvin.
For extremely small particles, the 
magnetization should 
reverse by uniform rotation of magnetization (Sec.~\ref{St_W}). 
For somewhat larger particles, non uniform
reversal mode occur like the curling mode. 
(Sec.~\ref{non_uniform}). 
For even larger particles, 
magnetization reversal occurs via a domain wall 
nucleation process starting in a rather 
small volume of the particle (Sec.~\ref{nucleation}). 

The influence of temperature on the magnetization reversal 
is reported in Sect.~\ref{chap_reversal_clas}. 
We discuss in detail the N\'eel, Brown and 
Coffey's theory of magnetization reversal by 
thermal activation (Sect.~\ref{chap_reversal_clas}).

Finally, 
Sect.~\ref{chap_MQT} shows that for very small systems or 
very low temperature, magnetization can reverse via 
quantum tunneling.
The boundary between classical and 
quantum physics has become a very attractive 
field of research. 
This section discusses detailed measurements which 
demonstrated that molecular magnets 
offer an unique opportunity to explore the 
quantum dynamics of a large but finite spin.
The discussion is focused on the Fe$_8$ molecular magnet
with $S = 10$ because it is the first 
system where studies in the pure
quantum regime were possible. 
We showed that the understanding 
of the environmental decoherence
is one of the most important issues, in particular
for future applications of quantum devices.
We then discussed tunneling in nanoparticles and showed how
one might give a definite proof of 
their quantum character at low temperature. 

In conclusion, the understanding of the magnetization
reversal in nano-structures requires the knowledge of many
physical phenomena, and are therefore particularly
interesting for the development of new fundamental 
theories of magnetism and in modeling new 
magnetic materials for permanent magnets or high 
density recording. Using the quantum character
of nano-structures for applications like
\lq Quantum Computers' will be one of the major
concerns of the next decades.

%%%%%%%%%%%%%%%%%%%%%%%%%%%%%
\begin{acknowledgments}
%%%%%%%%%%%%%%%%%%%%%%%%%%%%%
The author is indebted to A. Benoit, E. Bonet Orozco, I. Chiorescu,
K. Hasselbach, M. Jamet, D. Mailly, B. Pannetier and C. Thirion 
for their experimental contributions 
and the developement of the micro-SQUID technology.
The author acknowledge the collaborations with 
J.-Ph. Ansermet, B. Barbara, A. Caneschi, A. Cornia, N. Demoncy, 
B. Doudin, V. Dupuis, O. Fruchart, D. Gatteschi,   
O. Kubo, J.-P. Nozieres, H. Pascard, C. Paulsen, 
A. Perez, C. Sangregorio and R. Sessoli.
He is also indebted to W.T. Coffey, A. Garg, N. Prokof'ev, 
P. Stamp, A. Thiaville, I. Tupitsyn and J. Villain for many 
fruitful and motivating discussions. 
Finally,  O. Fruchart is specially acknowledged for his
critical review of the manuscript.
This work has been supported by
CNRS, DRET, MASSDOTS and Rh\^ one-Alpes.
%%%%%%%%%%%%%%%%%%%%%%%%%%%%%
\end{acknowledgments}
%%%%%%%%%%%%%%%%%%%%%%%%%%%%%

\end{document}